\documentclass{article}

\usepackage{arxiv}

\usepackage[utf8]{inputenc} 
\usepackage[T1]{fontenc}    
\usepackage{hyperref}       
\usepackage{url}            
\usepackage{booktabs}       
\usepackage{amsfonts}       
\usepackage{nicefrac}       
\usepackage{microtype}      
\usepackage{lipsum}
\usepackage{appendix}

	\usepackage{amsmath,tipa,textcomp,tikz,amssymb,amsfonts,amsthm,caption,subcaption,upgreek,url,forest}
	\usepackage{mathtools}
\usepackage{xfrac,cite}
	\usepackage{blkarray, bigstrut} 
	\usepackage{graphicx,lipsum}
\usepackage{blindtext}
\usepackage{wasysym}
\usepackage[]{algorithm2e}
\graphicspath{ {images/} }
	\begin{document}
	\title{A Comprehensive Survey of Navigation Systems for the Visual Impaired}
	\author{
  Roya Norouzi Kandalan\\
  Department of Electrical Engineering\\
  University of North Texas\\
  Denton, TX, 76207 \\
  \texttt{royanorouzi@my.unt.edu} \\
   \And
Kamesh Namuduri\\
  Department of Electrical Engineering\\
  University of North Texas\\
  Denton, TX, 76207 \\
  \texttt{kamesh.namuduri@unt.edu} \\}

	\maketitle

\begin{abstract}
Sight is essential for humans to navigate their surrounding environment independently. Tasks that are simple for the sighted are often close to impossible for the visually impaired. Throughout the years, many researchers dedicated their time and efforts to design and implement technologies and devices that can help the visually impaired to navigate in unknown areas independently. 
A navigation assistive system for the visually impaired is built on multiple components of which localization, navigation, obstacle avoidance, and human-machine interaction form the essential components. Individual components have been explored extensively in the literature and it is a daunting task to review it in its entirety. In this paper, a compact but comprehensive collection of available methods to build each component has been provided. The discussed methods may not have been implemented for a navigation assistive technology for visually impaired directly but they can be employed to build an individual building block. Given that every approach has its own challenges, methods to combat those challenges have been presented. The purpose of this paper is to provide researchers with a shortcut to review most of the available methods to build each component, provide them with the essential knowledge to understand the nuances in each component and give them a baseline to start with. 
	\end{abstract}
	\begin{keywords}\\
	Visually impaired, Localization, Navigation, Obstacle Avoidance 
	\end{keywords}

\section{Introduction}

According to the World Health Organization (WHO) \cite{whoo}, 253 million people struggle with vision impairment globally. Ninety percent of visually impaired people live in developing countries, which reduces the chance of getting adequate treatment and support. When visual impairment is compounded with poverty, an individual often fully depends on their family for survival. This poses a challenge for the blind to lead an independent life. Although  visually impaired people living in developed countries have access to many resources to assist them, they can not carry out all daily life tasks  independently. Tasks such as shopping for groceries in a nearby supermarket or finding a painkiller to alleviate a headache are onerous for many visually impaired individuals.
	
One needs to know their current location and destination to navigate in an environment. Navigation and path finding are complex tasks and obstacle avoidance is essential along the way, especially in dynamic environments. In a supermarket or other indoor environments, finding the way is even more arduous due to limitations of the Global Positioning System (GPS). GPS signals are weaker indoors as buildings attenuate the signal strength \cite {5_8}. Consequently, alternative approaches are necessary for addressing indoor localization. Although there are techniques available for indoor position estimation, finding a specific item in an aisle remains a challenge for a visually impaired person. Reading the dosage instructions or side effects of medicine requires the help of a sighted person even for a partially visually impaired individual. 

More than eighty percent of people who suffer from blindness lost their vision due to a preventable or curable disease. While medical scientists are trying to provide vaccines for diseases causing blindness, assistive technologies are needed to make life easier for those who can not see. Fortunately, assistive technologies for the blind have been explored for many years.

 In 1990, Passini et al. made an experiment with eight groups of sighted, partially sighted, and visually impaired individuals of different sexes, ages, and education and they reported the results in \cite{90_1} . The experiment aimed to test the individuals through basic way-finding tasks. Each task corresponded to a certain spatio-cognitive function. The test was performed in a layout with different levels of complexity of the path network. The layout designed not only to provide a variety of difficulty levels but also to limit the irrelevant factors which might impact the test result. Not surprisingly, the experiment's results showed that those who were suffering from visual impairment spent more time to complete the tasks. The results also confirmed that age and education have an impact on performance. Passini demonstrated the need for indoor navigation assistive technologies for visually impaired individuals and motivated many researchers to seek solutions.

 
This paper targets the technologies that are available in the literature and explains the essential pieces of a navigation assistive system. Section \ref{systemModel} provides the reader with the overall view of the components that constitute in a navigation assistive system. This section also considers the interactions among these  building blocks and justifies the necessity of each building block. In later sections each individual building block will be explained extensively along with the available methods to implement them. A navigation assistive system can divided in three main modules, the way-finding module, the obstacle avoidance module, and the human-machine interaction module. Section \ref{systemModel} illustrates the rationale behind this taxonomy.

To build the way-finding module, localizing the users and navigating them to the destination is necessary. There are many surveys available in the literature that addressed the localization and navigation problem. Surveys such as  \cite{S_3_1,S_6_1,S_7_1,S_9_1,S_9_2,S_9_3,S_10_1,S_11_2,S_13_1,S_15_1,S_15_2,S_15_3,S_15_4,S_16_1,S_17_1,S_17_2,S_18_1,S_18_4,S_19_1,S_19_2} focused on different methods, technologies, and algorithms for localization.

A comprehensive review of localization central theories is presented in \cite{S_11_2} and \cite{S_15_3}. Emergence of wireless sensor network (WSN) and the ubiquitous infrastructure of WiFi have made localization based on WSN an appealing topic. In that regard, an overview of localization techniques concentrating on utilizing WSN has been presented in \cite{S_7_1,S_9_1,S_9_2,S_9_3,S_10_1}. Using personal networks for localization is discussed in \cite{S_9_3}. 

Indoor localization method is not limited to the WSN. In addition to the wireless network based localization technologies,  a comprehensive spectrum of technologies based on vision, infrared, and ultrasound sensors are discussed in \cite{S_15_2,S_17_2,S_18_1}.

Localization methods can be categorized based on their core methods. This survey focuses on trilateration, fingerprinting, Bayesian filtering and crowdsourcing methods. Of these methods, trilateration is one that is extensively employed in location estimation. Reviews presented in \cite{S_6_1,S_9_1} delineate the nuances in this method. The fingerprinting method has become very popular in recent years. Broad demonstration of this method has been presented by \cite{S_15_1, S_16_1,S_17_1}. Bayesian filtering is essential in location estimation in order to mitigate the error. The review over different methods of Bayesian filtering has been provided by \cite{S_3_1}. Crowdsourcing improves localization by providing frequent update for the system. The impact of crowdsourcing in a localization system, has been fully discussed in \cite{S_15_4}.The comparison among different methods and technologies in respect to cost, robustness, and complexity has been executed in \cite{S_7_1,S_9_3,S_17_1}. 

Indoor navigation is not just about positioning. Navigating the user is the second part of this module. Pedestrian navigation methods and technologies have been explained extensively in \cite{S_13_2}. Precise heading estimation is vital for navigation. In that respect, \cite{S_16_2} provides a review of the main techniques of heading estimation. Step count and stride length estimation are the other necessary components of navigation process. Survey of different methods of step detection and stride estimation has been provided in \cite{S_16_3, S_8_1}. A survey over using Internet of Things (IoT) and data based approaches to help visually impaired navigate independently have been presented at \cite{S_19_2}. Comprehensive survey over all the navigation approaches to assist visually impaired has been provided by Strumillo et al. \cite{S_18_4}. 

Obstacle avoidance is the next module in a navigation assistive technology for a visually impaired individual. Many surveys that addressed obstacle avoidance for mobile robots are available. The similarities between a dynamic robot which needs motion command and a visually impaired individual who needs information and guidance to navigate indoor makes such surveys related. Surveys, \cite{S_18_2, S_1_1, S_89_1,S_16_4} mainly focus on obstacle avoidance using sensors such as sonar and LIDAR. 

Object detection problem is also a major concern for mobile robots. This issue has been discussed mainly in reviews which concentrate on object detection or tracking in an image or video. Research work presented in \cite{S_5_1,S_14_1,S_16_5,S_16_6} reviews the methods of object detection in a scene. A review of neural network based methods for object detection is discussed in   \cite{S_18_3}.

Efficient interaction with visually impaired user is vital in a navigation assistive system for a visually impaired individual. Different methods of interacting with the user has been explained extensively in \cite{S_9_4}.

In addition to the surveys which cover the challenges and the solutions to build a navigation assistive system partially, there are a few papers which concentrate on the available systems in the market or in the research labs \cite{S_8_2,S_13_3,S_17_3}. 

This paper aims to provide a concise overview of navigation methods available in the literature for the visually impaired. It would guide the future researchers in understanding the existing research in such navigation systems and provides insights into the methods that can be aggregated to design  an assistive navigation system. 

\subsection{Key Contributions}

The major contributions of this paper are summarized below:
\begin{itemize}
    \item \textbf{Localization}

\begin{itemize}
    \item Provides an overview of trilateration based methods, the mathematical background underlying trilateration concept,  and the required infrastructure for implementing this method with a variety of technologies.
    \item Discusses fingerprinting method as a solution to addressing the difficulties associated with trilateration methods.
    \item Discusses Bayesian filtering as an essential component in localization, mitigation of error associated with the system modeling and measurement and the related mathematical background.
    \item Discusses crowdsourcing as a solution for location estimation and updating the map and localization data.
\end{itemize}

\item \textbf{Navigation}
\begin{itemize}
    \item Explains a variety of methods to estimate the heading and the underlying mathematical concepts.
    \item Discusses available methods to detect the steps based on variety of technologies
    \item Describes the techniques to approximate the user's stride length
\end{itemize}
\item \textbf{Obstacle avoidance}
\begin{itemize}
    \item Explains the fundamentals of obstacle avoidance with range finder methods such as sonar and LIDAR
    \item Presents the advances made by employing a camera in obstacle avoidance and object detection
\end{itemize}
\item \textbf{Human-Machine Interaction}
\begin{itemize}
    \item Discusses human-machine interaction strategies for way-finding and obstacle avoidance
    \item Presents an overview of methods for providing the user with information and inputs
\end{itemize}

\end{itemize}

In connection with the above contributions, sections \ref{localization} and \ref{navigation} explore the location estimation and navigation techniques used in way-finding, respectively. Section \ref{ObstacleAvoidance} explains obstacle avoidance methods. Section \ref{hmi} explains methods for human-machine interactions. 

\section{System Overview}
\label{systemModel}
\begin{figure}[h]
	\includegraphics[width=0.5\textwidth]{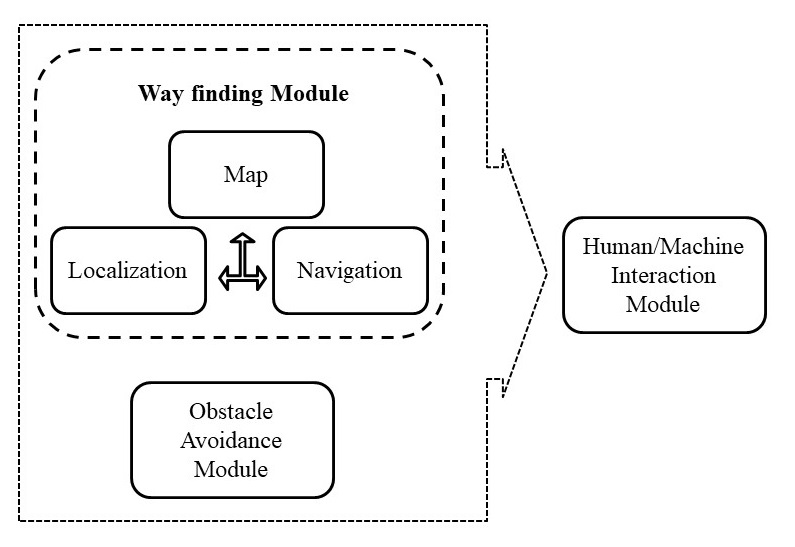}
	\caption{Overall System Model}
	\label{fig:SystemModel}
\end{figure}

The very first step to build a system is to design the system model. System models help depict a concise but precise representation of the system as well as the interaction of different components and it aids communication between the designer team and the development team. This paper aims to assist researchers who are interested in developing a navigation assistive technology for visually impaired users by providing them with the baseline to design such a system. For this matter, the overall system model for a navigation assistive system is depicted in figure \ref{fig:SystemModel} and each component as well as their interactions are explained briefly.

In order to develop a navigation assistive system for visually impaired individuals, one must study the elements that constitute it. Navigation is defined as the procedure of finding an initial location, determining the best path to the destination, and following it. While finding the initial location is not challenging for people who can see, it is almost impossible for visually impaired individuals to determine their initial location with no assistance.  
Recognizing the initial location is only the starting point in a navigation process. The next step is to decide the best route to the destination. The best route is defined based on user preference. For example, some users might select the shortest path while other may choose a path with fewer changes in direction. 

After path selection, the path must be followed. To follow a path, certain distances must be traveled and the direction must change in exact locations. An assistive navigation system measures the traveled distance by counting the number of a user's steps and learning the length of a user's stride. It also gathers information about the heading to direct the user based on their current location. Distance and heading information will help the user reach the destination.

For systems that aim to help individuals who might suffer from visual impairments, a series of instructions to follow a path is inadequate. Visually challenged individuals are at risk of being impeded by obstacles if the instructions have not been aggregated with comprehensive information about the area and the possible obstacles hindering the path in a dynamic environment. Consequently, a navigation assistive system, which aims to help visually impaired individuals navigate independently in indoor areas, must provide them with adequate information about the dynamic area on top of the instructions to follow the path.

By following a path and sidestepping the hurdles, the destination will be reached gradually for a visually impaired individual. 
According to the discussion above, a navigation assistive system that targets individuals with visual impairments needs to have three main modules: the \textit{Way-Finding Module}, the \textit{Obstacle Avoidance Module} and the \textit{Human-Machine Interaction Module}.

As mentioned earlier, a way-finding process consists of location estimation and navigation. Methods to address way-finding can be categorized based on the technologies that they employ, such as: radio frequency, sensors, and magnetic fields. 
The success of GPS for outdoor navigation encouraged the implementation of the same technique for indoor navigation. Indoor navigation using radio frequency is a wide research topic. The emergence of micro electro-mechanical systems (MEMS) also facilitates location estimation and navigation with the data provided through different sets of sensors. Another set of approaches is based on the unique magnetic map of an area that enables location estimation.

For way-finding, the initial location of the user must be known. Using the initial location of the user, a variety of navigation techniques can be employed to direct the user in reaching their destination. Methods to approach localization and navigation as the two main subsets of way-finding are explained in detail in the sections \ref{localization} and \ref{navigation}, respectively.  

When assisting an individual with visual impairment navigate an indoor area, providing information about the objects hindering the path is vital. Objects are in motion in real life scenarios and their position is constantly changing. The location of every hindrance has to be discovered in real time. The dynamic environment forces the system to detect the obstacles in real time and update the route frequently. As a consequence, implementing an obstacle avoidance module to update the route is necessary in the navigation assistive system for visually impaired individuals. Section \ref{ObstacleAvoidance} delineates methods to address this issue. 

While the navigation process is genuinely completed by providing the instruction to follow the path and information about obstacles impeding the way, without a concise means of communication, it will be impractical. Considering the challenges that users with visual impairments may face, screen instructions are not useful. Despite the fact that seeing ability is lost or limited in individuals with visual impairment, other senses may be assumed to be intact. Consequently, the system has to introduce other approaches that depend on abilities other than sight to receive information. In section \ref{hmi}, methods to interact with the visually impaired user are explained extensively.

\section{Localization}
\label{localization}

Localization is the first step in way-finding. To plan the path, the initial position of the user has to be known. Vision is crucial in location determination. In the absence of vision, assistive technologies must facilitate finding the initial location of the user for way-finding purposes. Strategies for localization that are available in the literature are briefly outlined in this section.
  
\subsection{Global Navigation Satellite System (GNSS)}

Global navigation satellite system (GNSS) or GPS eases outdoor navigation. GPS uses time difference of arrival (TDOA) of the signal received from at least four different satellites to find the exact position of the receiver \cite{91_2}. Trilateration is used to estimate the user's location. The use of GPS to aid visually impaired individuals in navigation was proposed by Loomis el al. \cite{85_1}, and Collins \cite{85_2} more than thirty years ago. Later, in 1989, Brusnighan made new experiments with the use of GPS to assist visually impaired users \cite{89_1}. In \cite{94_2}, Loomis et al. studied the use of differential global positioning systems to increase the accuracy in a positioning scenario. 

GPS is a promising approach for outdoor navigation \cite{00_1}\cite{12_4}. Electronic travel aid (ETA) and smart robot localization use GPS for outdoor localization. The smart robot functions like a guide dog to help a visually impaired person navigate. Yelamarthi et al. used GPS to determine the location of smart robots outdoors \cite{10_7}. For indoor scenarios, smart robots use radio frequency identification (RFID) tags for localization. RFID tags are discussed later in section \ref{RFID}.
Despite the accuracy of GPS signals outdoors, satellite coverage is not available indoors. Indoor structures attenuate satellite signals. The attenuated signals are undetectable for the receiver \cite{9_17}.  This puts a constraint on the use of GPS in indoor situations. This limitation forces scientists to explore new localization methods to address the location estimation challenge. One of the common techniques is \textit{Trilateration}.

\subsection{Trilateration}
Wireless based positioning is central to indoor navigation. Many researchers suggest its use in estimating the location of a user \cite{S_7_1}. A wireless local area network (WLAN)'s access points (AP) broadcast beacon frames periodically according to the IEEE 802.11 protocol. The transmitting period is about 10 ms. The beacon frames contain the AP's media access control (MAC) address \cite{12_10}. The MAC addresses that are provided to every mobile system enable the identification of multiple APs used for positioning. The position estimation is possible with triangulation and trilateration. 

Triangulation is another method used for range measurement. This method estimates the position of an object using the angle of the object relative to the reference points. Triangulation is suggested in \cite{12_13} to detect user's location. In this method, the user carries a white cane with infrared LEDs installed on it. The infrared cameras installed in an area detect the movement of the infrared light source to find the location of the white cane carrier. 

\begin{figure}[h]
	\includegraphics[scale=.5]{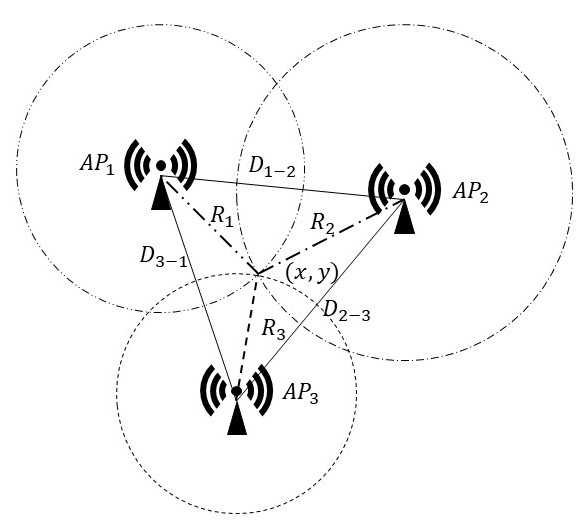}
	\caption{Trilateration}
	\label{fig:trilateration}
\end{figure}

In general, triangulation is more common in long distances than in indoor scenarios. Trilateration, on the other hand, performs better indoors. Trilateration is a range measurement technique, in which the position of an object is estimated by measuring the distance of the object to the access points. (See Figure \ref{fig:trilateration})
Common techniques to calculate the distance are time of arrival (TOA), time difference of arrival (TDOA), angle of arrival (AOA), and received signal strength (RSS).

\subsubsection{TOA}

This is one of the basic methods to find the distance from an AP to a mobile system. The TOA distance measurement method is based on the relation between the time and distance of propagation. The distance travelled and the propagation time are directly proportional to each other. In TOA-based systems, the distance between the mobile system and the access points will be calculated using the time it takes for the signal to travel one way. The distance to at least three access points is needed to estimate the position of a mobile system \cite{90_2}.

The position of the mobile system in a TOA algorithm can be computed by minimizing the sum of error functions. The error function is based on the least squares method. The function for every measuring unit $i$ is given by (\ref{eq:1}).

\begin{equation}\label{eq:1}
f_i(\pmb{x})=c(t_i-t)-\sqrt{(x_i-x)^2+(y_i-y)^2 }
\end{equation}

In the aforementioned equation, $c$ is the speed of light. The correct choice of $\pmb{x}=(x,y,t)$ makes the error go to zero. The sum of all the errors must be minimized to find the position of the mobile system. The sum of error functions are given in (\ref{eq:2}). The choice of $a_i$ indicates the received signal reliability at the measuring unit $i$.     
\begin{equation}\label{eq:2}
F(\pmb{x})=\sum_{i=1}^{N} a_i ^2f_i ^2(\pmb{x})
\end{equation}

TOA algorithms are prone to errors. One of the major error sources for this method is the multipath effect. The error caused by the multipath effect can be categorized in two groups: early-arriving multipath signal and attenuated line of sight (LOS). Early-arriving multipath signals correspond to the signals arriving immediately after the LOS signal and changing the location of the peak from the LOS signal. The attenuated LOS explains the situation in which the LOS signal is weakened too severely. In this situation, the LOS signal is lost in noise. To combat multipath disturbance, \cite{17_23} proposed defining a multipath delay profile. This multipath delay profile is analyzed to improve the accuracy in location estimation. In addition to the multipath effect, adaptive noise causes errors in the TOA algorithm. 

Adaptive noise even in the absence of a multipath effect can decrease accuracy \cite{87_1}. To combat the adaptive noise, a simple cross-correlator is used. In \cite{76_1} by Knapp et al , the simple cross-correlator was extended by a generalized cross-correlator.  

The ubiquitous presence of WiFi infrastructure makes it a convenient choice for location estimation systems. On the other hand, as mentioned earlier, wireless signal based systems are prone to error. Consequently, other types of signal have been used for location estimation method. As an example, the TOA method is not limited to RF signals, and has been implemented with a variety of signal types. Acoustic signals can be counted as an example of such signal types.

Acoustic signals are another common type of signal used with the TOA method for localization. In this approach, instead of utilizing the available WLAN networks, acoustic anchors are implemented in the area. The anchor network propagates the modulated beacon with location information. The beacons use highband acoustic signals. An indoor localization system called Guoguo, which has been proposed in \cite{13_1}, employs the TOA method in an acoustic signal network. Guoguo processes the information in real-time using a smart phone application. The prototype implemented based on this approach demonstrates a centimeter-level accuracy. 

Acoustic signal has been used in synergy with radio signals to improve accuracy. In \cite{12_9}, acoustic signal have been used to estimate the location of a system where different locations in an area have similar radio signatures. The synergy of acoustic signal and wifi signal has been utilized in \cite{12_1} and the results suggests that it can improve accuracy in NLOS localizing scenarios.  

Ultrasonic signal are another major type of signals to employ in location estimation modules. For instance, they were used in a localization system in \cite{3_2}. This work also used TOA to measure the distance of the system to the APs and estimate its location. To mitigate the impact of disturbances, and improve accuracy, this work proposed using fuzzy adaptive extended information filtering. 
For more information about TOA methods and approaches to combat its vulnerabilities, interested reader may read \cite{S_9_1}.

\subsubsection{TDOA}
While TOA is measuring the absolute arrival time, TDOA concentrates on the different times that the propagated signal reaches multiple measuring units. The difference in time of arrival is proportional to the distance of the measuring units to the mobile system. 

\begin{figure}[h]
	\includegraphics[scale=.5]{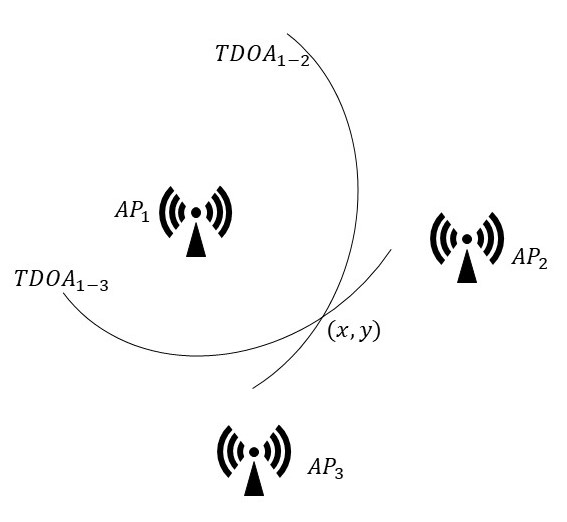}
	\centering
	\caption{Position Estimation using TDOA Measurements}
	\label{fig:tdoa}
\end{figure}

The TDOA measurements for every two receivers define a hyperbolic locus \cite{98_2}. The hyperbolic locus ($R_{i,j}$) follows (\ref{eq:3}) given that $i$ and $j$ are receivers at two fixed points. The location of each of these receiver points is indicated by $(x_i,y_i,z_i)$ and $(x_j,y_j,z_j)$ respectively.

\begin{multline}
 \label{eq:3}
R_{i,j}=\sqrt{(x_i-x)^2+(y_i-y)^2+(z_i-z)^2 }\\ 
- \sqrt{(x_j-x)^2+(y_j-y)^2+(z_j-z)^2 }
\end{multline}
The hyperbolic locus defines the area that the mobile system might be positioned at. The exact position of the mobile system is on the intersection of two hyperbolic loci. (See figure \ref{fig:tdoa}) 

In theory, there in no difference between the TOA and TDOA algorithms. Simulation results also confirm the idea that TOA and TDOA have practically the same performance \cite{12_11}. To improve accuracy of TDOA based location estimation, access points with two transmitters were introduced. The second transmitter gathers information about mobile systems and transmits this information to the APs, which will estimate the location of the mobile systems. This enhances the accuracy of position estimation without interrupting the network \cite {7_4}. 

The importance of the precise position estimation for visually impaired individuals can not be emphasized enough. Using the TDOA method in Ultra Wide Band (UWB) is a novel approach to improve the accuracy \cite{15_2}. UWB is another type of wireless communication. It became trendy after FCC made 3.1GHz- 10.6 GHz and 22GHz - 29 GHz available \cite{17_26}. In addition to improved accuracy, this method also has low installation complexity. The accuracy of UWB comes with a higher cost. 

\subsubsection{AOA}
Although the focus in methods such as TOA and TDOA is on the distance of the user to each access point, Angle of Arrival (AOA) (or as referred to in some literature as \textit{Direction of Arrival (DOA)}) focuses on the angle of the received signal on the receiver side. In this method receiving two signals is essential. Receiving more signals improves the accuracy of the estimation.

To implement this method, either a mechanically-agile directional antenna or an array of antennas is required. The angle of arrival is determined by finding the angle in which the signal strength is the highest in the former, or the antenna which samples the most strength in the latter infrastructure. 

\begin{figure}[h]
	\includegraphics[scale=.5]{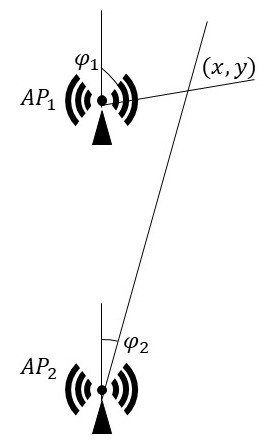}
	\centering
	\caption{Angle of Arrival (AOA)}
	\label{fig:aoa}
\end{figure}

As represented in figure \ref{fig:aoa} the angle of arrival from at least two known access points is needed to determine the location of the user. Given the distance between the two access points, the location of the user is determined through geometric relations.

To improve the accuracy of AOA estimation, Badawy et al.proposed a method using the cross-correlation function \cite{14_9}. The expenses associated with this method are high. Using off-the-shelf software-defined radio as suggested by \cite{12_14} reduces the cost of this approach. It also improves flexibility and makes deployment simple. Interested reader would find more details about aforementioned methods in \cite{13_10}. 

\subsubsection{RSS}

A major source of error in the TOA and TDOA algorithms is the multipath effect. It attenuates the signal strength and causes signal loss. In addition to the multipath effect, adaptive noise causes errors. Adaptive noise is a significant source of error even in the absence of the multipath effect. These challenges cause errors and reduce the accuracy of the estimated position. Furthermore, in indoor situations LOS is not available in all circumstances. To avoid the challenges associated with distance estimation using the above mentioned new approaches have been proposed. An alternative approach is to use  the received signal strength (RSS). 

The receiver measures the voltage of the received signal to find the RSS. In some research, instead of reporting the voltage of the received signal, its power is reported. The power of the RSS is the square of the amplitude of the received signal at the receiver.Measuring the power or the voltage of the received signal requires not only no specific tool but also no additional energy or bandwidth. The hardware for measuring voltage and power is implemented in every measuring unit. The low cost of hardware implementation and the convenience in using it, made this technique a ubiquitous approach in localization.

RSS is established on the relationship between signal attenuation and the distance. The signal attenuation happens during propagation. The major reason for this attenuation is path loss. In theory, the power decay of the received signal is proportional to the distance between the transmitter and the receiver. The path loss in free space is defined by (\ref{eq:rss_pathloss}) \cite{book_digital}.
\begin{equation} \label{eq:rss_pathloss}
PL_{FS}= (\frac{\lambda}{4\pi d} )^2
\end{equation}
Here $\lambda$ is the signal wavelength and $d$ is the distance between the transmitter and the receiver. 

As (\ref{eq:rss_pathloss}) demonstrates, the rate of power decay in the signal is inversely proportional to the distance squared between the transmitter and the receiver. Although this equation is theoretically correct, the empirical results show some discrepancies. The mismatch between the theory and the practice is caused by multipath effect and shadowing. 

Multipath signaling has an adverse impact on the accuracy of the RSS method. The desired signal is not the only signal detected by the measuring unit. The measuring unit detects multiple signals in a variety of amplitudes and frequencies. This can cause frequency-selective fading. The impact of frequency-selective fading can be alleviated with the spread-spectrum method \cite{book_digital} \cite{book_wireless}. The spread-spectrum method averages the power of the signal over a wide range of frequencies. The spread-spectrum method can address the multipath signaling issue to a reasonable extent but frequency selective fading is not the only source of error in the RSS method. Shadowing is another major source of inaccuracy.

Shadowing attenuates the power of the received signal in an indoor environment in addition to the multipath effect. In the literature, shadowing is defined as the situation in which the power of the signal deviates from its real average value \cite{book_wireless2}. Shadowing happens when the wave propagates in an environment with obstacles. Signal propagation in an indoor environment is attenuated by obstacles such as furniture and walls in the path of the signal. The shadowing effect is a random process. 

To improve the accuracy of position estimation using RSS measurement, an unsupervised learning algorithm has been proposed \cite{11_3}. This algorithm is based on the Gaussian Mixture model and Expectation Maximization (EM). In this research, the received signal is modeled as a Gaussian Mixture. The algorithm learns the parameters for the model from the transmitted packets. To learn the maximum likelihood estimate of these parameters, the EM method is used.

Another promising attempt to improve the accuracy of the position estimation with the RSS method is the on-line measurement of the AP's signal \cite{10_2}. In this method, the transmitted signal of every AP is measured by every other APs. This measurement provides useful information about error-prone sources in the environment. The real-time AP signal strength measurement is enriched with the impact of path fading, variations in signal strength caused by temperature, humidity, human mobility, or other changes in a dynamic indoor environment. This information can be utilized to generate a mapping between the RSS measurement and geographical locations. 

Using a more comprehensive indicator of the signal such as channel state information is helpful in improving the location estimation \cite{18_2}. 

Trilateration is a common approach to estimate the location of the user. The simple hardware requirements, inexpensive methods, and simple calculation makes it appealing to researchers. On the other hand, the inaccuracy caused by the multipath effect, shadowing, and noise encouraged researchers to explore different algorithms. 

\subsection{Fingerprinting}
As mentioned before, RSS is a popular method for localization. It is an inexpensive algorithm with little required equipment. On the other hand, WLAN is widespread all around the indoor environments. The massive distribution of WLAN access points in addition to the simplicity of the RSS measurement provides an appropriate groundwork for position estimation. Fingerprinting is one of the attractive techniques founded on this principle. 

Fingerprinting has an offline learning phase called \textit{radio map} construction. The radio map database is matched with the acquired data during the \textit{Fingerprint matching} phase.
\subsubsection{Radio Map}
Modeling of signal propagation in an indoor environment is a difficult task. To avoid modeling the environment with all of the complexities, a radio map can be utilized. The radio map is constructed by measuring and storing the RSS values at calibration points. These calibration points have known locations. The complexity of signal propagation in an indoor environment is avoided by using a radio map. Despite the simplicity of this task in theory, gathering and storing the RSS values relative to all the calibration points is laborious. 

During the radio map composition phase, the area of interest is divided into cells. The cell area definition is based on the floor plan. For each cell, the RSS value for every access point is measured on the calibration point over a specific time period. The RSS values are then stored in the database. 

Composing the database is not a one-time task. This process is cumbersome and varies over time. It will also change when there is an alternation in the environment or in the transmission power of the access point. Therefore, to keep up with the changes, the radio map has to be corrected periodically. 

A correct up-to-date radio map is the first phase of the position estimation using the fingerprinting method. 

\subsubsection{Fingerprint Matching}
In the fingerprint matching stage, the target finds the best matching cell for the measured RSS values. Finding the best match can be done through either deterministic or probabilistic methods. 
In deterministic methods, the approach is to find the best matching calibration point to the position of the user. The matched calibration point is determined by minimizing the difference of the RSS values measured by the APs in the on-line phase with those RSS values stored in the database. Suppose that an array of RSS values has been measured from each AP. The location of the user is detected by finding the spot in the database whose distance to the current location is minimum. This is a Euclidean distance minimization problem which is shown in (\ref{eq:deterministic})   \cite{9_8}. 
\begin{equation}\label{eq:deterministic}
\hat{x}=\underset{x_j}{\operatorname{argmin}}(\sum_{i=1}^{n} (r_i -\rho_i( x_j))^2)
\end{equation}
Aforementioned equation is the mathematical representation of the Euclidean distance minimization. In this equation, $r_i$ is the measured value of RSS in the on-line phase of the $i_{th}$ AP. The stored value of the measurements of the $i$-th AP in the $j$-th calibration point is presented with $\rho_i( x_j)$. Although in theory this approach is promising, to avoid measurement errors other methods have been suggested. One of the more reliable methods is the nearest neighbor algorithm.

In the \textit{k}-nearest neighbor (KNN) method, the value measured in the on-line phase is compared with the values measured for the calibration point in the dataset. The $k$ calibration points which have the closest value will be selected. In this method, \textit{k} is the design parameter. It will be selected based on the density of the radio map. In the \textit{KNN} method, the weight of all the calibration points are equal. However, the weighted \textit{KNN} (WKNN) method considers a different weight for each calibration point. The weight assignment is based on the distance of the calibration point to the approximated location of the measured RSS. 

The nearest neighbor method is not limited to the Euclidean distance. The use of Manhattan distance is suggested by \cite {17_8} for improved accuracy. The nearest neighbor method, whether using Euclidean distance or Manhattan distance, is a solution to a distance minimization problem. This solution makes location determination feasible. This method is considered a deterministic localization approach. Besides the deterministic approaches such as \textit{KK} and \textit{WKNN}, there are also probabilistic fingerprinting approaches used to estimate the position. 

In the probabilistic localization method, the conditional probability distribution of the RSS value will be estimated. The conditional probability distribution of the RSS value, given the location of the user, is plugged in to Bayes rule to estimate the location of the user. The probability of the location of the user given the RSS measurement is called the posterior probability distribution. In this method, the location of the user is the location $x$ which maximizes the posterior probability distribution (\ref{eq:post}).

\begin{equation}\label{eq:post}
\hat{x}=\underset{x}{\operatorname{argmax}}(p(x|r_1,...,r_n))
\end{equation}

 Aforementioned equation represents the calculation of the posterior probability distribution given the conditional probability of the RSS measurement. 
\begin{equation}\label{eq:bayes}
p(x|r_1,... , r_n)=\frac{p(r_1,...,r_n|x)p(x)}{p(r_1,...,r_n)}
\end{equation}

In equation (\ref{eq:bayes}), $p(x)$ is the prior probability distribution, which indicates the preference of the user to be in a location. If the user preference is equally distributed among all the locations this term will be a constant coefficient which can be removed from the calculation. The denominator of the fraction in (\ref{eq:bayes}) is a normalization factor because it is not proportional to the location. Considering these two simplifying assumptions, (\ref{eq:bayes}) can be rewritten as a maximum likelihood approximation (\ref{eq:like}).
\begin{equation}\label{eq:like}
\hat{x}=\underset{x}{\operatorname{argmax}}(p(r_1,...,r_n|x))
\end{equation}

To solve the maximum likelihood as depicted in equation, the knowledge of the distribution of the RSS values in all possible locations is required (\ref{eq:like}). Two well known probabilistic models for this purpose are the coverage area model and the path loss model. The former, as its name indicates, estimates the probability distribution through the coverage area of each AP. The latter, uses the logarithmic attenuation of the signal in respect to the distance from the AP. 

In addition to the deterministic and probabilistic approaches \cite{17_24}, machine learning methods have recently been applied to fingerprint-based position determination algorithms \cite{17_27}. Position determination is a classification problem that can conveniently be solved using Support Vector Machines (SVM). The accuracy of this method for position estimation in WLAN situations outperforms \textit{WKNN} as discussed in \cite{5_9}. Another machine learning based classification method to use with fingerprinting is the Neural network classifiers.The location estimation performance was improved by aggregating this method with a neural network classifier in \cite{17_22}. Deep neural network and convolutions neural network are the other neural network method which were used to address localization using fingerprinting method \cite {18_3,18_7}.

While fingerprinting is a reliable solution for position estimation problems, it is vulnerable to malicious attacks. These malicious attacks have an adverse impact on the localization process. When the attack strikes, the location information provided by the system is not reliable. In \cite{5_2}, Li et al. explains the attacks extensively and provides a solution for this matter. This research introduces an adaptive least squares estimator which is robust to attacks.

The fingerprinting method goes beyond just RSS values. This method works well with other types of signals. Sound, color, and motion are the other features that can be sensed with a cellular device through the microphone, camera, and inertial sensors. These signals were used in \cite{9_6} to find the location of the user. Fingerprinting is also used with Earth's magnetic field. The strength and variation of the magnetic field is used to build a map for fingerprinting and these data are used to estimate the user's location. For instance \cite{17_3} suggest using the magnetic field information which is attainable by an smart phone for localization. 

Fingerprinting principles can not only be applied to different types of signals, but they can also be put into use with the signal in different layers of communication. For instance, \cite{12_8} uses the channel frequency response(CFR) in the physical layer. In this research, an array of CFR values is associated with a specific location. This method reduces the associated error with location estimation to less than a meter. 

The fingerprinting method can be combined with other methods to improve accuracy. One such method is to use image-based localization. iMoon is a cellphone application presented in \cite{14_2} which uses fingerprinting to divide the area into partitions. Each partition contains a certain number of images. To find the exact location of the user, the image provided by the user will be matched with the image database of that partition. This method not only improves the precision of the location estimation, but it also reduces the computational power essential for image matching. 
 
While fingerprinting uses the available signals to determine the location, other methods such as Radio Frequency Identification (RFID) rely on specific equipment. Later in this section RFID tags have been explained in detail. 

\subsection{Bayesian Filtering} Location estimation is possible using the motion and measurement information. System measurement is prone to error. To overcome the error caused by noisy measurement Bayesian filters have been used extensively. The Bayesian approach is based on constructing the posterior probability distribution model of the system state based on the available measurements and the prior probability distribution model. In a recursive Bayesian approach, the filter is executed for every measurement. In every time instance, the system state can be predicted given the previous state of the system. When the measurement were obtained, the predicted model will be corrected based on the current measurements. The corrected probability distribution, is the prediction for the next time instance.

\begin{figure}
\begin{center}
	\includegraphics[scale=.4]{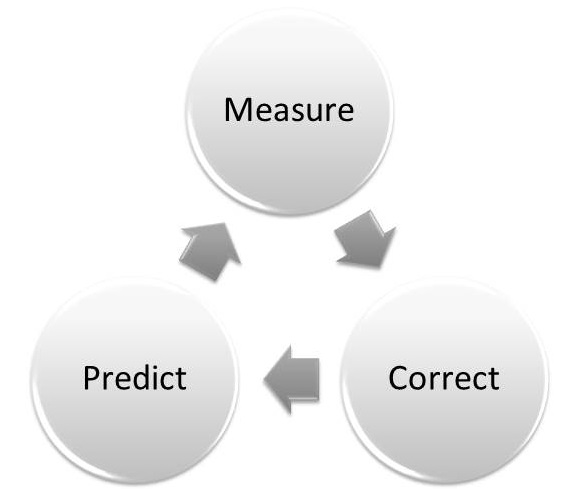}
	\caption{Bayesian filter cycle}
	\label{fig:filter}
\end{center}
\end{figure}

Figure \ref{fig:filter} represents the Bayesian filter cycle. In this cycle in time instance $t$, the prediction of the probability distribution of the $X_t$ is based on all of the observations $z_{\{{t-1}:1\}}$ from the beginning of time to the previous time instance.In other words, the probability distribution of the existence of system in state $X_t$ is estimated based on the all of the measurements.

In the next step, the new measurement will be introduced to the model. In general, every extra information, shrinks the probability distribution and reduces the uncertainty. Introducing the new measurement to the system, corrects the predicted probability distribution. The corrected prediction or the posterior probability distribution of time $t$ serves as the prior probability distribution of time $t+1$.

One of the early methods based on the recursive Bayesian approach is the \textit{Kalman filter}.

A Kalman filter is a well known approach for filtering and prediction proposed by Rudolf Kalman in 1960 \cite{kalman1960}. It has been extensively used for localization and navigation purposes. In localization and navigation systems, the signal is modeled with transition and measurement matrices as shown in (\ref{eq:kal1}), and  (\ref{eq:kal2}).
	\begin{align}
	x_{k+1}&= \boldsymbol{\Phi}_kx_k+G_k\boldsymbol{w}_k\label{eq:kal1}\\
z_k&=\boldsymbol{H}x_k+\boldsymbol{n}_k\label{eq:kal2}
	\end{align}

In (\ref{eq:kal1}), and (\ref{eq:kal2}), $\boldsymbol{\Phi}$ and $\boldsymbol{H}$ indicate the transition matrix and measurement matrix respectively. The process is not free of noise. The measurement noise and the process noise are added to the model via $\boldsymbol{w}_k$ and $\boldsymbol{n}_k$ respectively. The noise increase the distribution variance and widen the normal distribution.

In order to apply Kalman filter to shrink the probability distribution, the noise must be white noise with zero means. In addition to the noise, Kalman filter requires the system model to follow a normal distribution.

The covariance matrices for the process noise and the measurement noise are calculated as shown in (\ref{eq:kal3}) and (\ref{eq:kal4}):
	\begin{align}
	E[w_lw^T_m]= \boldsymbol{Q}\delta(l-m)\label{eq:kal3}\\
E[n_ln^T_m]= \boldsymbol{R}\delta(l-m)\label{eq:kal4}
	\end{align}

The prediction of probability distribution of the next state is based on the state transition matrix and the corrected probability distribution estimate of the current state, as shown in (\ref{eq:kal5}):
\begin{equation}\label{eq:kal5}
\hat{x}^-_{k+1}= \boldsymbol{\Phi}_k\hat{x}_k.
\end{equation}

The probability distribution of the measurement is the summation of the probability of observing a certain measurement for every possible system state (\ref{eq:kal6}):
\begin{equation}\label{eq:kal6}
\hat{z}^-_{k+1}= \boldsymbol{H}_k\hat{x}_{k+1}.
\end{equation}
If the variance matrix of the state is called $\boldsymbol{S}_{k}$ and defined as in (\ref{eq:kal7}),
\begin{equation}\label{eq:kal7}
\boldsymbol{S}_{k}=E[\delta\boldsymbol{x}\delta \boldsymbol{x}_k^T]
\end{equation}
then the prior estimate of the variance matrix of the next step is given by (\ref{eq:kal8})
\begin{equation}\label{eq:kal8}
\boldsymbol{S}^-_{k+1}= \boldsymbol{\Phi}\boldsymbol{S}_{k}\boldsymbol{\Phi}^*+\boldsymbol{G}\boldsymbol{Q}_{k}\boldsymbol{G}^*.
\end{equation}
The novelty of Kalman filtering is that it tunes the impact of the predicted probability distribution of the current state and the likelihood of the  measured value to determine the corrected probability distribution estimate of the variable as shown in (\ref{eq:kal9}):
\begin{equation}\label{eq:kal9}
\hat{x}_{k}=\hat{x}^-_{k}+\boldsymbol{K}_{k}(z_k-\boldsymbol{H}_k\hat{x}_{k}).
\end{equation}
In (\ref{eq:kal9}), $\boldsymbol{K}_{k}$ is the gain of the Kalman filter and using the transition and the measurement equations, (\ref{eq:kal1} and \ref{eq:kal2}), the gain of the Kalman filter is defined as in (\ref{eq:kal10})
\begin{equation}\label{eq:kal10}
\boldsymbol{K}_{k}=(\boldsymbol{S}^-_{k} \boldsymbol{H}_k^*)(\boldsymbol{H}_k\boldsymbol{S}^-_{k}\boldsymbol{G}_k^*+\boldsymbol{R}_k)^{-1}.
\end{equation}
Kalman filter reduces the uncertainty of the transition and measurement. Suppose the system measurement and transition are noisy, and one is more uncertain, the probability distributing of the system after applying Kalman filter has less uncertainty than either of the two. In other words, suppose $z_k$ and $u_{k}$ are the measurement and transition model with variance of  $S_u$ and $S_z$, respectively. The uncertainty of the estimated probability distribution after applying Kalman filter for this variable will be less than both  $S_u$ and $S_z$. Equation (\ref{eq:uncertainty}) shows the relationship between the uncertainties, given that $S$ indicates the uncertainty of the estimated value:
\begin{equation}\label{eq:uncertainty}
\frac{1}{S}=\frac{1}{S_z}+\frac{1}{S_u}.
\end{equation}
Although Kalman filter improves the accuracy of the measurements, it is limited to linear transition functions and Gaussian noise assumption. This limits the utility of the method to handle nonlinear functions.  \textit{Extended Kalman Filter (EKF)} attempts to approximate a non-linear function to a linear one to use Kalman filter.

In real applications, the system model is not a simple linear function. It limits the applications of the Kalman filter. The EKF overcomes this constraint, although it is not free of conditions.  The function requires to be differentiable to be compatible with EKF. In this method, to calculate the covariance matrix, the Jacobian of the transition matrix and measurement matrix must be calculated. The use of the Jacobian matrix in addition to the first order Taylor series, approximate the non-linear function with a linear function. The approximated linear function can use Kalman filter. Although this method facilitates the use of Kalman filter for nonlinear functions, it doesn't improve the accuracy. When the accuracy is not adequate the \textit{Unscented Kalman Filter (UKF)} is useful. 

Probability distribution models are the main representation of a random process. However, for all the random process there is not a simple way to represent the distribution analytically. To handle the complex probability distribution models, the distribution can be represented with infinite number of samples. UKF uses this method to approximate a probability distribution density of a random process. 

The UKF approximates the nonlinear function to make it compatible with the Kalman filter using sample representation. Sampling is an alternative for situations in which linearizing a nonlinear function using basic methods such as first order Taylor series is difficult or inaccurate. To study the probability distribution of the output of a non-linear system with noise, the nonlinear transition function is applied to the samples and the output is analyzed. The result approximates the characteristics of the probability distribution of the output of the nonlinear function. This method is prone to error if the number of samples is not adequate.

The UKF reduces the number of required points for function estimation to a few points called sigma points. In a UKF, a set of sigma points are computed. For every point a weight will be assigned. The output of the nonlinear function for these points will be determined. To determine the output, the non-linear function applies to the sigma points. Using the output of the nonlinear transition function for the sigma points, the probability distribution of the output could be estimated. This method prevents linearizing the function in close proximity to the mean. 

A Gaussian distribution is an indispensable prior assumption to use the Kalman filter or any of its descendants. This assumption is not valid in indoor scenarios. \textbf{Particle filter} method becomes effective to address the non-linearity caused by floor plan.

Location estimation accuracy improves by exploiting building geometry. Using the building floor plan reduces heading errors significantly. In indoor navigation scenarios, the floor plan adds constraints on movement. For one, all the paths connecting the two sides of a wall will be limited to those passing through the door. The same restriction applies to changing floors. The only possible case for changing floors is through the elevator or stairs. Map-aided navigation improves the accuracy by using the prior information obtained from the floor plan. Particle filter is used in probabilistic map-matching. Particle filter outperformed Kalman filter in location estimation as suggested by  \cite{2_1}.  A particle filter is a generic Monte Carlo algorithm to solve filtering problems. The filtering problem is to estimate the state of a dynamic system according to a partial observation. Some random noise exists in both the sensor measurement and the system actions. The goal of the particle filter is to use the partial observation to estimate the posterior distribution of the Markov process's state.

\begin{figure}
    \centering
    \includegraphics[scale=0.49]{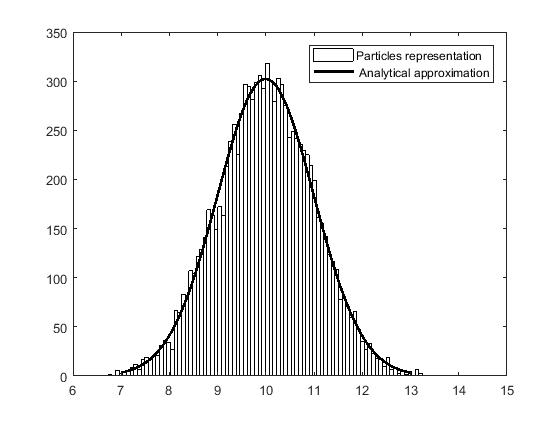}
    \caption{Particle representation of a Normal distribution}
    \label{fig:normalDist}
\end{figure}

Particle filters are based on the idea of representing the posterior distribution using a set of random state samples. Although this posterior distribution is approximated, it is non-parametric. For example a normal distribution representation with random state samples, represents more samples with higher weights close to the mean and less samples with smaller weights further from the mean,( figure \ref{fig:normalDist}). This representation of the normal distribution is non-parametric. Note that a normal distribution can be represented analytically with an exponential function of mean and variance as shown in (\ref{eq:normalDist}). 
\begin{equation}\label{eq:normalDist}
    P(x)=\frac{1}{\sigma\sqrt{2\pi}}e^{\sfrac{-(x-\mu)^2}{2\sigma^2}}
\end{equation}

Figure (\ref{fig:normalDist}) depicts a particle representation of a normal distribution with its mean and variance equal to ten and one, respectively. The height of each bar represents the population of the particles in that sub-region. The probability of a state is associated with the height or density of the particles in the sub-region corresponds to that state.

In contrast to a Kalman Filter which is only applicable to a linear system with Gaussian noise; no speculation about the system or the noise is required to apply particle filter. Although EKF and UKF overcome the non-linearity problem by linearizing the function, their dependence on noise with Gaussian distribution reduces their accuracy in estimation. In the particle filtering method, instead of using the parametric form to represent the distribution, a set of samples drawn from the corrected distribution is used. These samples, which are drawn from the corrected probability distribution of the previous state, are called particles. The particles are as denoted in (\ref{eq:particle1}):
\begin{equation}\label{eq:particle1}
\mathcal{X}_t :=x_t^{[1]},x_t^{[2]}, ..., x_t^{[M]}
\end{equation}
In the equation above, $\mathcal{X}_t$ is the set of particles and it has $M$ members. The members of the set $\mathcal{X}_t$ are the $x_t^{[m]}$s . Each of them is a state's sample in time $t$. The number of particles, $M$,  must be adequate for a reasonable approximation. 
As mentioned in UKF section, intuitively the nonlinear function of transformation can be estimated by observing the samples. The particle filter approximates the belief (probability distribution) of the system model, $bel(x_t)$, based on the probability distribution of the set of samples, $\mathcal{X}_t$.  

To estimate the probability distribution of the system state, the prior knowledge about that system state, $p(x)$ is necessary. To approximate the posterior distribution of the system, the system dynamic model must be known. The dynamic of the system in particle filter is represented by $u_t$. The dynamic model of the system allows the estimation of the probability of the current state of the system given the system dynamic and the previous state, (\ref{eq:actionModel}). 
\begin{equation}\label{eq:actionModel}
   p(x_t| u_{t-1}, x_{t-1})
\end{equation}
The dynamic model is necessary but not adequate to estimate the posterior probability distribution of the system. Similar to the Kalman filter, in particle filter sensor information is central. In particle filter, sensor model provides us with the likelihood of the getting a certain measurement, given that the system is in a certain state, (\ref{eq:sensorModel}).
\begin{equation}\label{eq:sensorModel}
    p(z_t|x_t)
\end{equation}
Equation (\ref{eq:sensorModel}) indicates that a certain measurement is more probable for a certain state of the system.  
Using the dynamic and sensor model and the actions and measurement information through out the time, this model estimates the likelihood, or the belief of the $x_t$. Equation (\ref{eq:like}) represents the likelihood relation \cite{5_10}.
\begin{equation}\label{eq:like}
x_t^{[m]} \sim p(x_t|z_{1:t},u_{1:t-1})
\end{equation}
Equation (\ref{eq:like}) denotes the probability distribution of the current state of the system given the series of previous actions and measurements. The obtained probability distribution demonstrates the density and weights of the samples in a certain sub-region of the state space. The population and the weight of the samples in a sub-region of the state space strengthen the hypothesis that the true state belongs to that sub-region. According to (\ref{eq:like}) the likelihood of the existence of the state $x_t^{[m]}$ as a member of the particle set,$\mathcal{X}_t$, is proportional to the posterior probability distribution of the system $p(x_t|z_{1:t},u_{1:t-1})$.  The validity of this hypothesis is proportional to the magnitude of $M$. In theory, the hypothesis is valid for $M \rightarrow \infty$. In practice $M$ is a finite number, and the error is negligible for a very large $M$. Increasing the size of samples set, increases the cost associated with using particle filter.

The particle filter is a descendant of the Bayesian filter. Recursivity is in the nature of the Bayesian filter. Particle filter follows the same principles. It constructs the current set of particles $\mathcal{X}_t$, based on the previous set of particles, $\mathcal{X}_{t-1}$. 
Algorithm ~\ref{alg:the_alg} represents the foundation of the particle filter.

\begin{algorithm}
\vspace*{10px}

\vspace*{10px}
\textit{ input $(\mathcal{X}_{t-1},u_t,z_t)$}\\
$\bar{\mathcal{X}_{t}}=\mathcal{X}_{t}= \emptyset$\\
 \For{$m=1$ to $M$}{
  sample $ x_t^{[m]}\sim p(x_t|u_t, x_{t-1}^{[m]})$\;
  weight  $ w_t^{[m]}\sim p(z_t| x_t^{[m]})$\;
 $\bar{\mathcal{X}_{t}}=\bar{\mathcal{X}_{t}}+\langle x_t^{[m]},w_t^{[m]}\rangle$}
\For{$m=1$ to $M$}{draw $i$ with probability distribution proportional$ \propto  w_t^{[i]}$\\
add $x_t^{[i]}$ to $\mathcal{X}_{t}$}
\textit{output to} $\mathcal{X}_{t}$
\vspace*{25px}
\vspace*{10px}
  \caption{Pseudo-code for particle filter}\label{alg:the_alg}
\end{algorithm}

Similar to other recursive algorithms, the input to the particle filter is the set of particles from the previous state, $\mathcal{X}_{t-1}$. The action data $u_t$ and the measurement information $z_t$ of the current state are the other inputs to this algorithm. In this algorithm, $\bar{\mathcal{X}_t}$ is a temporary particle set which is proportional to the initial $bel(x_t)$ according to (\ref{eq:like}). Particle filter as its other siblings, builds the distribution of $bel(x_t)$ based on the $bel(x_{t-1})$. 

In the current time, the hypothetical state $x_t^{[m]}$ is constructed based on the particles $x_{t-1}^{[m]}$ and the current action function $u_t$. Note that the Independence of the current status of the system toward the previous action functions $u_{1:t-1}$ is a result of the Markovian property of the system. Markovian property implies that the current state of the system is a result of the immediate previous action function and it is independent of the older actions( figure (\ref{fig:SystemModel}) .

\begin{figure}
    \centering
    \includegraphics[scale=0.49]{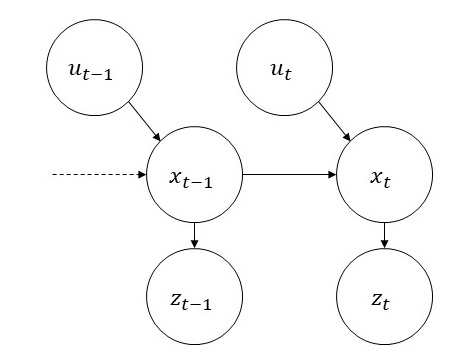}
    \caption{Model with Markovian property}
    \label{fig:Markovian}
\end{figure}

Figure \ref{fig:Markovian} represents the Markovian property in a system. As the figure represents, each hidden state of the system is the result of its immediate previous action function and it is independent of the older action functions. It also illustrates that the measurements are independent in regard to each others and each measurement is the result of its corresponding hidden state.

Line 4 of the algorithm \ref{alg:the_alg} represents the generation of the current hypothetical states based on previous set of states, $\mathcal{X}_{t-1}$, and the current action function $u_t$. The index $m$ states that this sample is derived form the $m^{th}$ particle of the particle set $\mathcal{X}_{t-1}$. The resulting set of particles is the prediction of the probability distribution of the current state or $\overline{bel}(x_t)$ distribution. The bar over the $bel$ symbol in $\overline{bel}(x_t)$ indicates that $\overline{bel}(x_t)$ is the predicted belief and correction based on the measurement has not executed yet. 

The weight of the particles must be calculated in next step as shown in line 5 of the algorithm \ref{alg:the_alg}. The weights are computed based on the current measurements $z_t$. Consequently, it is calculated by the probability of the occurrence of the measurement $z_t$ given the state is at particle $x_{t}^{[m]}$. As mentioned earlier, the measurements are only affected by the system state , (\ref{eq:sensorModel}). This property is called sensor independence. The resulting set of weighted particles indicates the corrected distribution of $bel(x_t)$. Line 6 of the algorithm \ref{alg:the_alg} represents how the set of particles, $\overline{\mathcal{X}_t}$, is updated with every pair of $\langle x_t^{[m]},w_t^{[m]}\rangle$. 

The novelty of the particle filter is in the \textit{re-sampling} section, which is shown in the second \textbf{For} loop in algorithm ~\ref{alg:the_alg} (line 8-12). In the re-sampling phase, $M$ particles are drawn from the $\bar{\mathcal{X}_t}$ with replacement. With the replacement, the probability of a particle being drawn more than once is not equal to zero. It means for every particle there is a chance to be drawn for a second time. The chance of being drawn for every particle is proportional to the weight of that particle. Re-sampling changes the distribution of the set of particles that form $\overline{{bel}}(x_t)$ to approximate the posterior $bel(x_t)$. The posterior $bel(x_t)$ is calculated according to (\ref{eq:posterior}). 

\begin{equation}\label{eq:posterior}
bel(x_t)=\eta p(z_t| x_t^{[m]})\overline{bel}(x_t)
\end{equation}
Above equation, represents the relationship between the posterior belief propagation of the particles with the prior belief propagation. Initially the weight of all the particles are equal. The measurements update the weight of the particle and reduce the importance of some of them while increasing the importance of a few other ones. In this equation $\eta$ is the normalization factor. 
\\
Method explained in algorithm \ref{alg:the_alg} is not the only approach to change the distribution of the set $\mathcal{X}_{t}$ to represent the posterior $bel(x_t)$. Another approach is to skip the re-sampling step and update the weight with (\ref{eq:weightUpdate}). 
\begin{equation}\label{eq:weightUpdate}
 w_t^{[m]}= p(z_t| x_t^{[m]}) w_{t-1}^{[m]}
\end{equation}

In (\ref{eq:weightUpdate}), each weight is updated according to its past value. To start with, the weight distribution of this algorithm is considered to be uniform or every weight is simply initialized to one at the beginning. 

In a localization problem, the particles are distributed over the whole area. However passing through walls is impossible. Consequently, the particles which are connected through walls with the current state will be eliminated. The same approach is applicable for the particles on two different floors. The particles connected on two floors are only sampled if the connection is possible through an elevator or stairs. This constraint is represented through (\ref{eq:wall}).

\begin{equation}\label{eq:wall}	
p(z_t| x_t^{[m]})=
\begin{cases}
		    0 \text{              no path between} x_t^{[m]} \text {and} x_{t-1}^{[m]}\\
	    p(z_t| x_t^{[m]})\text{otherwise}
	  \end{cases}
	\end{equation}

The particle filter has been used in literature extensively. Its performance and accuracy of location estimation have been studied and compared with other methods in \cite{4_6}. The comparison result showed that particle filter performs more accurately other available methods. 

As mentioned earlier, transferring from one floor to another floor only happens through stairways and elevators. The 2.5D map localization based on the particle filter is studied in \cite{8_5}. In this research, the height is considered in addition to the location coordinates. The change of height only possible in the vicinity of stairways. The 2.5D building description, which is based on defining a vertical position for an object with no depth, helps to avoid complexity. 

An extended particle filter (EPF) is proposed in \cite{16_3}. In this method, the following three collision scenarios have been considered: 
\begin{itemize}
 \item Close proximity to an obstacle 
 \item Passing through an obstacle due to heading error
 \item Passing through an obstacle due to stride error
\end{itemize}

The collision status of the current state in addition to the $n$ last states are used to define the weight of the particle in this research. The EPF proposed in this research detects collisions which might be missed because of the gap in crowdsourced data. Crowdsourcing will be explained in detail in the next section.

\subsection{Crowdsourcing} Generating a position estimation system can be difficult for two main reasons. First, a detailed map may not be available for the location of interest. Second, the area may be changing and keeping the map updated is difficult for the design team. Consequently, map-making is a time consuming procedure for the design team. To avoid such a burden, crowdsourcing the process is helpful.

Crowdsourcing uses the open collaboration of users to generate and update information. In this method, the data are provided by the users who are using the system. The data which is provided by the users will be processed and added to the system. The added data are available to all users. This cycle updates the system and provides users with a more detailed system which updates frequently. 

For a localization problem, the map needs to be detailed and periodically updated. When using crowdsourcing for map-making, each user provides the system with slightly advanced information about the location of interest, and the system adds this information to the map. This method helps to not only build a fully detailed map but also update it constantly. 

Crowdsourcing has been explored in recent years thanks to the emergence of smart phones with high quality built-in cameras. By integrating computer vision and crowdsourcing, Jigsaw \cite{14_10}, constructs the floor map based on the data from the user's mobile device. In iMoon, system builds a 3D map based on the pictures available on the internet \cite{15_3} and photos taken by users \cite{14_2} . For the initial map generation, iMoon uses the photos to build a 3D model. The 3D model is generated using the Structure from Motion method \cite {sfm}. After the initial model is generated, the information provided by users updates the model and adds more detail to it.

Building the floor map based on the video data from mobile users in addition to sensor measurements is proposed in \cite{15_9}. In this research, the relationship between the successive frames enhances the system performance.  

In addition to generating a 3D map, crowdsourcing is used in WiFi-fingerprinting. For instance, WicLoc \cite{15_4} is a WiFi fingerprinting with a crowdsourcing system for indoor localization. By using crowdsourcing, the task of surveying the site is avoided while the precision remains intact. HiMLoc \cite{13_4} is an indoor localization system based on smart phones which use crowdsourcing for WiFi fingerprinting. HiMLoc uses a crowdsourced WiFi fingerprinted map to update the weights in a particle filter.

In the previous subsections different methods to find the initial location of the user have been discussed. In section \ref{navigation}, a variety of approaches to chose to provide a series of instructions to direct the user to reach the destination have been explained. On the other hand, it would be incomplete to explain the localization methods and not to mention the methods to implement them. Wifi based methods have been explained extensively in this section. Wifi is not the only solution in location estimation.Two widely used technologies, RFID tags and BLEs, have been explained in detail in subsection \ref{tech}

\subsection{Overview of Technologies}\label{tech}

While many available methods are based on the Wifi technology, it is not the only solution for localization. Synchronization between the transmitter and the receiver, attenuation effect are just few examples of problems associated with using the Wifi for localization. In recent years, other technologies emerged to address the problems associated with Wifi technology while keeping the cost of infrastructure in an reasonable range. In this subsection, two widely used technologies for positioning that are more compatible with navigation assistive system for visually impaired individuals were explained. 

\subsubsection{RFID} \label{RFID} Radio Frequency Identification (RFID) is a method of localization based on radio waves. The RFID method is based on two parts: the reader and the tag. Each tag has a unique identification number. The reader uses the radio magnetic field to identify the RFID tag based on the tag identification. 

RFID tags are small circuits containing a microchip and an antenna. The antenna is actually a printed circuit board. Tags can be categorized in two groups based on their energy provider. The tags which rely on the energy emitted by the reader are passive tags. This type of tag is inexpensive and can be employed in cost-sensitive scenarios. The other type of tags have batteries installed in them. These kinds of tags propagate their identification in periodic signals. 

For position estimation purposes, the tags are attached to specific locations in the area. Each tag can carry information about the location and path conditions. This method is useful in both indoor and outdoor situations \cite{5_7}. These types of tags are the most common for localization purposes. The other type of tag which is useful for providing navigation assistance for visually impaired individuals is the cue tag. These kinds of tags can provide the user with a virtual road and guide the user along the way. In general, RFID tags provide the user with information about location and direction. 

The information obtained via the reader can be transferred to the user's cellphone. Smart phone applications are useful for conveying the information in a more convenient way to the user. Accessibility of cellphones facilitates call centers to help visually impaired people online. The call center can use the information from the tags to inform the user not only about the location but also the path to their destination\cite{7_7}. 

Installing the reader on a guide cane is a brilliant idea for visually impaired users \cite{ 7_8,9_9}. The cane in these studies is equipped with a tag reader and a Bluetooth transmitter. The reader transmits the acquired information the user. This information is conveyed to the visually impaired user through speech. Tag allocation is important for accuracy purposes. In this research, the tags' locations are in close proximity to achieve higher location estimation precision.

A different approach in using RFID tags for user localization is introduced in \cite{4_4}. In this approach, the user carries the RFID tag and the readers are placed throughout the area. In this technique, the location estimation is based on the information about the intensity of the signal propagated from the tag. This research uses the \textit{KNN} algorithm to calculate the location of the user. The tag used in this method is an active type. Active type tags require a battery to send signals periodically. Over time the battery loses power which causes variations in signal strength. 

\subsubsection{Bluetooth low energy} \label{ble}

The RSS value of the classic Bluetooth protocol has been used for localization \cite{5_1}. However, Bluetooth low energy (BLE) beacons provide a new means for position estimation. Not only are BLEs accurate and reliable, but they are also energy efficient. Although BLE works in the same frequency that classic Bluetooth does, it has some characteristics which make it more favorable for positioning purposes. These characteristics are energy efficiency, a short handshake procedure, and a high scan rate. A BLE beacon also utilizes a sleep mode which helps maintain low energy consumption. The device changes to awake mode only if a connection is initiated. The short connection time helps with the low energy consumption as well. When connected, the BLE transmits its unique identification number.

The fingerprinting method is also applied to the BLE beacons for position estimation. A BLE beacon can be thought of as a small WLAN AP without the need of a power connection. According to \cite {14_3}, BLE position estimation is more accurate in comparison with WLAN-based position estimation. The superiority comes from three main features: a channel hopping mechanism, high sampling rate, and low transmission power. 

The channel hopping mechanism employed in a Bluetooth device is less sensitive to interference. The channel interference problem is addressed through averaging. If the averaging does not solve the interference problem and the interference is too severe, the transceiver hops to another channel. BLE uses the hopping mechanism to avoid interference.

High sampling rate is another advantage of BLE over WLAN. BLE position estimation methods determine the position of the user with more samples. The higher number of available samples reduces the chance of error by averaging out the outliers. Outliers are generated by the multipath effect or interference. Averaging out the outliers improves the accuracy of localization. 

Low transmission power also makes BLE a better choice for position estimation. In addition to the energy efficiency, low transmission power reduces the multipath effect. Consequently, the receiver only hears the most powerful signal while the other signals generated by the multipath effect will be faded out.

To gain better accuracy, a two-level outlier detection method is introduced by \cite{16_5}. In this research, the first level outlier detection happens after the fingerprinting position estimation algorithm. The result is fed to an extended Kalman Filter. The second level outlier detection algorithm is applied to the output of the extended Kalman Filter. The second outlier detection algorithm is designed based on statistical testing. The two-level outlier detection algorithm deployed in this research enhances the robustness of the design. 

In close distances to the beacon, position is estimated using an attenuation model. Equation (\ref{eq:attenuation}) shows the logarithmic propagation model \cite{14_4}. 
\begin{equation}\label{eq:attenuation}
RSS(d)=RSS(d_0)+10 n \log(\frac{d}{d_0})+X_\sigma
\end{equation}

In aforementioned equation , $RSS(d_0)$ is the RSS value at the referenced distance $d_0$. The path loss is indicated by $n$. $X_\sigma$ is the Gaussian noise with zero mean and variance $\sigma^2$, which is added to the RSS value. Smoothing filters help with addressing the issue with unstable RSSI value for BLEs \cite{17_24}.    

Accuracy of position estimation is central for visually impaired individuals. One approach for enhancing accuracy is through combining the positioning methods with estimation filters. For instance, a Mount Carlo localization algorithm has been applied to a BLE technology based positioning system in \cite{17_21} which improved the localization performance. 

Table \ref{tab:localization} presents a summary over the available methods. 

{\renewcommand{\arraystretch}{1.8}
\begin{table*}
  \centering
\begin{tabular}{ |p{5cm}|p{.5cm}|p{5cm}|p{5cm}|  }
 \hline
 \multicolumn{4}{|c|}{Localization methods} \\
 \hline
 Method& Used for VI users & Advantages & Disadvantages\\
 \hline
 \hline
  GPS \cite{85_1,85_2,89_1,92_2,94_2,00_1,10_7,12_4,12_15} & X & Precise outdoor localization & Unavailable indoor\\
 \hline
   Trilateration ,
\begin{itemize}
\item $TOA$
\item $TDOA$
\item $AOA$
\item $RSS$
\end{itemize}
\cite{3_2,S_9_1,5_2,7_4,12_13,13_1,13_10,17_23,17_26,12_14,13_10,87_1,12_9,98_2,12_11,15_2,14_9,14_14,11_3,10_2,18_2}
& X &

\begin{itemize}
\item Convenient 
\item No extra infrastructure
\end{itemize}

&

\begin{itemize}
\item Sensitive to adaptive noise
\item Requires transmitter and receiver calibration 
\item Sensitive to shadowing 
\item Sensitive to multipath effect
\item Sensitive to frequency selective fading
\end{itemize} \\
 \hline
 Fingerprinting
 \begin{itemize}
     \item Radiomap
     \item Fingerprint Matching

 \end{itemize}
 
  \cite{9_8,17_8,17_24,17_27,5_9,17_22,18_3,18_7,5_2,9_6,17_3,12_8,14_2}
 & X &
 \begin{itemize}
     \item  Robust to multipath effect
     \item Robust to attenuation 
 \end{itemize}
&
\begin{itemize}
    \item Sensitive to dynamic environment
\end{itemize}
\\
 \hline
  Bayesian filtering 
 \begin{itemize}
     \item Kalman filter
     \item Extended Kalman filter
     \item Unsenced Kalman filter 
 \end{itemize}

 &  &
 \begin{itemize}
     \item  Improve the accuracy 
 \end{itemize}
   
&
\begin{itemize}
    \item Limited to Gaussian noise
    \item Limited to linear( or diffrentiable ) functions 
\end{itemize}\\
 
 \begin{itemize}
     \item Particle filter
 \end{itemize}
 
  \cite{2_1, 5_10,4_6,8_5,16_3,13_4,15_4,14_8,5_13,10_10,19_5}
 & X &
 \begin{itemize}
     \item  Improve the accuracy 
     \item Independent of Gaussian noise
     \item Compatible with non-linear function
 \end{itemize}
   
&
\begin{itemize}
    \item Computationally expensive
    \item Time consuming 
    \item Requires frequent re calibration
\end{itemize}\\
 \hline

  Crowdsourcing
  \cite{14_10,15_3,14_2,15_9,15_4,13_4}
 
 & X &
 \begin{itemize}
     \item Improve the accuracy 
     \item Provides frequent calibration
 \end{itemize}
   
&
\begin{itemize}
    \item Vulnerable to malfunctions 
\end{itemize}\\
 \hline
 \end{tabular}
  \caption{Summary of Localization Strategies (VI stands for "Visually Impaired") }
  \label{tab:localization}
\end{table*}}

Table \ref{tab:localization} summarizes the information which was provided in this section. Future researchers can combine the methods to ameliorate the accuracy of their design. The table also summarized method assuming that they have been implemented with WiFi. Each of the methods or the combination of them can be implemented with a variety of signals type such as acoustics, UWB, Cellular, etc ... and/or a variety of tools such as BLE, RFID tags, etc... . Detailed information about the hybrid designs have been provided in \cite{S_7_1}

\section{Navigation}
\label{navigation}

Localization is at the core of empowering a visually impaired person to navigate independently. However, localization is only a portion of the complex navigation procedure.  There are other pieces to complete this puzzle. The next part of the problem is to navigate the user in the area of interest.

Cognitive mapping is the primary tool of navigation for a visually impaired individual \cite{98_1}. Cognitive mapping refers to the spacial figure of an area that one has in their mind. The human brain can remember the arrangement of the objects in a known area. A visually impaired person is able to use the spacial arrangement of the objects in an area to find a path. This arrangement is based on the perception of distance. Visually impaired individuals who have lost their vision in adulthood have a better sense of distance. Those who were born without sight face more challenges in arranging a spacial image of the area of interest \cite{98_3}. 

Navigating in a new area is challenging for both sighted and visually impaired people. However, this task is more manageable for sighted people because of available signs. Despite the the fact that signs are helpful in path finding , it is yet an arduous task in unfamiliar and crowded areas such as airports, shopping malls, or conferences. Consequently, researchers suggested sign reader technologies, which help sighted users find their current location on the map and the best path to their destinations \cite{9_4}. 

On the other hand, path finding and sign reading is not possible for individuals with visual impairment. They depend on their other senses and navigation assistive technologies to find a path and read signs. Devices based on navigation assistive technologies have different functionalities. A group of these devices are built to read signs. Tjan et al. designed a hand held sensor to detect and read passive retro-reflective tags \cite{5_6}. The main objective of these devices is to read the signs for the user. Typically in this group of devices, detecting a sign is followed by the information about the current location. It also comes with information about the orientation of the user based on the viewing angle of the camera. Location estimation is feasible by calculating the distance of the user to the sign. In addition, it provides the user with movement instructions to reach the destination.  

First and foremost, this group of navigation assistive technologies depend on sign detection. Sign detection happens only if the detector's angle of view covers the marker. Therefore, optimizing the shape, color, and design of the marker is fundamental. These required features have been fully discussed in \cite{9_3}. It suggests that by employing the right color, shape, and design, the process of detecting the marker will be optimized and the marker will be detected more precisely. 

Another camera based navigation approach is to use image recognition based on photos of the area of interest. While sign detection doesn't required high computational power, image detection and image matching based navigation demand high computational power. The emergence of more powerful GPUs in recent years facilitates new approaches to solve the direction estimation problem using computer vision based solutions. 

In iMoon \cite{14_2}, the direction of the user is determined through the provided images. The images provided by the user in the current time will be matched with the database of images of the area of interest. The result provides the system with the user's heading.

Cameras are not the only means to help visually impaired individuals navigate independently. An alternative approach for navigation assistive technologies is to employ the \textit{pedestrian dead reckoning (PDR)} method \cite{1_2}.

\begin{figure}[h]
		\includegraphics[scale=.5]{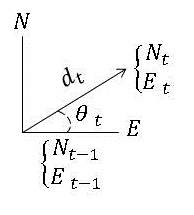}
		\centering
		\caption{Vector of displacement in North-East frame}
		\label{fig:disp}
	\end{figure}

PDR is a successive position estimation method. In this method, the initial location of the user in an environment of interest must be known. The initial location of the visually impaired user is obtained using the localization methods explained in the previous section. The displacement of the user in each transition is added to the previous position of the user to find the current position of the user. The displacement of the user can be estimated in different forms such as Cartesian coordinates or heading and distance form, although heading and distance are more common. Equation (\ref{eq:pdr}) shows the displacement calculation in the North-East frame.
	\begin{equation}\label{eq:pdr}
	\begin{cases}
	    N_t=N_{t-1}+d_t cos(\theta_t)\\
	    E_t=E_{t-1}+d_t sin(\theta_t)
	  \end{cases}
	\end{equation}
	In this $d_t$ indicates the vector of displacement in time $t$ and $\theta_t$ shows the heading of the user at $t$. Figure \ref{fig:disp} represent the same concept.
	
As figure \ref{fig:disp} demonstrates, suppose the user in time $t-1$ stands in $(N_{t-1}, E_{t-1}$ in the North-East frame. The user moves to  $(N_{t}, E_{t}$ at time $t$. The user's displacement is represented with a vector $d_t$ which has the magnitude of $d$ and angle of $theta$. 

In this method, the accurate estimation of the displacement is vital. Erroneous estimates will cause mismatch to accumulate. The error aggregation is prevented with correction methods. Without a correction algorithm, the drift in estimation will continue to build up. Consequently, the level of inaccuracy in each transition will be increased exponentially. Bayesian methods which were explained in previous section are useful in mitigating errors. 

\begin{figure}[h]
		\includegraphics[scale=.4]{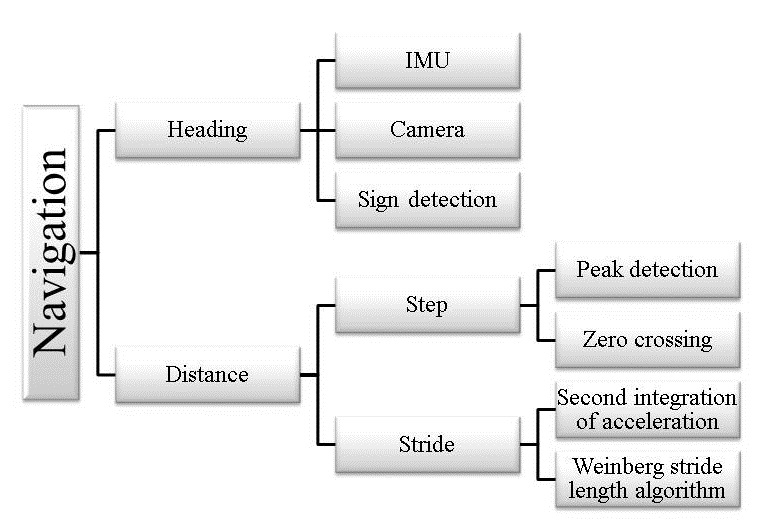}
		\centering
		\caption{Navigation process}
		\label{fig:navigation}
	\end{figure}

For PDR, the displacement of the user has to be determined for each transition. To generate the vector of displacement, information about the heading of the user and the distance passed in every transition are essential. Figure \ref{fig:navigation} represents a few popular methods to obtain these information. 

The heading and distance determination is explored in literature extensively. This section explains these two subjects respectively.

\subsection {Heading} 

Direction estimation is necessary for a navigation assistive technology. Although a sighted individual might take this for granted, estimating orientation precisely is a difficult assignment for a visually impaired person. An individual walking without vision, either visually impaired or blindfolded, is prone to veering off course \cite{7_12}. The most intuitive way to find the heading is to calculate the vector connecting two successive positions and define the heading as the angle of the vector \cite{12_13}. This simple method is only applicable in sensor rich areas in which the location estimation is absolute and the error is negligible. In a new area without installed sensors, navigation algorithms assist with detecting the heading. 

Navigation assistive devices use these navigation algorithms to detect the heading. In a navigation algorithm, transforming between reference frames is essential. The beginning of this subsection is dedicated to explaining the mathematical technique for the coordinate frame transformation to avoid further confusion. An interested reader is encouraged to read \cite{Inertial_Navigation} for a deeper explanation.

The coordinate frame for the mobile inertial sensor, the carrier, and the Earth are not the same. To have the right displacement vector, the displacement vector has to be transformed to the same coordinate frame as the location and destination. 

The displacement vector, $\boldsymbol{r}^k$, is shown with a bold letter with a superscript, which indicates the coordinate frame. The coordinate frame is where the component of the displacement vector belongs to (See (\ref{eq:vector})). 
	
\begin{equation}\label{eq:vector}
	\boldsymbol{r}^k=
	\begin{bmatrix} 
	x^k \\
	y^k \\
	z^k 
	\end{bmatrix}
\end{equation}

The displacement vector $\boldsymbol{r}^k$ can be represented in other frames. To transfer a vector to a different frame, the appropriate transformation matrix is needed. Vector $\boldsymbol{r}^k$ exists in the k-frame. It transforms to the r-frame as shown in (\ref{eq:transform}):

	\begin{equation}\label{eq:transform}
	\boldsymbol{r}^m=R^m_k\boldsymbol{r}^k
	\end{equation}

In (\ref{eq:transform}), $R^m_k$ represents the transformation matrix between the k-frame and r-frame. The superscript shows the target frame and the subscript shows the source frame. The transformation is only correct if the subscript matches the displacement vector superscript. \\ The inverse of the transformation matrix transfers the vector from r-frame to the k-frame as shown in (\ref{eq:inverseTransform}): 

\begin{equation}\label{eq:inverseTransform}
	\boldsymbol{r}^k=(R^m_k)^{-1}\boldsymbol{r}^m=R^k_m\boldsymbol{r}^m
\end{equation} 

If the r-frame and the k-frame are mutually orthogonal, then the corresponding transforms matrices $R^m_k$ and $R^k_m$ are orthogonal. Consequently, as the properties of the orthogonal matrices imply, the inverse of the orthogonal matrix is equal to its transposed form. (See (\ref{eq:invTrans})) For a transform matrix to be orthogonal, all of its row vectors must be orthogonal with respect to each other.

	\begin{equation}\label{eq:invTrans}
	R^m_k=(R^k_m)^{-1}=(R^k_m)^{T}
	\end{equation} 

In order to find the displacement vector in any reference frame, a source of data relative to that frame is needed. In the absence of vision, another source of data is needed to make the approximation. Mobile sensors are convenient providers of such data. Mobile inertial sensors carried by users are necessary for direction estimation. These sensors help find the user's heading, speed, and displacement. Sensors such as accelerometers, magnetometers, and gyroscopes are the most commonly used inertial sensors. Tri-axial accelerometers, tri-axial magnetometers, and tri-axial gyroscopes provide information in 9 degrees of freedom \cite {9_2,3_3}. These sensors are available both as stand-alone sensors or embedded into other smart devices. 

These sensors are not free from errors. The placement of the sensors in respect to the user's body changes the measurements. The availability of the sensors in different forms enables researchers to explore them both in fixed positions and dynamic positions. Some types of sensors have a fixed placement in respect to the walking direction, such as headsets. However,the placement of other sensors can be dynamic or unknown. For instance, a watch that is worn on the wrist. The wrist has a dynamic movement which makes the direction estimation complex. The placement of the sensor can be detected using acceleration data \cite{7_9}. Different sensor placements correspond to the different coordination frames. 

Inertial sensors make measurements in respect to a frame. The main frames used in the literature are:
	\begin{itemize}
	  \item Earth-Centered Frame 
	  \item Local Level Frame
	  \item Body Frame
	\end{itemize}

Earth-Centered Frame and Local Frame are common coordinate frames which are used to express the position of an object. In an Earth-Centered Inertial Frame the center of the Earth is the origin. The z-axis is pointing to the rotational conventional terrestrial pole (CTP). The x-axis points toward the vernal equinox. It defines in equatorial plane. The y-axis follows the right hand rule. The Earth-Centered Earth-Fixed Frame has the same properties. The only difference between the Earth-Centered Inertial Frame and the Earth-Fixed frame is that in the Earth-Fixed frame the x-axis passes through the Greenwich meridian. 

The Local-Level Frame on the other hand is the frame which has its y-axis point to the North and the x-axis point to the east. The z-axis follows the right hand rule and points upwards. It also called the navigation frame and is the most common frame in the literature. This frame is sensitive to the geographical location and the latitude effects its rotation rate.

\begin{figure}[h]
		\includegraphics[scale=.5]{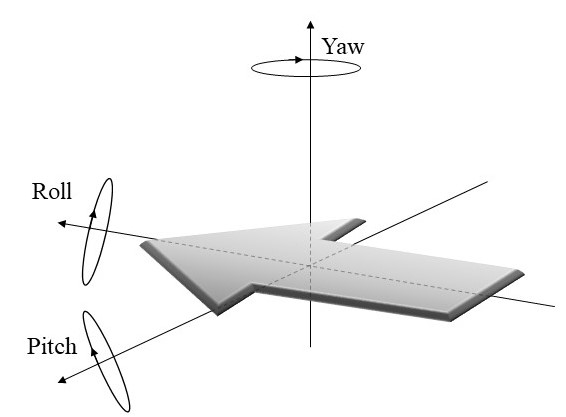}
		\centering
		\caption{Rotation in respect to the body frame}
		\label{fig:qua}
	\end{figure}

Body frame is another frequently used frame defined in the literature. It defines the frame whose origin is at the center of gravity of the object. In this frame, the y-axis points toward to the forward direction. The x-axis is the side to side direction. As a result of the right hand rule the z-axis points toward the vertical direction. For simplicity, the motion of the object in body frame is described with \textit{yaw},\textit{ pitch} and \textit{roll}.Yaw or azimuth rotation defines as the rotation of the object in respect to its z-axis by an angle $\gamma$. Pitch or rotation defines the rotation of the object with angle $\beta$ in respect to the y-axis. Roll or bank rotation gradually indicates the rotation of the object in respect to its x-axis with angle $\alpha$. (See Figure \ref{fig:qua}) 

In every problem, different frames might be defined to ease the calculation. It is necessary to have coordination transformation to move from one frame to another. The main techniques used to transform the frames are 
	\begin{itemize}
	  \item Direction Cosine Matrix 
	  \item Rotation Angle (Euler) 
	  \item Quaternions 
	\end{itemize} 

These methods introduce the transformation matrix, $R^m_k$, to transform the origin frame to the target coordinate frame. The subscript defines the origin frame and the superscript indicates the target frame. 

In general, despite the fact that orientation is a 3D quantity, it cannot be shown as a vector is 3D space. To overcome this issue one of the aforementioned methods is used to describe it in relation to a reference frame \cite{ASEN3200}. The oldest method used to describe the rotation of the body in a reference frame was Direction Cosine Matrix (DCM). 

DCM transforms the displacement vector from one frame to the other frame. This transformation represents the rotation or change in the orientation. The transform matrix in DCM method is the dot product of the unit vectors of each reference frame as shown in (\ref{eq:DCMmatrix}): 

	\begin{equation}\label{eq:DCMmatrix}
	R^m_k=
	\begin{bmatrix} 
	i_k.i_m && j_k.i_m && k_k.i_m   \\
	i_k.j_m && j_k.j_m && k_k.j_m \\
	i_k.k_m && j_k.k_m && k_k.k_m 
	\end{bmatrix}
	\end{equation}

Euler angles on the other hand are defined based on the sequence of rotation in respect to the fixed reference frames. To transfer a displacement vector from frame $k$ to the frame $m$ three rotations are needed. For instance, the first rotation is in respect to the z-axis with angle of $\gamma$. The next rotation is with the angle of $ \beta$ in respect to the x-axis, and the third rotation in respect to the y-axis is the angle $\alpha$. $\alpha$, $\beta$, and $ \gamma$ are Euler angles. Each of these rotations are executed through the appropriate DCM. To reduce the complicity, $R^z_k$ is defined as the DCM for rotation in respect to z-axis (See (\ref{eq:EulerMatrix1})). $R^y_z$ and $R^m_y$ are defining the rotation in respect to the y-axis and the x-axis, respectively(See (\ref{eq:EulerMatrix2}) and (\ref{eq:EulerMatrix3})). 

\begin{equation}\label{eq:EulerMatrix1}
	R^z_k=
	\begin{bmatrix} 
	cos(\gamma) && sin(\gamma) && 0   \\
	-sin(\gamma) && cos(\gamma) && 0 \\
	0 && 0 && 1
	\end{bmatrix}
	\end{equation} 
	\begin{equation}\label{eq:EulerMatrix2}
	R^y_z=
	\begin{bmatrix} 
	1 && 0 && 0   \\
	 0&& cos(\beta) && sin(\beta) \\
	0 && -sin(\beta) && cos(\beta)
	\end{bmatrix}\end{equation} 
	\begin{equation}\label{eq:EulerMatrix3}
	R^m_y=\begin{bmatrix} 
	cos(\alpha) &&0&& -sin(\alpha)    \\
	0 && 1 && 0\\
	sin(\alpha) &&0&& cos(\alpha) 
	\end{bmatrix}
	\end{equation} 

The final matrix of transformation is the dot product of all the three matrices represented in (\ref{eq:EulerMatrix1} - \ref{eq:EulerMatrix3}) as shown in (\ref{eq:EulerMatrix4}): 
	\begin{multline} \label{eq:EulerMatrix4}
	R^m_k =  R^m_y R^y_z R^z_k =\\
	\begin{bmatrix} 
	\begin{smallmatrix}cos(\alpha)cos(\gamma)-\\sin(\beta)sin(\alpha)sin(\gamma) \end{smallmatrix}&&\begin{smallmatrix}cos(\alpha)sin(\gamma)+\\cos(\gamma)sin(\beta)sin(\alpha)\end{smallmatrix}&&\begin{smallmatrix}-cos(\beta)sin(\alpha)\end{smallmatrix} \\
	-cos(\beta)sin(\gamma)&&cos(\beta)cos(\gamma)&&sin(\beta) \\
	\begin{smallmatrix}cos(\gamma)cos(\alpha)+\\cos(\beta)sin(\alpha)sin(\gamma)\end{smallmatrix} &&\begin{smallmatrix}sin(\alpha)sin(\gamma)-\\cos(\alpha)cos(\gamma)sin(\beta)\end{smallmatrix}&&\begin{smallmatrix}+cos(\beta)cos(\alpha)\end{smallmatrix} 
	\end{bmatrix}
	\end{multline}
	
\begin{figure}[h]
		\includegraphics[scale=.5]{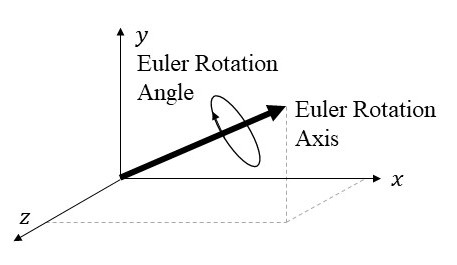}
		\centering
		\caption{Rotation Axis and Angle in Euler Theorem}
		\label{fig:euler}
	\end{figure}

	Given that the rotation is small, the approximation shown in (\ref{eq:approximation}) is common in the literature. 
	\begin{equation} \label{eq:approximation}
	sin(\theta) \approx \theta , cos(\theta) \approx 1
	\end{equation}
	The Euler theorem states that the rotation of a object with one fixed point can be explained with rotation angle about a rotation axis. (See figure \ref{fig:euler}) 
	
This theorem led to the inception of \textit{Quaternions}. Quaternions is an example of a hyper-complex system. This hyper-complex system explains the movement of an object in the space. Quaternions make a 4D system of numbers with basis of $1, i, j, k$ \cite{Quaternion}. The quaternions are as shown in (\ref{eq:Qua1}):  
	\begin{equation} \label{eq:Qua1}
	Q=\{ q_0+q_1i+q_2j+q_3k\}
	\end{equation}
	The basis used in (\ref {eq:Qua1}),$i$, $j$ and $k$, are square root of $-1$. ( See (\ref{eq:Qua2}))
	\begin{equation} \label{eq:Qua2}
	i^2=j^2=k^2=ijk=-1
	\end{equation} 
	The quaternions can be shown in a form of a constant and a vector as represented in (\ref{eq:Qua3}):
	\begin{equation} \label{eq:Qua3}
		Q= q_0+q_1i+q_2j+q_3k = (q_0,\boldsymbol{q}) 
	\end{equation} 
The rotation of a vector $p$ is represented with quaternions method in (\ref{eq:Qua4}):
	\begin{equation} \label{eq:Qua4}
		p{\prime}=qpq^{-1} 
	\end{equation} 
The quaternions' rotation can be represented as a rotation matrix $R$.
	\begin{equation}\label{eq:QuaMatrix}
	R=\begin{bmatrix} 
	1-\frac{2(q_2^2+q_3^2)}{||q||^{2}} &&\frac{2(q_1q_2-q_3q_0)}{||q||^{2}}&& \frac{2(q_1q_3+q_2q_0)}{||q||^{2}}    \\
	 \frac{2(q_1q_2+q_3q_0)}{||q||^{2}}&& 1-\frac{2(q_1^2+q_3^2)}{||q||^{2}}  && \frac{2(q_2q_3-q_1q_0)}{||q||^{2}}\\
	\frac{2(q_1q_3-q_2q_0)}{||q||^{2}} &&\frac{2(q_2q_3+q_1q_0)}{||q||^{2}} && 1-\frac{2(q_1^2+q_2^2)}{||q||^{2}}
	\end{bmatrix}
	\end{equation}  
$||q||$ is the norm of the quaternions and it is calculated as represented in (\ref{eq:Qua5}): 
\begin{equation} \label{eq:Qua5}
		||q||=\sqrt{q_0^2+q_1^2+q_2^2+q_3^2}
	\end{equation} 
Using the transforming matrix $R$, introduced in (\ref{eq:QuaMatrix}) vector $p$ transforms to $p'$ as represented in (\ref{eq:Qua6}): 
	\begin{equation} \label{eq:Qua6}
		p{\prime}=Rp 
	\end{equation} 
When frame $m$ rotates in respect to the frame $k$, the transformation matrix will involve time variable functions. The derivative of the transformation matrix in respect to time can be approximated with the skew-symmetric matrix of the angular velocity, $\Omega^m_{km}
$. (See (\ref{eq:angualrV})) 
	\begin{equation} \label{eq:angualrV}
		\dot{R^m_k}=R_m^k\Omega^k_{mk}
	\end{equation} 

In pedestrian navigation, the body frame rotates in respect to the Earth centered frames. A gyroscope measures the angular velocity of the body frame which moves in respect to an inertial space. The integration of the angular velocity provides the heading. 

The local magnetic field affects the yaw, roll and pitch \cite{8_2}. This research presents an algorithm to estimate the orientation of the object based on the data acquired from the accelerometer and magnetometer. The proposed algorithm uses the magnetic data for determining the rotation of the object in respect to the vertical axis. Use of the quaternions reduces the computation cost while keeping the accuracy at a satisfactory level. 

Quaterniona are also used in \cite{14_5} combined with a Kalman Filter. A Kalman Filter is an algorithm which uses a series of measurements, which are prone to noise, to estimate the signal value. The estimated value is more accurate than the measurement data. Consequently, the Kalman Filter estimates the quaternions error. The magnetic field angular rate also updates with the quaternions parameters. To reduce error in this research, a gradient of the acceleration has been used. 

The Extended Kalman Filter is another Bayesian estimation filter combined with the quaternions in \cite{11_4}. In this research, the full 3D estimation of the user's heading makes the system robust to the sensor movements. 

\begin{figure}[h]
		\includegraphics[scale=.5]{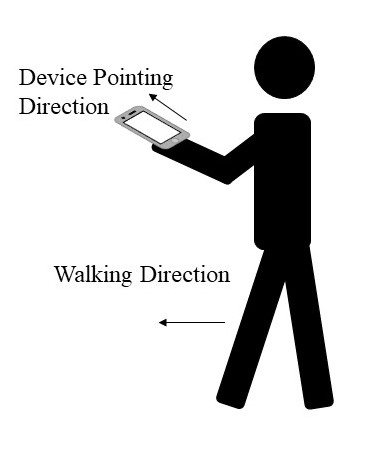}
		\centering
		\caption{Misalignment between the Device Orientation and the Walking Direction}
		\label{fig:misalignment}
	\end{figure}

In recent years the emergence of smart devices encourage the researchers to use their built-in sensors for convenience. The downside of using mobile devices is that they are examples of devices with unknown locations. Unlike the shoe-mounted devices whose forward direction is the same as the sensor frame direction, the mobile devices have unknown locations. Consequently, incomplete information about the sensor placement and movement can cause error in direction estimation. 
	 Considering the information provided by the built-in sensors depends on taking the different scenarios of how they carried into account. An individual might use a cellphone to take picture, make a call, or reply to a text. When not in use, people are likely to place their phone in their trousers pocket, belt bag, backpack, or purse \cite{5_11}. In each of these placements, there is a misalignment between the phone orientation and the user direction (See Figure \ref{fig:misalignment}).
	 
Misalignment between the hand held device orientation and the walking direction or so called heading offset estimation is an open research problem. The orienttion of the cellphone must be approximated to estimate the misalignment. This estimation is central to acquire acceptable accuracy\cite{15_5}. To estimate the orientation of the cellphone in respect to the walking direction different methods have been proposed.

One of the common methods to solve this problem is to use \textit{Principle Component Analysis (PCA)} method \cite{15_5_2,15_5_3, 12_12}. In this method, the projection of the acceleration data on the navigation frame will be calculated. The result is used to estimate the body frame acceleration. To find the forward direction, two assumptions based on the study of bio-mechanics have been made. The first assumption expresses that the variance of the horizontal acceleration is maximum in the forward path. The second assumption is that the minimum value of the variance is in the lateral direction. Given these two assumptions, the forward direction is the unit vector $ u^{ne}$ in the North-East coordinate frame which maximizes (\ref{eq:pca}). 
	
	\begin{equation}\label{eq:pca}
	\underset{\Vert u\Vert =1}{max}(\sum ^n_{i=1}\langle u^{ne}, a^{ne}(i)\rangle^2)
	\end{equation}
	In this $\langle,\rangle $ indicates the scalar product in North-East frame. The number of the samples taken during a step or a stride is indicated with $n$. 

PCA is not the only method to determine the misalignment. Another method to estimate the misalignment is \textit{Forward and lateral accelerations modeling (FLAM)} \cite{15_5_4}. This method is based on the patterns of filtered acceleration. FLAM approximates the direction of the user by matching the estimated acceleration and the predetermined models. The acceleration model has both the forward and lateral predetermined accelerations. Equation (\ref{eq:flam1}) represents the predetermined forward acceleration.
	 
\begin{equation}\label{eq:flam1}
	a_{f}(t)=\sum^n_{i=1}a_i sin (tb_i+c_i)
	\end{equation}
	In this, ${a_i,b_i,c_i}$ represent the model's coefficients. These values change corresponding to the mode of walking and the type of user. Equation (\ref{eq:flam1}) represents the forward acceleration, the same approach is applicable to the lateral acceleration. 
	
The angle which maximize the correlation of the estimated acceleration and the predetermined model is considered as the pedestrian walking direction. Equation (\ref{eq:flam2}) shows the correlation between the between the predetermined acceleration and the estimated acceleration. 
	\begin{equation}\label{eq:flam2}
	h(\theta)=\sum_{i}a_N (i) cos(\theta)+a_E (i) sin(\theta)
	\end{equation}
	In (\ref{eq:flam2}) $a_N, a_E$ are the image of the acceleration vector in North-East coordinate frame on North and East vectors respectively. This maximize when the $\theta$ is equal to heading angle $\theta_h$. The angle for the two predetermined accelerations will be calculated and the final angle is the weighted sum of the acquired values. 
	
\textit{Frequency analysis of Inertial Signals (FIS)} is another method of estimating the heading direction\cite{15_5_6}. This method focuses on maximizing the spectral density of the energy of the accelerometer. This method analyzes the patterns of human movement in the frequency domain. This research assumed that human movement, although complex, has a periodic nature. 

\subsection {Distance} 

To navigate the visually impaired user in an area of interest not only is direction needed but also the distance. In the literature, distance estimation is based on the count of the steps and the length of the strides. \\ Step and stride identification is essential for distance determination. The source of data required for step and stride estimation is provided through the Inertial Measuring Units (IMU)s. As mentioned in heading subsection, the IMU can be used as a single sensor attached to the user or as a sensor implemented in a smartphone. 

The step detection has been explored extensively in the literature \cite{S_13_2}. The nature of the human step is a periodic action. As a result, the step cycle can be used as an indicator of the steps. In this approach the focus is to find the repetitive pattern of the motion signal. As such, repetitive motion examples allow researchers to find the pattern using \textit{peak detection} and \textit{zero crossing} methods.

Step determination based on the peak detection is intuitive. The peak is obvious in the acceleration data in the vertical axis\cite{5_14}. The peak is the result of the heel strike on the ground. The peak might not be unique based on the placement of the sensor. The force of the strike can generate local maximums. To avoid the error generated through the sensor bouncing, more complex algorithms have been exploited.

To avoid some of the complexity associated with peak detection, zero crossing has been used for sensors attached to the body. Zero crossing is a less expensive approach to detect the steps. The motion signal will be equal to zero periodically during human movement.

A waist-worn PDR proposed in \cite{11_4}, uses zero crossing to detect steps. This research detects step based on the movement of the pelvis. The acceleration signal consists of the rise and fall of the pelvis. The extra peaks detection might cause some error in the step detection. To overcome this problem, the system only detects one peak for every zero crossing of the acceleration data. 
 
The motion signal, as mentioned earlier, has a periodic nature. This periodic nature is robust to the sensor placement and will sustain regardless of the sensor site. On the other hand, the placement of the data affects the maximum and the minimum of the motion signal. To overcome this issue, the result of cross-correlation of the motion signal with the template data stored in a pretest phase is calculated.The template has the information of the step or stride stored in it and the result of the cross-correlation will show when a complete step has happened. 

While the motion signal in the time domain is used to calculate the cross-correlation of the signal with the pre-analysis data, the frequency representation of the signal is used to find the dominate frequency of the walking. The frequency spectrum of the motion signal has strong peaks at the walking frequency.

Gait pattern is also used to detect the slope of the path. Using a Gaussian Mixture Model classifier , \cite{9_14} detects if the user is inclining or declining. 

All the aforementioned methods focus on the periodic nature of the motion signal. These methods are useful for step segmentation and detecting swing and stance phase. As their names imply the swing phase is when the foot is moving and the stance phase is when the foot is planted on the ground. If the sensor is placed on the foot, then the stance phase can be detected by determining the time which foot is planted on the ground. This method is mainly used to find the step count. 

In the stance detection method, the amount of time in which the foot is placed on the floor is estimated. To find this time, a minimum and a maximum for the stance phase is defined. On the other hand, this maximum and minimum are defining the boundary of a threshold. 

The step detection methods can also be categorized based on the sensor which provides the data. In general, the acceleration data and gyroscope data are used to detect the steps. The gait segmentation using a gyroscope is studied in \cite{5_12}. According to this research, each gait consists of heel-off, toe-off, heel-strike, and foot planted. This research detects the steps when the angular velocity is less than a threshold. A magnetometer is also a source of data for step detection. In step detection using the magnetometer data, the DC component of the signal will be removed using a high pass filter. The remaining signal which fits in the threshold indicates the steps\cite{10_12}. 

PDR is prone to accumulative error. The measurement errors are inevitable when reading sensor data. More accurate sensors are used in delicate scenarios such as aviation and marine dead reckoning. These sensors are heavy and expensive. Consequently, it is not reasonable to use those for pedestrian navigation. Micro Electro-Mechanical Systems (MEMS) are common in PDR. These sensors are inexpensive and easy to carry. On the other hand they are subject to error.

The measured value of acceleration or the angular velocity is integrated to find the traveled distance or the orientation respectively. The integration in an open loop causes significant error. For example, an error in reading the acceleration data results in cubic growth of the time error. The cubic time error is the result of integration over the acceleration to find the corresponding distance. To suppress the error, some constraints have to be considered to close the integration loop. As mentioned before, correction algorithms are essential to close the integral and prevent accumulative error. 

Error correction methods are essential to estimate the location of the user and navigate them throughout the path accurately. As mentioned earlier, the motion signal has a cyclic pattern. If the sensor is mounted on the body frame, in stance phase it stays stationary. During the stance phase the velocity is zero in theory. The stance phase detection makes the use of \textit{Zero Velocity UPdate (ZUPT)} feasible. 

ZUPT is one of the most common methods that puts constraints on the integral and prevents accumulative error. Although the name, ZUPT, implies that the velocity in the stance phase is zero, in practice a none zero value might be read. As the velocity of the sensor in this phase is known any nonzero value in this phase considered as error. In practice, a narrow threshold is defined for the stance phase. 

In literature, ZUPT is used in two major approaches. In the conventional PDR method, the data provided by IMU is used to estimate the position by integrating the velocity over the time frame. In this method, ZUPT is used so velocity will be manually set to zero when the sensor is in stance phase. It suppresses the incorrect growth of the velocity when integrating the acceleration over time in every time frame \cite{14_7}.

In addition to the traditional PDR estimation, ZUPT has also used in combination with the estimation filters. In an estimation filter, the ZUPT is used as an update for the filter. The nonzero measured value of the velocity indicates a difference with the expected velocity measurement. This difference can serve as error and it is fed to the filter as an error measurement. 
For example, the difference in measured value in stance phase with the expected zero value for the velocity has been fed to the EKF in \cite{14_8, 5_13,10_10}. 

ZUPT is based on the fact that in the stance phase the feet are still, consequently the value of the velocity is zero. The same argument is valid for the angular velocity of the feet in a stance phase. This leads to the generation of the \textit{ Zero Angular Rate Update (ZARU)}. 

ZARU relies on the fact that the tri-axis gyroscope data for motionless feet on the ground is equal to zero. The measurement for the angular velocity in the stance phase is supposed to be zero. Any other value for the measurement is error and can be used in a Bayesian filter. 

For instance, \cite{14_8} not only uses the error associated with the velocity for an EKF but also fed the error of angular velocity to it. 

In addition to the ZUPT and ZARU, which use the velocity and angular velocity respectively, \textit{Zero UNrefined Acceleration update (ZUNA)} uses the acceleration in the stance phase. According to the \cite{14_8}, the measured acceleration during the stance phase is supposed to be equal to the average acceleration in the start point. The average acceleration in this research is defined as the average of the measured acceleration in the first three seconds of motion. The mean value for the acceleration stored and the difference between the measured value and the expected value has been fed to the EKF as error. 

Body movement is not the only source of constraint which can be used to put the integral in a closed loop. The type of path that an individual takes to reach to its destination adds a constraint as well. \textit{Heuristic Drift Reduction (HDR)} first introduced in \cite{9_13} to reduce drift in estimating the location of a vehicle. The idea was that the man-made paths are almost straight. In this approach the likelihood of the vehicle driving in a straight path is estimated. Given the probability of a straight line is high, it corrects the gyroscope measurements. 

The same assumption is true for individuals who are walking in an indoor area with many straight paths and corridors. For pedestrian navigation, HDR estimates the likelihood of human movement in an straight line. Same as in the original algorithm, if the probability of movement in the straight line is high, fluctuations in the gyro measurement will be corrected. 

Although in the original paper, \cite{9_13}, the gyro signals has been filtered with a binary I-controller, in pedestrian navigation a straight line is detected by analyzing successive measurements and studying the changes in the measurement \cite{10_11}. 
Step detection defines the number of steps as discussed earlier. To find the user's displacement, the length of the stride has to be known. Length of the stride is a function of multiple variable such as height of the user, walking speed, and the path condition.

It is intuitive to define a default value for the stride length and run the algorithm. The error of misplacement is fed to the algorithm for correction as suggested by the \cite{14_2}.

The displacement can also be obtained finding the second integration of the acceleration as shown in (\ref{eq:secondA}): 
	\begin{equation}\label{eq:secondA}
	stride=\frac{a\times t^2}{2}
	\end{equation}
An small DC error in the method suggested by the (\ref{eq:secondA}) causes a significant error. To avoid the error correction algorithms have been used. ZUPT as mentioned earlier is one of those correction algorithm. 

ZUPT algorithm is one of the reliable approaches in the stride length detection\cite{9_10}. Finding the length of the stride with this method is based on integrating the acceleration in the time frame between the two consecutive zero velocity. The error of the measurement in this method as mentioned before will be corrected before calculation the displacement. As a result the method is robust to the DC error.

In this method, first the acceleration which was measured in the body frame, $a^b$ must be transferred to the local level $L$. The transformation happens using the rotation matrix $R^L_b$, as shown in (\ref{eq:aTran}):
	\begin{equation}\label{eq:aTran}
	\boldsymbol{a}^L=R^L_b\boldsymbol{a}^b
	\end{equation}
The velocity of the body in respect to the local frame ($\boldsymbol{v}^L$) in the North and East axes obtained by the integration of the acceleration $\boldsymbol{a}^L$. As aforementioned, the integrated velocity suffers from drift. Using the ZUPT method, the drift will be suppressed. The vector of displacement in North-East frame is obtained by integrating the corrected velocity in local frame in one step time. To achieve the length of the stride, the magnitude of the vector of displacement has to be calculated as shown in (\ref{eq:magnitudeDisplacement}):
	\begin{equation}\label{eq:magnitudeDisplacement}
	Stride=\sqrt{\Delta P_{k(north)}^2+\Delta P_{k(east)}^2}
	\end{equation}
As in (\ref{eq:magnitudeDisplacement}), the length of the stride is calculated for every step. The user height and speed impact the length of the stride. The length of stride can vary significantly based on the speed of the user. Slow movement is associated with shorter strides while long strides comes with higher speed. The length of the user's leg provides useful information to reduce the error of determining the stride length. This famous method is known as the Weinberg stride length algorithm \cite{2_3}. 

\begin{figure}[h]
		\includegraphics[scale=.35]{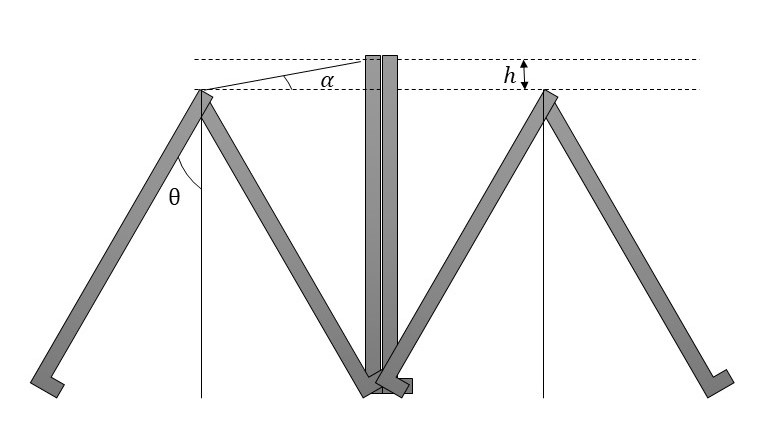}
		\centering
		\caption{Body motion}
		\label{fig:stride}
	\end{figure}

The leg length doesn't change during the movement and the knee only bends when the foot is not on the ground. As figure \ref{fig:stride} shows, the hip and consequently the upper body moves along the vertical axis during movement.
It can be proven with the geometry of the similar triangles that angle $\alpha$ is equal to the angle $\theta$. Consequently, the length of a stride can be obtained as represented in (\ref{eq:lengthStride}):
	\begin{equation}\label{eq:lengthStride}
	stride=\frac{2\times h}{tan(\alpha)}\approx \frac{2\times h}{\alpha}
	\end{equation}
It is obtained by double integrating the acceleration in z-axis. The approximation used in (\ref{eq:lengthStride}) is valid for small angle $\alpha$. In this $h$ is the vertical displacement of the upper body. This can be simplified more by using the empirical value obtained for the $\alpha$ and treating it as an constant. Using this simplification, the stride can be calculated using (\ref{eq:lengthStride2}).
	\begin{equation}\label{eq:lengthStride2}
	stride\approx \sqrt[4]{a_{max}-a_{min}}\times K
	\end{equation}
In equation (\ref{eq:lengthStride2}) $a_{max}$ and $-a_{min}$ are the maximum and the minimum of the acceleration in the z-axis in a single stride respectively. $K$ is the unit conversion constant. 

 Navigation is not always followed after localization. By emergence of powerfull computational devices, fingerprinting and crowdsourcing were also used in finding patterns in navigation to guide the user without estimating the initial location\cite{17_3,17_5}.

This section has been summarized in the table \ref{tab:navigation}. As this table represents a variety of methods are available in the literature for navigation systems. Although this is a very interesting topic, not many researchers have been expanded on this for visually impaired users.

{\renewcommand{\arraystretch}{1.8}
\begin{table*}
  \centering
\begin{tabular}{ |p{5cm}|p{.5cm}|p{5cm}|p{5cm}|  }
 \hline
 \multicolumn{4}{|c|}{Navigation methods} \\
 \hline
 Sub-module & Used for VI & Advantages & Disadvantages\\
 \hline
\hline
   Heading Determination & & & \\
\hline
\hline
IMU  \cite{12_13, 9_2,3_3,7_9,8_2,14_5}& X&\begin{itemize}
    \item Convenient 
    \item Requires low computational power
\end{itemize} & \begin{itemize}
    \item Measurement error
    \item Sensor error
\end{itemize}\\
\hline

 Sign \cite{9_4,5_6,9_3} & X&\begin{itemize}
    \item Provides detailed information  
    \item Requires low computational power
\end{itemize} & \begin{itemize}
    \item Sensitive to the angle of the detector 
\end{itemize}\\
\hline
 Camera  \cite{14_2,19_5} & X& \begin{itemize}
    \item No essential infrastructure  
\end{itemize}  &\begin{itemize}
    \item Delay in detection
    \item Requires high computational power
\end{itemize} \\
 \hline
\hline
   Step Detection & & & \\
\hline
\hline
Pick detection \cite{5_14}& & \begin{itemize}
    \item Convenient 
    \item Requires low computational power
\end{itemize} & \begin{itemize}
    \item Prone to error
    \item Complex error mitigation methods
\end{itemize}\\
\hline
Zero crossing  \cite{11_4,9_14,5_12,10_12,14_7,14_8,5_13,10_10,9_13,10_11}& X&\begin{itemize}
    \item Inexpensive  
    \item Convenient
\end{itemize} & \begin{itemize}
    \item Sensitive to the angle of the detector 
    \item Sensitive to sensor placement
\end{itemize}\\
\hline
\hline
   Stride estimation & & & \\
\hline
\hline
Integration  over speed or velocity \cite{14_2,9_10}  & X& \begin{itemize}
    \item Convenient 
\end{itemize} & \begin{itemize}
    \item Sensitive to DC error
    \item Sensitive to speed variation
\end{itemize}\\
\hline

 Weinberg Algorithm \cite{2_3}  & & \begin{itemize}
    \item Robust to errors
\end{itemize} & \begin{itemize}
    \item Sensitive to user height
    
\end{itemize}

\\
\hline
 \end{tabular}
  \caption{Summary of Navigation Strategies, \\ VI stands for Visually Impaired }
  \label{tab:navigation}
\end{table*}}

\section{Obstacle Avoidance}
\label{ObstacleAvoidance}
Localization and navigation are useless for a visually impaired individual if it does not involve an obstacle avoidance module. The problem with obstacle avoidance is that it can not be hard coded into a map. 

Traditionally, visually impaired individuals use a white cane to avoid obstacle. White canes are useful but they are not free of issue. First they can only detect objects by hitting them, as a result their range of object detection is very small. In addition, suspended objects can not be detected using a white cane. Using a cane requires a training course which cost time and money. It also reserve one of the user's hands for exploring the area. \\ Guide dogs are also common to help the person avoid obstacles. A guide dog finds its way and the person follows the dog. Training a guide dog is difficult, time consuming, and expensive. A trained dog costs between 15k to 20k USD and only functions for five years. In addition, the dog requires nurture and care. To overcome this problem, different methods for obstacle avoidance have been proposed in the literature. The most convenient ones are: 
\begin{itemize}
  \item Sonar
  \item Laser Range Finder
 \item Mono Camera 
\item Stereo Camera.
\end{itemize}

\subsection{Sonar}
 Nature has always inspired curious scientists. One of the early endeavors to avoid obstacles in the path of a visually impaired individual was to imitate bats' echolocation.
Bats are not generally sightless, but they hunt small species in the night and to detect prey they use an interesting method called echolocation. In this method, bats project high frequency sound waves while they are flying. The echo of sound waves are reflected from objects provides them with information about the surrounding area. Bats can estimate the location of objects based on the time it takes for sound wave to travel back to them\cite{bats}. 

Echolocation methods inspired researchers to make a mobility aid for individuals with visual impairment using acoustic signals. This leads to the idea using sonar  \cite{91_1} to detect the object and create a floor plan \cite{17_4}. 

In general, sonar is a range finding method based on sound reflection. In this method an acoustic transmitter and receiver are needed. First the transmitter emits a short acoustic signal. The timer starts counting and stops when the receiver detects the reflection of the acoustic signal. There is a time cap which if the timer exceeds that it will turn off. The distance of the sensor to the object is calculated by multiplying the speed of sound in that climate by half of the time of flight of the sound. The time is divided in two because the signal travels to the object and returns. The following equation,(\ref{eq:sonar1}), shows the formula to calculate the range. 

	\begin{equation}\label{eq:sonar1}
	Dis=c\frac{t}{2}
	\end{equation}

In (\ref{eq:sonar1}), $c$ is the speed of sound. The speed of sound is a function of the medium it travels through. It varies in different climates and locations based on temperature and humidity. With a good approximation, air can be treated as an ideal gas. Given that the medium is an ideal gas, the speed of the sound is calculated as represented in (\ref{eq:speedSound}): 

	\begin{equation}\label{eq:speedSound}
	c=\sqrt{nRT} \sfrac{m}{s}
	\end{equation}
In this, $n$ is the heat capacity ratio. At room temperature, based on the kinetic energy, $n$ is equal to $1.4$. Here $R$ is the ratio of the molar gas constant by the mean molar mass of the dry air which is equal to $287 \frac{m^2}{s^2K}$. The dependency of the speed of sound on temperature is reflected by $T$ in (\ref{eq:speedSound}) and it simplifies to (\ref{eq:speedSoundSimple}) if all the values are substituted. 

\begin{equation}\label{eq:speedSoundSimple}
	c=20\sqrt{T} \sfrac{m}{s}
\end{equation}

According to (\ref{eq:speedSoundSimple}), at $68^{\circ}F$, the speed of the sound equals to 1126.3 $\sfrac{f}{s}$. The speed of sound depends significantly on the temperature. Consequently, a measurement difference of $10^{\circ}$ degree in the temperature leads to about $1\%$ error in the calculated distance. 

\begin{figure}[h]
		\includegraphics[scale=.6]{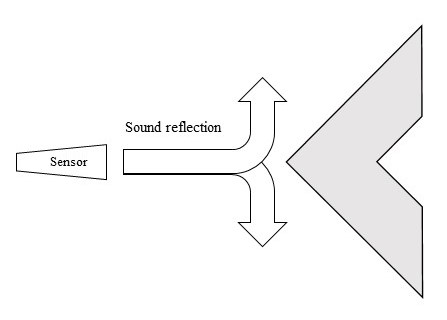}
		\centering
		\caption{Acoustic signal reflection of a object's corner}
		\label{fig:objectCorner}
\end{figure}

Ultrasonic sensors are inexpensive and system implementation is not generally arduous. But this method has some drawbacks. Ultrasonic sensors are limited to the reflection of the sound. The geometry of the object can mislead the user. For example, if the object has a corner facing the user as shown in the figure \ref{fig:objectCorner}, the signal will be reflected in a way which will not be detected by the system. Consequently, the sensor detects would incorrectly detect no obstacle. 

\begin{figure}[h]
		\includegraphics[scale=.6]{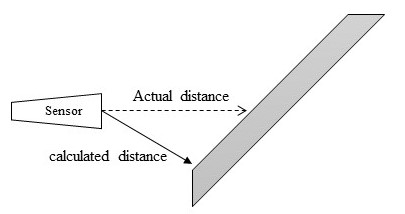}
		\centering
		\caption{Error in Calculation of Distance Due to the Angle of the Sensor to the Object}
		\label{fig:objectAngle}
	\end{figure}

Distance errors caused by the angle of the object is another flaw in this method. For the object shown in figure \ref{fig:objectAngle}, the sensor detects the closest point of the object. As is represented in figure \ref{fig:objectAngle}, although the actual distance of the user with the object is as shown by the dashed line, the sensor detects the solid line as the distance to the object. 

\begin{figure}[h]
		\includegraphics[scale=.6]{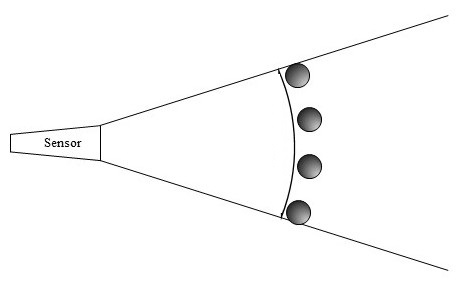}
		\centering
		\caption{Equal Distance Calculation of Objects in Different Locations in Respect to the Sensor}
		\label{fig:equalResponce}
	\end{figure}

The other drawback encountered with using sonar for obstacle avoidance is that the sensor detects the object, but it is unable to detect the exact position of it as shown in figure \ref{fig:equalResponce}. As the figure shows, the sensor can correctly detect every object according to the reflected response, but it can not distinguish among them \cite{1_3}.

To use ultrasonic sensors, the user is required to constantly scan the area to estimate the location of the obstacle. An extra needed effort from user is to approximate the shape of the obstacle to find a path to avoid it. To solve this problem different methods have been proposed. Using an array of ultrasonic sensors facing different directions within a short distance is a solution to these problems and it has been explored in different research articles \cite {7_2}. For instance, in \cite{7_11} an array of ultrasonic sensor has been implemented on the user's jacket.The array of the sensors can scan a wider angle and provide more precise information about the surrounding area. It can also estimate the shape and location of the obstacle more precisely. 

Using an array of sensors is an appealing approach to gather more information about the environment. Using four sensors proposed in \cite{13_7}, the researchers were able to cover not only the area around the user but also to specifically focus on the area in front of the user where usually the white cane is used for obstacle avoidance. Depending on the low cost of the additional sensors, this research proposes a simple and mobile system to reduce the user's reliance on the white cane.

\begin{figure}[h]
		\includegraphics[scale=.5]{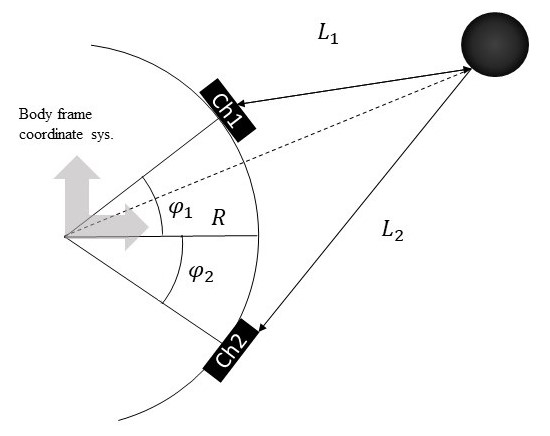}
		\centering
		\caption{Obstacle Detection Based on the Geometry of the Sensors in Ring Architecture}
		\label{fig:geoRing}
	\end{figure}

The effective area of the sensor array depends on the design of sensors' location in respect to each other and in respect to the body. Jung et al proposed using a ring architecture, a sensor array with 16 pairs of transmitters and receivers \cite{7_14}. This architecture allows the system to estimate the precise location of the object based on the geometry of the nearest sensors detecting the obstacle (See figure \ref{fig:geoRing}).

As it is represented in figure \ref{fig:geoRing}, the distance of the object to the adjacent sensor is calculated through the time it takes the signal sent from a sensor in channel one to reach the object and reflect. The reflection is detected by both channel one and two. The corresponding distance between channels one and two and the sensors is $L_1$ and $L_2$ respectively. Given that the radius of the ring architecture and the angles $\phi_1$ and $\phi_2$ are known, the coordinates of the object in the body frame is derived through (\ref{eq:ring1}) and (\ref{eq:ring2}). 

\begin{align}
	(x-R cos(\phi_1)^2+ (x-R sin(\phi_1)=&L_1^2\label{eq:ring1}\\
(x-R cos(\phi_1)^2+ (x-R sin(\phi_1)=&L_1^2\label{eq:ring2}
\end{align}

The aforementioned equations have two answers. The positive value is the coordinate in the direction of the signal propagation and the negative value is in the opposite direction. 

Although having an array of sensors is a decent approach to overcome the problems associated with the ultrasonic sensors, it causes some problem itself. In \cite{4_3} six ultrasonic sensors were implemented in an array to cover a wider angle. Each of the sensors transmit an ultrasonic ping for 10 $\mu$S and detect the reflection of the ping. The small distance between the sensors in the array and the short period between the pings causes cross talk between the sensors. This cross talk misleads the measurements. To avoid the error, the sensors emit the ultrasonic signal in way that no adjacent sensor emits it simultaneously. The sensors divid into two groups of nonadjacent sensors. All of the sensors in a group emit the signal simultaneously and there is a waiting period before the other group sends and receives the signal. 

Ultrasonic sensors have been attached to other tools to help visually impaired individuals. Guide canes are tools which have been improved with ultrasonic sensors to better inform the user. Ulrich et al propose the cane supplied with an array of ultrasonic sensors which can detect the location of the obstacle with adequate precision \cite{1_1}. The ultrasonic sensors cover $120^{\circ}$ in front of the user. Any obstacle in this angle can be detected using the sensors implemented on the cane. The cane is also equipped with a wheel in the bottom. The user pushes the cane slightly. When sensors on the cane detect an obstacle on in the way it will compute the best way to avoid the obstacle and rotates in that direction. The wheel steering generates a noticeable force which the user feels in their hand. The force guides the user in that direction to avoid the detected obstacle.

Installing the sensors on the objects that the user can wear adds a level of convenience. In research studied by Sadi et al the sensors have been attached to the glasses \cite{14_12}. This research uses only one ultrasonic sensor and covers up to three meters in front of the user in $60^{\circ}$ width. The focus of this research is to reduce the cost and the weight associated with the device to make it accessible and available to visually impaired individuals. The ultrasonic sensors have been attached to a hat in \cite{15_13}. The result of testing the design and implementation of this device was positive. 

NavGuide \cite{18_1} installed six ultra sonic sensors on the user's shoes. In addition, a wet floor detector was also installed installed. NavGuide is able to detect the obstacle and provide the user with appropriate feedback. It also warns the visually impaired user about wet floors. NavGuide also prioritizes the information to prevent overwhelming the user with unnecessary information.

A person suffering from visual impairment has the ability to move but depends on the sensors to move. Similarly, robots also have the kinetic energy to move but they can not move without the information provided by sensors to detect obstacles and avoid them. This similarity in the problem statements, motivates the researcher to use the obstacle avoidance techniques employed in robot navigation for the assistive devises whose functionality is to guide the user through obstacles and provide information about the surrounding area. For instance, NavBelt \cite{94_1}, is made of an array of sensors attached to the belt to scan the surrounding area. In this work the information about the surrounding is provided through an array of ultrasonic range finder sensors.

Sonar uses acoustic signals to detect the object. Light signals can also be useful in obstacle avoidance. Obstacle avoidance based on the light waves will be explained in the next subsection. 
\subsection{Laser Range Finder}
Laser range finder, or LIDAR is another method of obstacle avoidance used frequently in the literature  \cite{14_15,10_13,15_10,8_8,12_18,17_9,9_15,15_11,17_8B}. LIDAR stands for \textit{ Light Detection and Ranging}. It follows the same principle as the sonar. In robotics, LIDAR is used extensively \cite{10_13}. The similarity of the obstacle avoidance problem for robots and people with visual impairment encourage the researchers to employ this technique in assistive technologies for visually impaired individuals. In LIDAR, instead of employing acoustic signals, they are substituted by the electromagnetic signals which are emitted by the object. The emitted signal is a narrow focused infrared beam. When the signals make contact with the object, a part of the energy of the system will be absorbed, and a portion will be reflected. The reflection phase and intensity is detected by the sensor and in this method it provides a 3D map of the area.

Similar to sonar, a transmitter and receiver is required in LIDAR.  The transmitter emits the ray of light and starts the counter. The receiver detects the reflected signal. The system calculates the distance between the object and the sensor by multiplying the speed of the light with the half of the time which takes the light to hit the object and reflect. In LIDAR the time divides in half as the light beam travels the distance twice. Equation (\ref{eq:TOF1}) represents the formula to calculate the distance. 

	\begin{equation}\label{eq:TOF1}
	Dis=c_{light}\frac{t}{2}
	\end{equation}
At first, (\ref{eq:TOF1}) is similar to (\ref{eq:sonar1}) which was used to calculate the distance of an object to the senor using sonar. The difference between the two equations lays in the the speed of the emitted signal, $c$. In (\ref{eq:sonar1}), $c$ is the speed of sound, which as (\ref{eq:speedSound})indicates, depends on the temperature and the medium. In (\ref{eq:TOF1}), $c_{light}$ is the speed of light. Speed of light is calculated using the following equation. 
	\begin{equation}\label{eq:speedLight}
	c_{light}=\frac{1}{\sqrt{\epsilon_0\mu_0}}
	\end{equation}

In (\ref{eq:speedLight}), $\epsilon_0$ is the electric constant and $\mu_0$ is the magnetic constant. 
Unlike the speed of sound, speed of light is approximately about $3\times 10^8$ regardless of the temperature. 

As mentioned earlier, the ultrasonic sensors have trouble with detecting the exact location of the objects. An array of ultrasonic sensors addressed this issue to some acceptable extent. To cover a wider area around the user, different scanners are mounted to the laser to sweep the area in multiple directions. These sensors differs from each other is range they can cover and in their resolutions. Consequently, each sensor is capable of scanning a different range and resolution. Combining the result of all of these sensors provides the user with the necessary information of a wider area surrounding them. The location of the obstacle in the body frame is calculated using equations (\ref{eq:laser1}) and (\ref{eq:laser2}). 
\begin{align}
	X_i=R_i\times cos(i)\label{eq:laser1}\\
Y_i=R_i\times sin(i)\label{eq:laser2}
\end{align}

The value for $i$ in (\ref{eq:laser1}) and (\ref{eq:laser2}) depends on the range and the resolution of the laser. For instance, in \cite{15_10} $i\in(-5,185)$. In this research, to avoid missing the obstacle in close distance or if the object is short, the sensor is mounted in a tilted position. The sensor positioning addresses the issue of missing the short obstacles. The distance to the object in this approach is calculated in reference to the angle of the sensor in the body frame. Equations (\ref{eq:laser3}) and (\ref{eq:laser4}) represent the impact of the slope of the view of the laser in this research. 

\begin{align}
	X_i=R_i\times cos(i)\times cos(\alpha)\label{eq:laser3}\\
	Y_i=R_i\times sin(i)\times cos(\alpha)\label{eq:laser4}
\end{align}

In the aforementioned equations, $\alpha$ is the angle of lean in this research. Kineckt is a ubiquitous technology which employs infrared beams for obstacle avoidance purposes\cite{17_9}. 

LIDAR are not free of drawbacks \cite{8_8}. For example it has a limited range. To overcome this problem \cite{12_18} borrowed the idea from bi-focal glasses and proposed using two sensors of short range and long range simultaneously. In this research, two sensors are employed concurrently. The short range sensor covers the area of 20 cm to 150 cm away from the user and the long range sensor includes the area of 1m to 5m from the user. Consequently, the combination of the two sensors cover the areas as close as 20cm up to areas as far as 5 meters. Although this research employs two sensors, the cost of the design has not changed significantly. 

Ultrasonic and LIDAR are used in combination to increase the precision \cite{15_11}. Depth finding cameras are made of LIDAR sensors. Employing them in addition to the ultrasonic sensors is proposed in \cite{17_8B} to increase the accuracy of detecting small obstacles. Adding cameras to the obstacle avoidance modules improves the detection significantly. This topic will be explained in the next subsection.

\subsection{Camera} Camera based methods are prevalent in obstacle avoidance systems. Based on the type of camera which is used, different approaches can be implemented for obstacle avoidance purposes. The primary type of camera to use in the obstacle avoidance scenarios is the monocular camera. As the name of the method conveys, it only uses one camera. Different algorithms have been proposed to detect the obstacle or to find the distance of the camera to the obstacle. 

\begin{figure}[h]
		\includegraphics[scale=.5]{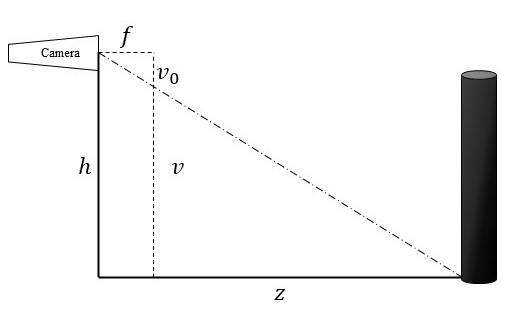}
		\centering
		\caption{Distance Calculation with Monocular Cameras}
		\label{fig:Dis1}
	\end{figure}

The intuitive method to detect the distance of the camera to the obstacle in these approaches is to use the focal distance of the camera \cite{7_15}. The focal distance of the camera is a constant value. Figure \ref{fig:Dis1} represents the geometry of the sensor in respect to the object \cite{17_2}.
As shown in figure \ref{fig:Dis1}, $f$ is the focal distance of the camera. The height of the camera is indicated with $h$. $v_0$ and $v$ are the coordinate of the image and the coordinate of the object respectively. Suppose that the object is on the ground and the camera is pointed at $90^{\circ}$. In this situation, the distance of the object to the sensor, $z$ as shown in the figure\ref{fig:Dis1}, is calculated using (\ref{eq:disCam}). 

	\begin{equation}\label{eq:disCam}
	z=\frac{f\times k_v\times h}{v-v_0}
	\end{equation}
In (\ref{eq:disCam}), $k_v$ represents the pixel density. 

Object detection using monocular cameras has been explored extensively in the literature. Comparing the images to match the template for a specific object is one of the early methods to detect objects \cite{92_2,93_1}. To increase the speed of matching and reducing the required computational power a variety of distance measurement tools and transforms have been used \cite{0_2}. One of these transforms is Hausdroff Distance, which is a measurement tool to find the dissimilarity of two sets. The two sets $P$ and $Q$ are finite point sets as represented in (\ref{eq:p}) and (\ref{eq:q}) respectively: 
\begin{align}
	P=\{ p_1,..., p_m\}\label{eq:p}\\
Q=\{ q_1,..., q_m\}\label{eq:q}
\end{align}

Given that the two sets of $P$ and $Q$ are finite, the Hausdroff Distance measures the dissimilarity as shown in (\ref{eq:hausdroff}): 
\begin{equation}\label{eq:hausdroff}
	H(P,Q)= max(h(P,Q),h(Q,P))
\end{equation}
In (\ref{eq:hausdroff}), $h(P,Q)$ and $h(Q,P)$ are the directed Hausdroff Distances. $(h(P,Q)$ is calculated as represented in (\ref{eq:directed}):
\begin{equation}\label{eq:directed}
	h(P,Q)=\underset{p\in P}{max} (\underset{q\in Q}{min}||p-q||)
\end{equation}

As (\ref{eq:directed}) implies, $h(P,Q)$ and $h(Q,P)$ are not necessary equal. The use of Hausdroff Distance in combination with texture segmentation is proposed in \cite{98_5}. In this research the texture of the local region of the image is compared with the distinct regions to find the obstacles. 

Using symmetry is another well explored method to detect objects in an image. If all the points in an image are identical in respect to a line the image is symmetric. Marola et al. proposes an algorithm to define approximately symmetric images \cite{89_3}. Approximate symmetry can be an indicator of an object in an image.
   
Color segmentation is another means to detect obstacles proposed in the literature. The intuition behind this method is that the area of interest can be divided into sets of obstacles and ground. To detect obstacles based on the color segmentation, three main condition have to be satisfied. 

\begin{itemize}
  \item Obstacle Appearance is different from the ground
  \item The ground is flat
  \item Obstacles are connected to the ground  
\end{itemize}

Assuming that the appearance of the obstacle differs from the ground facilitates the obstacle detection in this method. The difference in the appearance of the obstacle from the ground can be in the color, texture, intensity, etc. 

To detect the obstacle reliably, the selected features have to be adequately distinct in multiple environments. It is also important that the selected attributes require little computational power and are relatively fast. These facilitate the system to detect the obstacle in real-time.

 Color segmentation has been proposed in \cite{0_3} which satisfies the aforementioned conditions and the requirements. In comparison with other attributes, color provides more information. In this method, first the input image is filtered using a Gaussian filter. Filtering the image with a Gaussian filter reduces the level of noise in the image. The next step is to change the image from RGB (Red-Green-Blue) format to the HSI (Hue-Saturation-Intensity) format. The HSI model reduces the sensitivity of the hue and saturation band by separating the color information into the two components of color and intensity. The third step is to define the reference area as the area in front of the user. The hue and intensity values for the pixels inside the area of reference are histogrammed into two graphs. The histogram representation has the advantages of requiring little memory and being computationally fast. In the last step to detect the obstacle, the image pixels are compared to the hue and intensity histograms. If the bin value of the hue or intensity, or the pixel's hue or intensity's value are below the threshold, the pixel is categorized as an obstacle. If none of these conditions are met, then the pixel is categorized as part of the ground.

Using monocular cameras, objects can also be detected with the deformation grid method. In this method a set of vertices, $V$, are initially located on the figure at time $t$. Each vertex, $v$, is connected to four other vertices. Each two adjacent vertices, $v_i$ and $v_j$, are connected through edges. The set of all the edges used in the image is indicated with $E$. Each vertex, $v_i\in \boldsymbol{R^2}$,in this method has a territory, $T_i$. The territory associated with the vertex, $v_i$, is defined as represented in (\ref{eq:territory}):

\begin{equation}\label{eq:territory} 
T_i=\{ x \in \boldsymbol{R^2} |\ ||x-v_i^0||< k_t\rho\}
\end{equation}
In (\ref{eq:territory}), $v_i^0$ indicates the $i$th vertex at time $0$ and $\rho$ is the minimum length of the edges which are connected to the $v_i$ at time $0$ as represented in (\ref{eq:rho}): 
\begin{equation}\label{eq:rho} 
\rho= min(e^0 (||v_i,v_j||), \forall v_j\in N_i
\end{equation}
In the aforementioned equation,(\ref{eq:rho}), $N_i$ is the $i_{th}$ vertex in $V$.

\begin{figure}
    \centering
    \begin{subfigure}[b]{0.3\textwidth}
        \includegraphics[width=\textwidth]{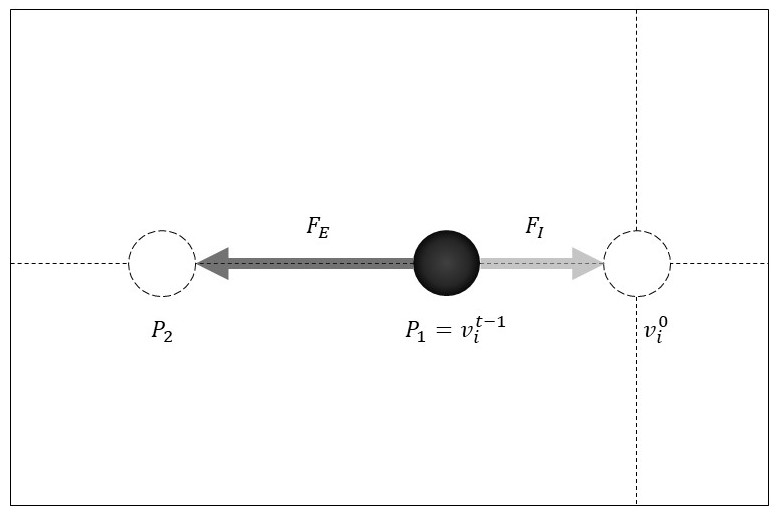}
        \caption{Motion of $v_i$}
        \label{fig:V1}
    \end{subfigure}
    ~ 
    \begin{subfigure}[b]{0.3\textwidth}
        \includegraphics[width=\textwidth]{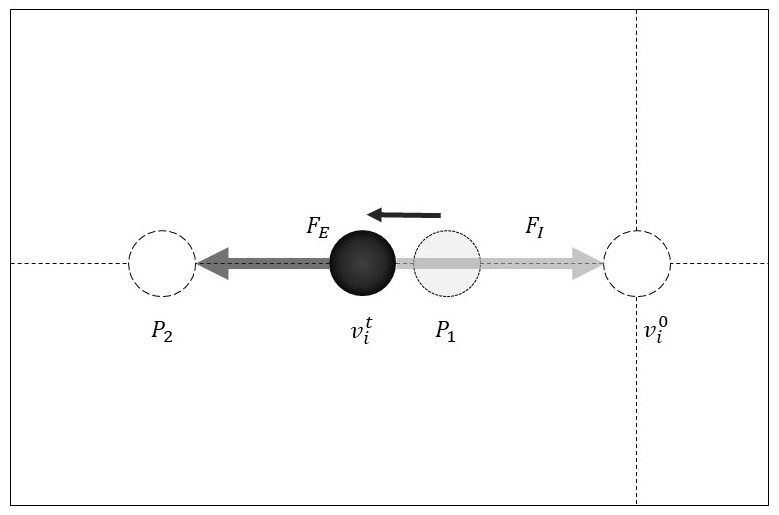}
        \caption{Equilibrium State of $v_i$}
        \label{fig:V2}
    \end{subfigure}
    ~ 
\end{figure}

The vertices in this method have a floating property. It means the vertices can move independently in the next frame. The motion happens based on the position of the pixel in the next frame in respect to the position of the vertex in the current frame. In this method, the motion of the vertices in consecutive frames is explained by defining internal and external forces as shown in figure \ref{fig:V1}.

In figure \ref{fig:V1} $P_1$ is the pixel associated with the vertex $v_i^{t-1}$. Suppose that pixel $P_1$ travels to pixel $P_2$ at time $t$. Two forces are defined to explain the motion of the pixel. The external force $F_E$ and the internal force $F_I$. $F_E$ is the force to push the $v_i$ to move toward the $P_2$ and $F_I$ is the force which tries to keep $v_i$ at its initial position. Vertex $v_i^t$ is the equilibrium point for the vertex $v_i$, where the summation of the forces is equal to zero as shown in figure \ref{fig:V2} (See (\ref{eq:equilibrium})). 
\begin{equation}\label{eq:equilibrium} 
v_i^t=\hat{x} \in \boldsymbol {R^2} s.t. ||F(\hat{x})||=0
\end{equation}
$F$ in (\ref{eq:equilibrium}) is calculated as represented in (\ref{eq:sumF}):
\begin{equation}\label{eq:sumF} 
F(x)= \lambda F_E(x) + F_I(x)
\end{equation}
$\lambda$ is a controlling constant in (\ref{eq:sumF}). Internal and external forces, $F_I$ and $F_E$, are calculated as shown in (\ref{eq:fi}) and (\ref{eq:fe}) respectively: 

\begin{equation}\label{eq:fi} 
F_I(x)= -k_I(x-v_i^0)
\end{equation}
\begin{equation}\label{eq:fe} 
F_E(x)= -k_E(x-P_2)
\end{equation}

 $k_I$ and $k_E$ are the modulus elasticity associated with the $F_I$ and $F_E$ respectively. The perspective projection properties causes the magnitude of the motion vertices associated with the objects approaching to the camera to increase. The motion cause the grid of vertices to deform. The deformation of the grid can be used to detect objects. In this method, severely deformed regions define when deformation is higher than a certain level. It indicates that the object is in such a close location in respect to the camera that risk of collision is high \cite{15_12}. The advantage of the deformation grid method in comparison with other methods is that only two consecutive frames are needed to detect the obstacle. In \cite{17_11}, a function of deformation for each of the vertices has been defined to improve the obstacle detection. The function of deformation is used to find the shape of the change in the vertices. The shape of variation is used to estimate the risk of collision with an obstacle. A vision-based mobile indoor assistive technology proposed by Li et al. employs on board camera with Kalman filter based method to mitigate detect obstacles \cite{19_5}. 
 
While monocular vision based techniques rely on only one camera, stereo vision is a method which employs two cameras to detect the obstacle and estimate the distance of the objects to the cameras. This method has been widely used in obstacle avoidance and object detection. The accuracy of the method is adequate to even detect pedestrians or bicyclists using on-board cameras \cite{6_11}. Stereo cameras follow the same principles that a pair of eyes do. In this method, both cameras are implemented facing the same direction. They are placed adjacent to each other at a short distance. The cameras' are positioned in such a way that their focal axes are in parallel.

In this implementation each camera sees a slightly different view. The difference is caused by the short distance between the cameras' locations. This distance allows for the calculation of the location of the objects using triangulation. To solve the geometry of the triangle and find the distances, the exact position of the cameras in respect to each other and the focal distance of the cameras must be known. Implementation and calibration of the stereo cameras are critical to find the precise location of the obstacles. 

Stereo vision finds the distance to an object by comparing the two images. The main problem in calculating the distance to the object is finding the matching features between the two images. Features must remain consistent with respect to the point-of-view of each camera and the lighting. Feature matching, using a correlation method, has been explained in \cite{98_6}. Correlation method has the following five steps: 
\begin{itemize}
\item Geometry correlation 
\item Image transform
\item Area correlation
\item Extreme extraction
\item Post filtering
\end{itemize} 

In the first step, the image will be wrapped in a standard form, which reduces the distortion in the image. In the image transform step, the image will be transformed to a more appropriate form. Later, in the third step, the search window will sweep the whole image and compare small areas in the images. The forth stage is where the disparity map will be constructed. The highest correlation between the left and right images are depicted in the disparity image. The last step of the procedure is to clean up the noise in the disparity image. 

Stereo vision is difficult because finding the matching features between the images can be cumbersome. Generally, for every point in either of the images there are many possible matches in the other image. The problem is many of these possible matches are similar and it adds a level of ambiguity to the matching process. In stereo vision, the geometry of the cameras are known which can help with limiting the number of potential matches. This leads to \textit{Epipolar geometry}.

\begin{figure}[h]
		\includegraphics[scale=.4]{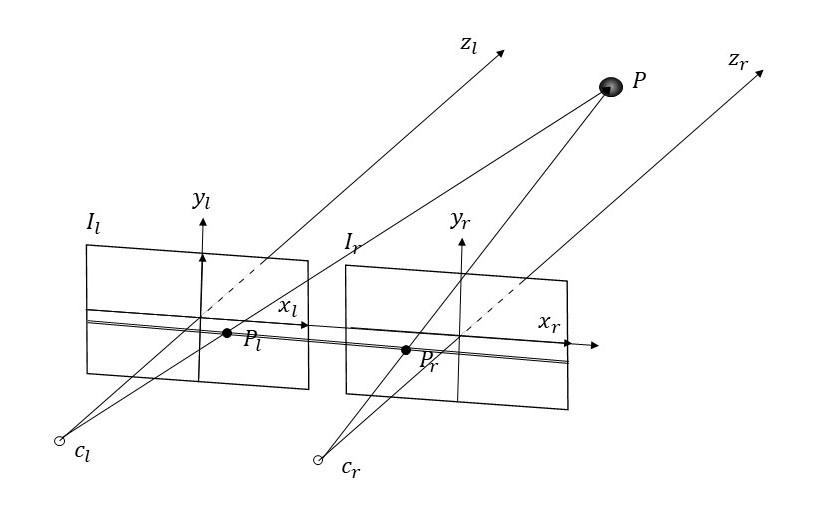}
		\centering
		\caption{Epipolar Plane}
		\label{fig:epipolar}
	\end{figure}

Stereo correlation is mainly based on epipolar geometry. As shown in figure \ref{fig:epipolar}, the center of the two cameras is indicated with $c_l$ and $c_r$. The line connecting the two centers is called the baseline ($b$). Point $P_l$ is the image of point $P$ in the left image. It indicates that point $P$ is on the line $\vec{c_lP_l}$. The three points of $c_l$, $c_r$, and $P_l$ make a plane which is called an epipolar plane. The intersection of the epipolar plane and the image planes generates the epipolar line. This epipolar line is where the projection of point $P$ in the right image can be found. This method limits the possible matches for point $P$ to only the potential matches which exist on the epipolar line. This algorithm is not limited to the scenarios with aligned cameras. An epipolar plane can be constructed in situations in which the cameras are not aligned \cite{stereoVision}.

Given that the pairs are matched, the distance of the object to the camera can be estimated. In figure \ref{fig:epipolar}, $P_l$ and $P_r$ are matching pairs. In this situation the distance of point $P$ to the camera is calculated using:
\begin{equation}\label{eq:depthEstimation} 
Dis=b.\frac{f}{P_1-P_2}
\end{equation}
In the aforementioned equation, $b$ is the baseline and $f$ indicates the focal length of the cameras \cite{14_11}. The distance information provided through the stereo vision procedure enables the system to construct the 3D image of the environment. The dense 3D map, which is made from the disparity map, is useful in detecting the obstacles and estimating their distances to the user. Rodr et al. proposed a polar grid representation to find the potential obstacle \cite{12_16}.

To establish the polar grid in this research, a semicircle is defined. The semicircle covers the area in front of the camera. Suppose the 3D point cloud, which is made by the information provided through the stereo vision, is on top of the semicircle. The polar grid is mapped on this semicircle. The point cloud will be projected onto the semicircle. Consequently, the slots of the semicircle which are the bins in the polar grid will be filled with the number of points from the point cloud on the top of them. In the other words, the 3D point cloud is pressed to make a histogram of the points in a semicircle plane. This 2D histogram is analyzed to detect the obstacle. Where the bins in the histogram has more points the possibility of existence of the obstacle is higher.

The polar grid representation of the obstacle inherits the idea from the Vector Field Histogram \cite{91_3}. In this method, the VFH is updated continuously to generate the 2D cartesian histogram model. To reduce the complexity of the date the 2D histogram is reduced to a one dimensional polar histogram. The values on the polar one dimensional histogram depicts the density of the obstacle. This method is not limited to stereo vision and it is compatible with all the depth estimation methods such as sonar and LIDAR \cite{9_15}. The data reduction in this method helps to provide user with brief information about the surrounding area. In the research proposed by Meers et al. the compact information is conveyed to the user through pulsing gloves \cite{5_5}. 

The ground plane estimation is based on the \textit{Random Sample Consensus (RANSAC)} \cite{12_16}. Estimating models for a set of data which is not refined and has outliers is not free of errors. In unrefined data sets, the outliers have to be eliminated. RANSAC is an iterative non-linear method to estimate the model and simultaneously omit the outliers \cite{12_15}. 

RANSAC is a repetitive technique; however, the process only iterates a few times to reach an acceptable result. In this method, the model is initially estimated based on a random subset of data. At the beginning, the members of this subset are considered to be inliers. The distance of every other point of the data set to the estimated model will be calculated and the points which have a distance less than a determined threshold will be considered as the inliers. The points which have a distance more than threshold will be considered as the outliers. Given that the initial set of the inliers has changed according to the measurements, a new model will be calculated. The new error of inliers, based on the distances, will be calculated in this stage. The whole process will be iterated for a certain number of repetition to find the best model with the least number of errors.

In order to hasten the disparity map creation and make the procedure more efficient, the Rank Transform and Census transform has been used widely in the literature \cite{9_16}. Rank transform is a method of measuring local intensity while Census transform is a summary of the local spacial structure. 

 In these methods, instead of depending on the absolute value of the intensity the relative value of intensity has been used \cite{94_3}. Suppose $P$ is a pixel and $I(P)$ indicates the intensity of this pixel. If a windows with the diameter of $d$ is defined around the pixel $P$, $N(P)$ shows the set of pixels in the windows. In the Rank transform, the intensity of each pixel in the window, is compared with the $I(P)$ ( See (\ref{eq:rank})).
\begin{equation}\label{eq:rank} 
	\xi (P, P')=\begin{cases}
	    1 \ \ \ \ \ \ I(P)>I(P')\\
	   0 \ \ \ \ \ \ I(P)\leq I(P')
	  \end{cases}
\end{equation}
The non-parametric transform for this windows is calculated as depicted in (\ref{eq:Nparametric}):
\begin{equation}\label{eq:Nparametric} 
\Upxi(P)= \underset{P'\in N(P)}{\bigcup} (P',\xi (P, P'))
\end{equation}
The Rank transform consequently is calculated as shown in (\ref{eq:rank2}): 
\begin{equation}\label{eq:rank2}
R(P)=||\{ P' \in N(P)\ |\ I(P')<I(P)\}||
\end{equation}
The rank transform as represented in (\ref{eq:rank2}) is not intensity. It counts the number of pixels that have an intensity higher than the intensity of the pixel $P$. \\ Census transform, $R_t(P)$, maps the pixels surrounding the pixel $P$ into a bit string representation. In this transform, $R_t(P)$ is the set of pixels around pixel $P$ that have intensity less than the intensity of the pixel $P$. The Census transform is calculated using (\ref{eq:census}).
\begin{equation}\label{eq:census} 
R_t(P)= \underset{[i,j]\in D}{\bigotimes} (P,P+[i,j])
\end{equation}
In the aforementioned equation, $\bigotimes$ indicates concatenation and $D$ is the set of displacement, which is calculated as shown in (\ref{eq:D}): 
\begin{align}\label{eq:D} 
N(P)&=P\bigoplus D \\
P\bigoplus D&=\{ p+d | p \in P , d\in D\}
\end{align}
In (\ref{eq:D}), $\bigoplus$ is the Minkowski sum. The similarity of two images using Census transform can be calculated easily with Hamming distance. The relative intensity dependence makes a system robust to the outliers. 

The faster version of Census transform, differential transform, proposed in \cite{13_9}, preserves the robustness of the Census transform to the illumination while making it computationally faster. In this method, instead of binary comparison of the pixels the difference of the intensity is used. 

The emergence of fast computing units and huge data repositories opened a new domain in obstacle detection based on computer vision. In this category of techniques, different decedents of Neural Network architecture have been used to detect the obstacle. In the Neural Network based methods, the obstacle has to be defined for the system in advance. Different object recognition techniques, such as Convolutional Neural Network or Deep Neural Network, are used to find the object in the image and estimate the position of the obstacle in respect to the user. Meers et al. provide an approach to provide the user with a fast and reliable obstacle avoidance module by employing both fast CNN and YOLO \cite{17_2}. 

This section finishes with table \ref{tab:obstacle} which summarizes the information provided in this section. Advantages and disadvantages associated with each technology mentioned in the table. 

{\renewcommand{\arraystretch}{1.8}
\begin{table*}
  \centering
\begin{tabular}{ |p{5cm}|p{.5cm}|p{5cm}|p{5cm}|  }
 \hline
 \multicolumn{4}{|c|}{Obstacle avoidance methods} \\
 \hline
Technology & Used for VI & Advantages & Disadvantages\\
 \hline
\hline
   Sonar \cite{91_1,17_4,1_3,7_2,7_11,13_7, 7_14,4_3,1_1,14_12,15_13,18_1,9_15} & X & \begin{itemize}
    \item Convenient 
    \item Simple algorithm
\end{itemize} & \begin{itemize}
    \item Sensitive to angle of object
    \item Incapable of detecting exact location
\end{itemize}\\

\hline

   Lidar  \cite{14_15,10_13,15_10,8_8,12_18,17_9,9_15,15_11,17_8B}& X  & \begin{itemize}
    \item Precise  
    \item Fast responce
\end{itemize} & \begin{itemize}
    \item Sensitive to color and transparency of the object
    \item Limited range
\end{itemize}\\
\hline

    Camera 
   \begin{itemize}
    \item Mono camera 
    \item Stereo camera
\end{itemize} \cite{ 7_15, 17_2, 92_2, 93_1, 0_2, 98_5, 89_3, 0_3, 15_12, 17_11, 6_11, 98_6, 14_11, 12_16, 9_16, 94_3, 13_9, 13_9, 17_2} &  X &\begin{itemize}
    \item Precise 
    \item Detailed information 
\end{itemize}  & \begin{itemize}
    \item Complex algorithm
    \item Require high computational power
\end{itemize}\\
\hline

 \end{tabular}
  \caption{Summary of Obstacle avoidance Strategies, \\ VI stands for Visually Impaired}
  \label{tab:obstacle}
\end{table*}}

\section{Human-Machine Interface for Assistive Systems}
\label{hmi}
Given all of the technologies available to locate and navigate a visually impaired person that have been discussed so far, there is still the problem of how the user is able to interface with any implemented system. The next few sections will examine  how users wear or hold an assistive system, how users  input information or commands to these systems, and the different methods for giving feedback to the visually impaired users.

\subsection{Device allocation}
One of the design constraints developers must consider is how their assistive system will be held or carried by the end user. Either the users need to have the system mounted on their body or article of clothing or the system needs to be implemented on a portable object such as a cane. This section will look at the different approaches that researchers have taken to provide a means for transporting the navigation system. 

\subsubsection{Wearable}
Having the device be mounted to the user's body has been extensively used in the literature. This method helps to avoid errors associated with direction difference between the device and moving direction. It also prevents the complexity of device position estimation caused by unknown device position. The main categories in this subsection are the following items. 
\begin{itemize}
  \item Head Mounted
  \item Hand Mounted
  \item Chest Mounted
\end{itemize}
\paragraph{Head Mounted}
The first wearable device category to explore is head mounting. Devices located on the head have certain advantages such as being in close proximity to the ears for audio feedback and they can be integrated into common accessories such as glasses or hats. 

Headsets are one type of accessory that can have an assistive technology built in or be used as part of a large system for feedback. Two researchers from the Karadeniz Technical University in Turkey created one such system by integrating ultrasonic obstacle avoidance technologies onto a pair of headphones \cite{16_8}. Another method is to mount the system onto a helmet capable of carrying the weight of the system. One such system used a Microsoft Kinect as a sensor packaged on top of a helmet worn by the visually impaired user that could identify faces from a distance and inform the user \cite{16_9}.

Smart glasses can be used as head mounted systems for the visually impaired. While smart glasses such as the Google Glass failed to take the market by storm, there is continued research in the idea. A common feature in most smart glass systems is to place a camera in the position of the user’s eyes. These could be monocular cameras meant for image processing \cite{15_15,16_10,17_12} or stereoscopic cameras capable of distance measurements \cite{16_11}. Read2Me is a smart glass system that uses OCR on text appearing in the video feed to read the information to the user \cite{16_10}. ThirdEye is a shopping assistive system that uses video from a pair of smart glasses to find the requested item on a shelf and a glove based device direct the user to picking up the item \cite{17_12}.

\paragraph{Hand Mounted}
Hand, wrist, and finger mounted devices are able to take advantage of the natural sensitivity of these places to provide haptic feedback to the visually impaired users. As mentioned above in the ThirdEye system, vibrating gloves have been used in a variety of settings to transmit different types of information to users. Such information may include  distance \cite{16_10,16_12}, commands from a third party \cite{16_12}, or identifying a specific item \cite{17_12} of interest. A sensor can also be attached to a glove mounted system. For example,  the color identify gloves from International Islamic University Chittagong in Bangladesh has color sensors in the palm and uses the rest of the system as a mounting place for the micro-controller \cite{16_13}.

Electronic braille is a surface that can be refreshed to write different information in braille to a visually impaired user. Finger-Eye is a system that had a finger worn camera with an electronic braille surface on top for the user, OCR is used to identify the letters observed by the camera which are transmitted to the braille interface \cite{16_14}.

\paragraph{Chest Mounted}
The chest provides a large area where devices can be placed, hidden in a vest, or hung from the neck. Some researchers have explored using belt mounted systems for placing cameras or ultrasonic sensors as well as motor for haptic feedback\cite{16_15,17_13}. EyeVista is another assistive device wherein all of the parts of the system; the micro-controller, battery, sensors, motors; are incorporated into a vest that allows visually impaired runners to stay on course, avoid other runners, and identify the end of the track \cite{17_14}.

Assistive systems can also be hung from the user's neck as another hands free way of carrying the device. Researchers in Singapore created one such system that involved depth cameras and an embedded processor suspended from the user's neck \cite{17_15}. This system also utilized an attached belt for the haptic feedback to the user \cite{17_15}.

\paragraph{Other Mounting Techniques}
Finally, there have been other attempts at placing the system on other parts of the user. Shoes have been used for both mounting sensors and for providing forms of haptic feedback \cite{14_13}. One of the more unique approaches was to mount the assistive system to a guide dog \cite{17_16}. While initially seeming redundant, the camera and audio system used deep learning to describe what the dog was doing to the visually impaired person to give them more insight about the dog's behavior\cite{17_16}.

\subsubsection{Non-wearable devices}
In addition to the devices which have been attached to the body there are devices which are attached to tools used by visual impaired people. Some well explored examples has been explained in this section. 
\paragraph{Handheld}
Not all electric travel assistive systems are mounted on the user's clothing or accessories. The visually impaired users can instead carry the device. The first common approach is to adapt the historical white-cane for use with new technologies.

\paragraph{Cane Based}
A white-cane can be fitted with different sensors and feedback devices to help assist the visually impaired user. For outdoor navigation, multiple groups have integrated a GPS module into the cane for better localization \cite{15_16,15_17}. Obstacle avoid technology can also be added by attaching distance sensors such as ultrasonic sensors to the cane \cite{15_16,15_17,17_17}.

\paragraph{Cellphone Based}
Many assistive technologies are also realized as smart phone applications. Since the modern smart phone contains a camera and most of the common sensors used for navigation, they are suited for many applications and do not require the user to purchase a new device. The previously discussed Read2Me application also has an implementation that replaces the smart glasses with an Android phone \cite{16_10}. There are also examples of similar programs being designed to help identify medicines, read signs, and read text messages for the visually impaired using smart phones and OCR \cite{16_16,17_18,15_14}. In addition to reading, object detection is another useful feature that can be implemented on a smart phone. VisualPal is one such program that uses a neural network to identify objects and reports the result verbally\cite{15_18}.

\subsection{User Input}
Holding or wearing an assistive device is just one piece of the user interface problem. Each type device or system will need some input from the user be it in the form of a menu, a voice command, or just to be pointed in the right direction. The next section will look at some of the approaches different research groups have tried to allow the visually impaired to use their assistive systems.

\subsubsection{Pointing}
Many designers are able to simply the use of their systems by limiting the required user input and functionality to specific tasks. From smart glasses to medicine identification apps, the preferred method of input is to point the device in the direction the user wants the device to perform its function \cite{15_15,16_10,16_16,17_18}. 

\subsubsection{Menu}
Smart phone based assistive systems for partially visually impaired users are also known use a touch screen menus for input. The previously mentioned VisualPaul uses large buttons on the screen to navigate through the program's options \cite{15_18}. In a similar way, the Read2Me application has two large buttons for the top and bottom of the screen for selecting the user's spoke language and to access the camera \cite{16_10}. 

\subsubsection{Speech and Gesture}
The proliferation of smart home devices such as Google's Alexa and Amazon's Echo which are able to respond to the voice commands of users show another form of input that a visually impaired user can have with an assistive technology. One team of researchers were able to use an android smartphone to received and translate voice commands from a visually impaired user to the assistive device \cite{16_12}. HABOS is a similar system that allows visually impaired users shop online without assistance of another person thanks to a combination of haptic feedback technologies and voice commands \cite{15_9}.

Gesture recognition provides another type of input that can be approached from different angles. One method is to use a subsonic tone and identify gestures based on the Doppler Effect \cite{14_13}. GRIB is a cellphone based program that attempts to recognize gestures using the data measured by the IMU on a smart phone \cite{14_14}. There are also alternative typing methods based on gesture recognition using a computer's touchpad \cite{12_19}.

\subsection{Feedback}
The last topic to be covered is how any of these systems communicate information back to the visually impaired users. Since giving visual information is clearly not the best choice, designers can choose to send information via the sense of touch or sound.
\subsubsection{Haptic}
A feedback sensation that is related to a sense of touch is called haptic feedback. A common example of haptic feedback is the small vibration a smart phone does when a key is pressed. This feedback gives the user the information about the button being pressed. The next few sections will explore some of the forms of haptic feedback used in assistive systems for the visually impaired.
\paragraph{Vibration}
Vibration is a popularly employed method of haptic feedback thanks to the simplicity and ubiquity of small electronic motors. Small motors can be embedded in a smart cane, and the vibrations can be used to alert the user of oncoming obstacles or to direct the user left or right \cite{15_17,17_17}. Vibration feedback has also been used with gloves such as the ThirdEye shopping system that uses vibration to direct the users hand towards the desired item \cite{17_12}. Other researchers have combined vibrating gloves with computer visual to inform the users of obstacles in regions of interest in front of them by changing the vibration intensity of each finger \cite{16_10}.
	
While the hands are very sensitive to tactile stimulation, vibrations can also be felt through other parts of the body. Researchers have also experimented with placing haptic feedback devices onto belts and strapped onto arms \cite{17_15,17_13}. Vibrating footwear has also been explored as a means of giving directions to the visually impaired user with different areas of the foot vibrating to indicated the action the user should take such as turning or stopping \cite{14_13}.

\paragraph{Electronic Braille}
As mentioned in the section on wearable devices for the hand, electronic braille is a type of physical feedback with a small refreshable braille display is used to convey information to a user. The Finger-Eye wearable text reader explored in that section is just one example of electronic braille being used. There are other systems that use much larger electronic braille displays to show more information at one point in time \cite{17_15}. These displays also do not have to be mounted directly on the finger, they can instead be placed on another device or mounted on a belt \cite{17_15}.
\subsubsection{Auditory}
In cases where more detailed information need to be conveyed to the user, researchers will often choose to use audio feedback to be able to include those details. This can be in the form of text-to-speech converts or pre-recorded audio.
\paragraph{Text-To-Speech}
Text-to-speech (TTS) engines allow computers to reproduce the phonetics of text as wave forms. This allows arbitrary textual information to be converted to sounds for visually impaired users. From the previously discussed examples Read2Me, VisualPal, and Voice Helper all use TTS to give feedback to the user in the form of audio \cite{16_10,15_18,15_14,19_5}. TTS methods have also been used to help navigation by reading informative signs, street signs, and billboards to the user as they travel \cite{16_17,17_18}.
\paragraph{Other types of Audio}
Some navigation systems only need to get the user limited audio feedback information that can save on processing time by pre-recording the information. For simple systems, audio feedback can come in the form of a tone that warns the user of a detected object or an acknowledgment of a command \cite{17_17,16_12}. Alternatively, the pre-recorded information could tell the user directions to take, information about a detected object such as its color, or warn them of the direction of an incoming obstacle \cite{15_17,16_13,16_8}.

Some devices have also been designed to take advantage of the human brain’s ability to interpret distance and position based on how a sound is heard. This is referred to as the Head Related Transfer Function (HRTF) and is used in 3D audio applications for both visually impaired and fully signed people. Researchers have tried both converting distance measurements into 3D audio and using information from Stereo vison to create 3D audio to help navigate the visually impaired \cite{17_19,17_20}.

Table \ref{tab:HMI} summarize the information provided in section \ref{hmi}. The tables presents an abstract of available method with some of their advantages and disadvantages.

{\renewcommand{\arraystretch}{1.8}
\begin{table*}
  \centering
\begin{tabular}{ |p{5cm}|p{.5cm}|p{5cm}|p{5cm}|  }
 \hline
 \multicolumn{4}{|c|}{Human-machine interaction methods} \\
 \hline
 Sub-module & Used for VI & Advantages & Disadvantages\\
 \hline
\hline
   Device Allocation & & & \\
\hline
\hline
Wearable  \cite{16_8,16_9,15_15,16_10,16_11,17_12,16_12,16_13,16_14,16_15,17_13,17_14,17_15,14_13,17_16}& X&\begin{itemize}
    \item Robust to direction error 
    \item No device position estimation
\end{itemize} & \begin{itemize}
    \item Weight
    \item Appearance 
\end{itemize}\\
\hline

 Non-wearable \cite{15_16,15_17,17_17,16_10,16_16,17_18,15_14,15_18} & X&\begin{itemize}
    \item Freedom of movement  
\end{itemize} & \begin{itemize}
    \item Sensitive to the position of device
    \item Complex device position estimation algorithm
\end{itemize}\\

\hline
   User Input & & & \\
\hline
\hline
Pointing \cite{15_15,16_10,16_16,17_18}& X& \begin{itemize}
    \item Convenient 
\end{itemize} & \begin{itemize}
    \item Limited
\end{itemize}\\
\hline

Menu \cite{15_18,16_10}& X& \begin{itemize}
    \item Multiple applications 
\end{itemize} & \\
\hline
Speech and Gesture \cite{16_12,15_9,14_13,14_14,12_14,19_5}& X& \begin{itemize}
    \item Detailed information 
\end{itemize} & \begin{itemize}
    \item Requires high computational power
\end{itemize}\\
\hline
\hline
   Feedback & & & \\
\hline
\hline
Haptic \cite{15_17,17_17,17_12, 16_10,17_15,17_13,14_13,19_5}  & X& \begin{itemize}
    \item Fast 
    \item Familiar for VI user
\end{itemize} & \begin{itemize}
    \item Limited
\end{itemize}\\
\hline

Auditory \cite{16_10,15_18,15_14,16_17,17_18,17_17,16_12,15_17,16_13,16_8,17_19,17_20} & X  & \begin{itemize}
    \item Detailed information
\end{itemize} & \begin{itemize}
    \item Requires high computational power
    
\end{itemize}

\\
\hline
 \end{tabular}
  \caption{Summary of Human-Machine Interaction Strategies, \\ VI stands for Visually Impaired}
  \label{tab:HMI}
\end{table*}}

\section{Conclusions and lessons learned}
Navigation is quite a challenging task for visually impaired people. Assistive technologies have been explored extensively in the literature over a long period of time. This article provides the reader with a comprehensive review of the existing  methods. It includes localization and navigation technologies, obstacle avoidance strategies, and human-machine interaction methods that form the core of the assistive devices for indoor navigation.

This work attempts to ease the way for future researchers who are interested in investigating assistive technologies for visually impaired people. Studying for this matter provided the authors with insight into the vast area of research which one needs to know to begin working on assistive technologies. This article provides the future researchers with a summary of available methods in addition to a simplified explanation of their complex algorithmic foundation.This work mainly focused on the methods which are applicable in visually impaired navigation assistive technology. While there are a variety of models available in the literature in this work methods have been critically analyzed through a visually impaired applicability lens. 

\begin{figure}[ht]
    \centering
    
    \includegraphics[scale=.5]{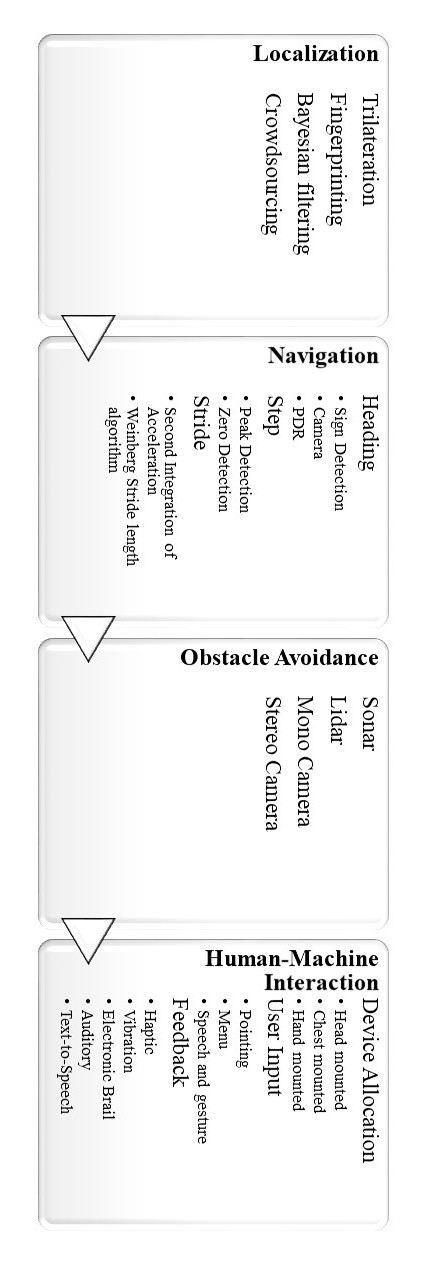}
    \caption{Summary of the Existing Strategies}
    \label{fig:summary}
\end{figure}

For a quick overview, future researcher might consider figure \ref{fig:summary} to determine the method they would like to implement in their project. The first block is the localization block which is the most complex building block to handle when designing a navigation assistive technology for visually impaired individuals. The extensive explanation for each of these methods has been included in section \ref{localization}. Hybrid methods which combine a subset of localization methods in recent years were appealing to the researchers. The hybrid methods have the potential to improve the accuracy and precision of location estimation. 

The next step for a researcher interested in this area is to find the best method to implement to assist a user with the navigation. The navigation building block has smaller components as depicted in the figure \ref{fig:summary}. The type of sensor that the researcher would like to use and its allocation impacts the choice of method in this step.

Localization and navigation have been deeply explored in the literature due to their wide applications. This paper provides a summary of the methods and technologies which can be applied for a visually impaired user. While navigation assistive technologies for visually impaired are meaningless without determining the initial location of the user, not many navigation assistive technology have been proposed for Visually impaired user which determine the initial location of the user. 

In addition, navigation systems for visually impaired user mainly suffer from accumulated error. Measurement error and the complex movement decreases the accuracy and precision of estimation. Using error mitigation methods in measurement such as Bayesian filters would improve the accuracy. Emergence of powerful computational units facilitates using more complex algorithm for location determination and navigation assistance. 

Similar to other problems, complexity makes finding an optimum solution more difficult but it provides us with more conditions to use toward finding the satisfactory solution. In designing the indoor navigation assistance for visually impaired individual, indoor area adds conditions which can be used to improve the precision of prediction of the path that user is taking.  

The third choice to make is to pick a compatible approach for obstacle avoidance. While Sonar and LIDAR based methods have satisfactory performance level, emergence of smart phones and consequently the availability of cameras in addition to the proliferation of image detection methods using convectional neural networks made cameras the most appealing technology for obstacle avoidance. CNN presents impressive performance in object detection in resent years. While its been employed in other application extensively, it didn't attract many researchers in navigation assistive technology design. 

The final step in this project is to decide the method that the device uses to interact with the user. The system designer must consider device allocation, receiving user input and providing a user with the feedback. Natural language processing can improve the performance of a navigation assistive system for visually impaired individual significantly. 

Aggregating these pieces would form the skeleton for a navigation assistive technology for visually impaired individuals. This article is an attempt to assist future researchers in understanding the essential research required for designing and implementing an assistive technology for visually impaired individuals.

	\bibliographystyle{IEEEtran}
	\bibliography{ReviewP}

\begin{thebibliography}{100}
\providecommand{\url}[1]{#1}
\csname url@samestyle\endcsname
\providecommand{\newblock}{\relax}
\providecommand{\bibinfo}[2]{#2}
\providecommand{\BIBentrySTDinterwordspacing}{\spaceskip=0pt\relax}
\providecommand{\BIBentryALTinterwordstretchfactor}{4}
\providecommand{\BIBentryALTinterwordspacing}{\spaceskip=\fontdimen2\font plus
\BIBentryALTinterwordstretchfactor\fontdimen3\font minus
  \fontdimen4\font\relax}
\providecommand{\BIBforeignlanguage}[2]{{%
\expandafter\ifx\csname l@#1\endcsname\relax
\typeout{** WARNING: IEEEtran.bst: No hyphenation pattern has been}%
\typeout{** loaded for the language `#1'. Using the pattern for}%
\typeout{** the default language instead.}%
\else
\language=\csname l@#1\endcsname
\fi
#2}}
\providecommand{\BIBdecl}{\relax}
\BIBdecl

\bibitem{whoo}
\BIBentryALTinterwordspacing
WHO, ``Blindness and vision impairment prevention.'' [Online]. Available:
  \url{http://www.who.int/blindness/world_sight_day/2017/en/}
\BIBentrySTDinterwordspacing

\bibitem{5_8}
G.~Dedes and A.~G. Dempster, ``Indoor gps positioning - challenges and
  opportunities,'' in \emph{VTC-2005-Fall. 2005 IEEE 62nd Vehicular Technology
  Conference, 2005.}, vol.~1, Sept 2005, pp. 412--415.

\bibitem{90_1}
\BIBentryALTinterwordspacing
R.~Passini, G.~Proulx, and C.~Rainville, ``The spatio-cognitive abilities of
  the visually impaired population,'' \emph{Environment and Behavior}, vol.~22,
  no.~1, pp. 91--118, 1990. [Online]. Available:
  \url{https://doi.org/10.1177/0013916590221005}
\BIBentrySTDinterwordspacing

\bibitem{S_3_1}
V.~{Fox}, J.~{Hightower}, , D.~{Schulz}, and G.~{Borriello}, ``Bayesian
  filtering for location estimation,'' \emph{IEEE Pervasive Computing}, vol.~2,
  no.~3, pp. 24--33, July 2003.

\bibitem{S_6_1}
R.~Peng and M.~L. Sichitiu, ``Angle of arrival localization for wireless sensor
  networks,'' in \emph{2006 3rd Annual IEEE Communications Society on Sensor
  and Ad Hoc Communications and Networks}, vol.~1, Sep. 2006, pp. 374--382.

\bibitem{S_7_1}
H.~{Liu}, H.~{Darabi}, P.~{Banerjee}, and J.~{Liu}, ``Survey of wireless indoor
  positioning techniques and systems,'' \emph{IEEE Transactions on Systems,
  Man, and Cybernetics, Part C (Applications and Reviews)}, vol.~37, no.~6, pp.
  1067--1080, Nov 2007.

\bibitem{S_9_1}
I.~Guvenc and C.~Chong, ``A survey on toa based wireless localization and nlos
  mitigation techniques,'' \emph{IEEE Communications Surveys Tutorials},
  vol.~11, no.~3, pp. 107--124, rd 2009.

\bibitem{S_9_2}
I.~Amundson and X.~D. Koutsoukos, ``A survey on localization for mobile
  wireless sensor networks,'' in \emph{Mobile Entity Localization and Tracking
  in GPS-less Environnments}, R.~Fuller and X.~D. Koutsoukos, Eds.\hskip 1em
  plus 0.5em minus 0.4em\relax Berlin, Heidelberg: Springer Berlin Heidelberg,
  2009, pp. 235--254.

\bibitem{S_9_3}
Y.~{Gu}, A.~{Lo}, and I.~{Niemegeers}, ``A survey of indoor positioning systems
  for wireless personal networks,'' \emph{IEEE Communications Surveys
  Tutorials}, vol.~11, no.~1, pp. 13--32, First 2009.

\bibitem{S_10_1}
\BIBentryALTinterwordspacing
J.~Wang, R.~K. Ghosh, and S.~K. Das, ``A survey on sensor localization,''
  \emph{Journal of Control Theory and Applications}, vol.~8, no.~1, pp. 2--11,
  Feb 2010. [Online]. Available:
  \url{https://doi.org/10.1007/s11768-010-9187-7}
\BIBentrySTDinterwordspacing

\bibitem{S_11_2}
M.~Z. {Win}, A.~{Conti}, S.~{Mazuelas}, Y.~{Shen}, W.~M. {Gifford},
  D.~{Dardari}, and M.~{Chiani}, ``Network localization and navigation via
  cooperation,'' \emph{IEEE Communications Magazine}, vol.~49, no.~5, pp.
  56--62, May 2011.

\bibitem{S_13_1}
\BIBentryALTinterwordspacing
G.~Han, H.~Xu, T.~Q. Duong, J.~Jiang, and T.~Hara, ``Localization algorithms of
  wireless sensor networks: a survey,'' \emph{Telecommunication Systems},
  vol.~52, no.~4, pp. 2419--2436, Apr 2013. [Online]. Available:
  \url{https://doi.org/10.1007/s11235-011-9564-7}
\BIBentrySTDinterwordspacing

\bibitem{S_15_1}
Q.~D. {Vo} and P.~{De}, ``A survey of fingerprint-based outdoor localization,''
  \emph{IEEE Communications Surveys Tutorials}, vol.~18, no.~1, pp. 491--506,
  Firstquarter 2016.

\bibitem{S_15_2}
L.~{Mainetti}, L.~{Patrono}, and I.~{Sergi}, ``A survey on indoor positioning
  systems,'' in \emph{2014 22nd International Conference on Software,
  Telecommunications and Computer Networks (SoftCOM)}, Sep. 2014, pp. 111--120.

\bibitem{S_15_3}
D.~{Dardari}, P.~{Closas}, and P.~M. {Djurić}, ``Indoor tracking: Theory,
  methods, and technologies,'' \emph{IEEE Transactions on Vehicular
  Technology}, vol.~64, no.~4, pp. 1263--1278, April 2015.

\bibitem{S_15_4}
\BIBentryALTinterwordspacing
Z.~Yang, C.~Wu, Z.~Zhou, X.~Zhang, X.~Wang, and Y.~Liu, ``Mobility increases
  localizability: A survey on wireless indoor localization using inertial
  sensors,'' \emph{ACM Comput. Surv.}, vol.~47, no.~3, pp. 54:1--54:34, Apr.
  2015. [Online]. Available: \url{http://doi.acm.org/10.1145/2676430}
\BIBentrySTDinterwordspacing

\bibitem{S_16_1}
S.~{He} and S.~.~G. {Chan}, ``Wi-fi fingerprint-based indoor positioning:
  Recent advances and comparisons,'' \emph{IEEE Communications Surveys
  Tutorials}, vol.~18, no.~1, pp. 466--490, Firstquarter 2016.

\bibitem{S_17_1}
A.~{Yassin}, Y.~{Nasser}, M.~{Awad}, A.~{Al-Dubai}, R.~{Liu}, C.~{Yuen},
  R.~{Raulefs}, and E.~{Aboutanios}, ``Recent advances in indoor localization:
  A survey on theoretical approaches and applications,'' \emph{IEEE
  Communications Surveys Tutorials}, vol.~19, no.~2, pp. 1327--1346,
  Secondquarter 2017.

\bibitem{S_17_2}
P.~{Davidson} and R.~{Piché}, ``A survey of selected indoor positioning
  methods for smartphones,'' \emph{IEEE Communications Surveys Tutorials},
  vol.~19, no.~2, pp. 1347--1370, Secondquarter 2017.

\bibitem{S_18_1}
C.~{Laoudias}, A.~{Moreira}, S.~{Kim}, S.~{Lee}, L.~{Wirola}, and
  C.~{Fischione}, ``A survey of enabling technologies for network localization,
  tracking, and navigation,'' \emph{IEEE Communications Surveys Tutorials},
  vol.~20, no.~4, pp. 3607--3644, Fourthquarter 2018.

\bibitem{S_18_4}
\BIBentryALTinterwordspacing
P.~Strumillo, M.~Bujacz, P.~Baranski, P.~Skulimowski, P.~Korbel, M.~Owczarek,
  K.~Tomalczyk, A.~Moldoveanu, and R.~Unnthorsson, \emph{Different Approaches
  to Aiding Blind Persons in Mobility and Navigation in the ``Naviton'' and
  ``Sound of Vision'' Projects}.\hskip 1em plus 0.5em minus 0.4em\relax Cham:
  Springer International Publishing, 2018, pp. 435--468. [Online]. Available:
  \url{https://doi.org/10.1007/978-3-319-54446-5_15}
\BIBentrySTDinterwordspacing

\bibitem{S_19_1}
\BIBentryALTinterwordspacing
F.~Zafari, A.~Gkelias, and K.~K. Leung, ``A survey of indoor localization
  systems and technologies,'' \emph{CoRR}, vol. abs/1709.01015, 2017. [Online].
  Available: \url{http://arxiv.org/abs/1709.01015}
\BIBentrySTDinterwordspacing

\bibitem{S_19_2}
\BIBentryALTinterwordspacing
M.~Sattarian, J.~Rezazadeh, R.~Farahbakhsh, and A.~Bagheri, ``Indoor navigation
  systems based on data mining techniques in internet of things: A survey,''
  \emph{Wirel. Netw.}, vol.~25, no.~3, pp. 1385--1402, Apr. 2019. [Online].
  Available: \url{https://doi.org/10.1007/s11276-018-1766-4}
\BIBentrySTDinterwordspacing

\bibitem{S_13_2}
R.~Harle, ``A survey of indoor inertial positioning systems for pedestrians,''
  \emph{IEEE Communications Surveys Tutorials}, vol.~15, no.~3, pp. 1281--1293,
  Third 2013.

\bibitem{S_16_2}
K.~Gade, ``The seven ways to find heading,'' \emph{Journal of Navigation},
  vol.~69, no.~5, p. 955–970, 2016.

\bibitem{S_16_3}
F.~{Woyano}, S.~{Lee}, and S.~{Park}, ``Evaluation and comparison of
  performance analysis of indoor inertial navigation system based on foot
  mounted imu,'' in \emph{2016 18th International Conference on Advanced
  Communication Technology (ICACT)}, Jan 2016, pp. 792--798.

\bibitem{S_8_1}
B.~C. Carter, M.~Vershinin, and S.~P. Gross, ``A comparison of step-detection
  methods: how well can you do?'' \emph{Biophysical journal}, vol.~94, 2008.

\bibitem{S_18_2}
A.~Shitsukane, W.~Cheriuyot, C.~Otieno, and M.~Mgala, ``A survey on obstacles
  avoidance mobile robot in static unknown environment,'' \emph{International
  Journal of Computer (IJC)}, 03 2018.

\bibitem{S_1_1}
S.~B.~Modi, P.~Chandak, V.~Sagar~Murty, and E.~Hall, ``Comparison of three
  obstacle-avoidance methods for a mobile robot,'' \emph{Proceedings of SPIE -
  The International Society for Optical Engineering}, vol. 4572, 10 2001.

\bibitem{S_89_1}
C.~H.R~Everett, ``Survey of collision avoidance and ranging sensors for mobile
  robots,'' \emph{Robotics and Autonomous Systems}, vol.~5, p. 5–67, 05 1989.

\bibitem{S_16_4}
R.~Ismail, Z.~Omar, and S.~Suaibun, ``Obstacle-avoiding robot with ir and pir
  motion sensors,'' \emph{IOP Conference Series: Materials Science and
  Engineering}, vol. 152, p. 012064, 10 2016.

\bibitem{S_5_1}
R.~J. {Radke}, S.~{Andra}, O.~{Al-Kofahi}, and B.~{Roysam}, ``Image change
  detection algorithms: a systematic survey,'' \emph{IEEE Transactions on Image
  Processing}, vol.~14, no.~3, pp. 294--307, March 2005.

\bibitem{S_14_1}
H.~S. Parekh, D.~G. Thakore, and U.~K. Jaliya, ``A survey on object detection
  and tracking methods,'' 2014.

\bibitem{S_16_5}
P.~H.L, ``A survey on moving object detection and tracking techniques,''
  \emph{International Journal Of Engineering And Computer Science}, 04 2016.

\bibitem{S_16_6}
G.~Cheng and J.~Han, ``A survey on object detection in optical remote sensing
  images,'' \emph{CoRR}, vol. abs/1603.06201, 2016.

\bibitem{S_18_3}


\bibitem{S_9_4}
D.~{Dakopoulos} and N.~G. {Bourbakis}, ``Wearable obstacle avoidance electronic
  travel aids for blind: A survey,'' \emph{IEEE Transactions on Systems, Man,
  and Cybernetics, Part C (Applications and Reviews)}, vol.~40, no.~1, pp.
  25--35, Jan 2010.

\bibitem{S_8_2}
\BIBentryALTinterwordspacing
J.~Zhang, S.~K. Ong, and A.~Y.~C. Nee, ``Navigation systems for individuals
  with visual impairment: A survey,'' in \emph{Proceedings of the 2Nd
  International Convention on Rehabilitation Engineering \& Assistive
  Technology}, ser. iCREATe '08.\hskip 1em plus 0.5em minus 0.4em\relax Kaki
  Bukit TechPark II,, Singapore: Singapore Therapeutic, Assistive \&
  Rehabilitative Technologies (START) Centre, 2008, pp. 159--162. [Online].
  Available: \url{http://dl.acm.org/citation.cfm?id=1983222.1983264}
\BIBentrySTDinterwordspacing

\bibitem{S_13_3}
K.~M. Varpe and M.~P. Wankhade, ``Survey of visually impaired assistive
  system,'' 2013.

\bibitem{S_17_3}
W.~Elmannai and K.~Elleithy, ``Sensor-based assistive devices for
  visually-impaired people: Current status, challenges, and future
  directions,'' \emph{Sensors}, vol.~17, p. 565, 03 2017.

\bibitem{91_2}
J.~S. Abel and J.~W. Chaffee, ``Existence and uniqueness of gps solutions,''
  \emph{IEEE Transactions on Aerospace and Electronic Systems}, vol.~27, no.~6,
  pp. 952--956, Nov 1991.

\bibitem{85_1}
J.~M. Loomis, ``Digital map and navigation system for the visually impaired,''
  1985.

\bibitem{85_2}
C.~C. Collins, \emph{On Mobility Aids for the Blind}.\hskip 1em plus 0.5em
  minus 0.4em\relax Dordrecht: Springer Netherlands, 1985, pp. 35--64.

\bibitem{89_1}
D.~A. Brusnighan, M.~G. Strauss, J.~M. Floyd, and B.~C. Wheeler, ``Orientation
  aid implementing the global positioning system,'' in \emph{Proceedings of the
  Fifteenth Annual Northeast Bioengineering Conference}, Mar 1989, pp. 33--34.

\bibitem{94_2}
\BIBentryALTinterwordspacing
J.~M. Loomis, R.~G. Golledge, R.~L. Klatzky, J.~M. Speigle, and J.~Tietz,
  ``Personal guidance system for the visually impaired,'' in \emph{Proceedings
  of the First Annual ACM Conference on Assistive Technologies}, ser. Assets
  '94.\hskip 1em plus 0.5em minus 0.4em\relax New York, NY, USA: ACM, 1994, pp.
  85--91. [Online]. Available: \url{http://doi.acm.org/10.1145/191028.191051}
\BIBentrySTDinterwordspacing

\bibitem{00_1}
Q.~Ladetto, ``On foot navigation: continuous step calibration using both
  complementary recursive prediction and adaptive kalman filtering,'' I.~G.
  2000, Ed., 2000.

\bibitem{12_4}
\BIBentryALTinterwordspacing
S.~Kammoun, G.~Parseihian, O.~Gutierrez, A.~Brilhault, A.~Serpa, M.~Raynal,
  B.~Oriola, M.-M. Macé, M.~Auvray, M.~Denis, S.~Thorpe, P.~Truillet, B.~Katz,
  and C.~Jouffrais, ``Navigation and space perception assistance for the
  visually impaired: The navig project,'' \emph{IRBM}, vol.~33, no.~2, pp. 182
  -- 189, 2012, numéro spécial ANR TECSANTechnologie pour la santé et
  l'autonomie. [Online]. Available:
  \url{http://www.sciencedirect.com/science/article/pii/S1959031812000103}
\BIBentrySTDinterwordspacing

\bibitem{10_7}
K.~Yelamarthi, D.~Haas, D.~Nielsen, and S.~Mothersell, ``Rfid and gps
  integrated navigation system for the visually impaired,'' in \emph{2010 53rd
  IEEE International Midwest Symposium on Circuits and Systems}, Aug 2010, pp.
  1149--1152.

\bibitem{9_17}
R.~{Mautz}, ``The challenges of indoor environments and specification on some
  alternative positioning systems,'' in \emph{2009 6th Workshop on Positioning,
  Navigation and Communication}, March 2009, pp. 29--36.

\bibitem{12_10}
V.~Gupta and M.~Rohil, ``Bit-stuffing in 802. 11 beacon frame: Embedding
  non-standard custom information,'' vol.~63, pp. 6--12, 02 2013.

\bibitem{12_13}
\BIBentryALTinterwordspacing
L.~A. Guerrero, F.~Vasquez, and S.~F. Ochoa, ``An indoor navigation system for
  the visually impaired,'' \emph{Sensors (Basel)}, vol.~12, no.~6, pp.
  8236--8258, Jun 2012, sensors-12-08236[PII]. [Online]. Available:
  \url{http://www.ncbi.nlm.nih.gov/pmc/articles/PMC3436027/}
\BIBentrySTDinterwordspacing

\bibitem{90_2}
B.~T. Fang, ``Simple solutions for hyperbolic and related position fixes,''
  \emph{IEEE Transactions on Aerospace and Electronic Systems}, vol.~26, no.~5,
  pp. 748--753, Sep 1990.

\bibitem{17_23}
A.~{Zayets} and E.~{Steinbach}, ``Robust wifi-based indoor localization using
  multipath component analysis,'' in \emph{2017 International Conference on
  Indoor Positioning and Indoor Navigation (IPIN)}, Sep. 2017, pp. 1--8.

\bibitem{87_1}
G.~C. Carter, ``Coherence and time delay estimation,'' \emph{Proceedings of the
  IEEE}, vol.~75, no.~2, pp. 236--255, Feb 1987.

\bibitem{76_1}
C.~Knapp and G.~Carter, ``The generalized correlation method for estimation of
  time delay,'' \emph{IEEE Transactions on Acoustics, Speech, and Signal
  Processing}, vol.~24, no.~4, pp. 320--327, Aug 1976.

\bibitem{13_1}
K.~Liu, X.~Liu, and X.~Li, ``Guoguo: Enabling fine-grained smartphone
  localization via acoustic anchors,'' \emph{IEEE Transactions on Mobile
  Computing}, vol.~15, no.~5, pp. 1144--1156, May 2016.

\bibitem{12_9}
\BIBentryALTinterwordspacing
H.~Liu, Y.~Gan, J.~Yang, S.~Sidhom, Y.~Wang, Y.~Chen, and F.~Ye, ``Push the
  limit of wifi based localization for smartphones,'' in \emph{Proceedings of
  the 18th Annual International Conference on Mobile Computing and Networking},
  ser. Mobicom '12.\hskip 1em plus 0.5em minus 0.4em\relax New York, NY, USA:
  ACM, 2012, pp. 305--316. [Online]. Available:
  \url{http://doi.acm.org/10.1145/2348543.2348581}
\BIBentrySTDinterwordspacing

\bibitem{12_1}
R.~Nandakumar, K.~Chintalapudi, and V.~N. Padmanabhan, ``Centaur: locating
  devices in an office environment,'' in \emph{MobiCom}, 2012.

\bibitem{3_2}
and and and, ``Ultrasonic self-localization and pose tracking of an autonomous
  mobile robot via fuzzy adaptive extended information filtering,'' in
  \emph{2003 IEEE International Conference on Robotics and Automation (Cat.
  No.03CH37422)}, vol.~1, Sep. 2003, pp. 1283--1290 vol.1.

\bibitem{98_2}
C.~Drane, M.~Macnaughtan, and C.~Scott, ``Positioning gsm telephones,''
  \emph{IEEE Communications Magazine}, vol.~36, no.~4, pp. 46--54, 59, Apr
  1998.

\bibitem{12_11}
R.~Kaune, ``Accuracy studies for tdoa and toa localization,'' in \emph{2012
  15th International Conference on Information Fusion}, July 2012, pp.
  408--415.

\bibitem{7_4}
\BIBentryALTinterwordspacing
T.~Manodham, L.~Loyola, and T.~Miki, ``A novel wireless positioning system for
  seamless internet connectivity based on the wlan infrastructure,''
  \emph{Wireless Personal Communications}, vol.~44, no.~3, pp. 295--309, Feb
  2008. [Online]. Available: \url{https://doi.org/10.1007/s11277-007-9373-1}
\BIBentrySTDinterwordspacing

\bibitem{15_2}
\BIBentryALTinterwordspacing
A.~S. Martinez-Sala, F.~Losilla, J.~C. Sánchez-Aarnoutse, and J.~García-Haro,
  ``Design, implementation and evaluation of an indoor navigation system for
  visually impaired people,'' \emph{Sensors}, vol.~15, no.~12, pp.
  32\,168--32\,187, 2015. [Online]. Available:
  \url{http://www.mdpi.com/1424-8220/15/12/29912}
\BIBentrySTDinterwordspacing

\bibitem{17_26}
S.~{Yang} and B.~{Wang}, ``Residual based weighted least square algorithm for
  bluetooth/uwb indoor localization system,'' in \emph{2017 36th Chinese
  Control Conference (CCC)}, July 2017, pp. 5959--5963.

\bibitem{14_9}
\BIBentryALTinterwordspacing
A.~Badawy, T.~Khattab, D.~Trinchero, T.~M. Elfouly, and A.~Mohamed, ``A simple
  aoa estimation scheme,'' \emph{CoRR}, vol. abs/1409.5744, 2014. [Online].
  Available: \url{http://arxiv.org/abs/1409.5744}
\BIBentrySTDinterwordspacing

\bibitem{12_14}
H.~C. Chen, T.~H. Lin, H.~T. Kung, C.~K. Lin, and Y.~Gwon, ``Determining rf
  angle of arrival using cots antenna arrays: A field evaluation,'' in
  \emph{MILCOM 2012 - 2012 IEEE Military Communications Conference}, Oct 2012,
  pp. 1--6.

\bibitem{13_10}
D.~Bartlett, \emph{Radiolocation technologies}, ser. The Cambridge Wireless
  Essentials Series.\hskip 1em plus 0.5em minus 0.4em\relax Cambridge
  University Press, 2013, p. 59–87.

\bibitem{book_digital}
B.~Sklar, \emph{Digital Communications: Fundamentals and Applications}.\hskip
  1em plus 0.5em minus 0.4em\relax Upper Saddle River, NJ, USA: Prentice-Hall,
  Inc., 1988.

\bibitem{book_wireless}
T.~Rappaport, \emph{Wireless Communications: Principles and Practice},
  2nd~ed.\hskip 1em plus 0.5em minus 0.4em\relax Upper Saddle River, NJ, USA:
  Prentice Hall PTR, 2001.

\bibitem{book_wireless2}
A.~Goldsmith, \emph{Wireless Communications}.\hskip 1em plus 0.5em minus
  0.4em\relax New York, NY, USA: Cambridge University Press, 2005.

\bibitem{11_3}
\BIBentryALTinterwordspacing
A.~Goswami, L.~E. Ortiz, and S.~R. Das, ``Wigem: A learning-based approach for
  indoor localization,'' in \emph{Proceedings of the Seventh COnference on
  Emerging Networking EXperiments and Technologies}, ser. CoNEXT '11.\hskip 1em
  plus 0.5em minus 0.4em\relax New York, NY, USA: ACM, 2011, pp. 3:1--3:12.
  [Online]. Available: \url{http://doi.acm.org/10.1145/2079296.2079299}
\BIBentrySTDinterwordspacing

\bibitem{10_2}
\BIBentryALTinterwordspacing
H.~Lim, L.-C. Kung, J.~C. Hou, and H.~Luo, ``Zero-configuration indoor
  localization over ieee 802.11 wireless infrastructure,'' \emph{Wireless
  Networks}, vol.~16, no.~2, pp. 405--420, Feb 2010. [Online]. Available:
  \url{https://doi.org/10.1007/s11276-008-0140-3}
\BIBentrySTDinterwordspacing

\bibitem{18_2}
W.~Kui, S.~Mao, X.~Hei, and F.~Li, ``Towards accurate indoor localization using
  channel state information,'' \emph{2018 IEEE International Conference on
  Consumer Electronics-Taiwan (ICCE-TW)}, pp. 1--2, 2018.

\bibitem{9_8}
F.~Seco, A.~R. Jimenez, C.~Prieto, J.~Roa, and K.~Koutsou, ``A survey of
  mathematical methods for indoor localization,'' in \emph{2009 IEEE
  International Symposium on Intelligent Signal Processing}, Aug 2009, pp.
  9--14.

\bibitem{17_8}
C.~Li, Z.~Qiu, and C.~Liu, ``An improved weighted k-nearest neighbor algorithm
  for indoor positioning,'' vol.~96, pp. 1--13, 05 2017.

\bibitem{17_24}
M.~M. {Rahman}, V.~{Moghtadaiee}, and A.~G. {Dempster}, ``Design of
  fingerprinting technique for indoor localization using am radio signals,'' in
  \emph{2017 International Conference on Indoor Positioning and Indoor
  Navigation (IPIN)}, Sep. 2017, pp. 1--7.

\bibitem{17_27}
S.~{Alraih}, A.~{Alhammadi}, I.~{Shayea}, and A.~M. {Al-Samman}, ``Improving
  accuracy in indoor localization system using fingerprinting technique,'' in
  \emph{2017 International Conference on Information and Communication
  Technology Convergence (ICTC)}, Oct 2017, pp. 274--277.

\bibitem{5_9}
\BIBentryALTinterwordspacing
M.~Brunato and R.~Battiti, ``Statistical learning theory for location
  fingerprinting in wireless lans,'' \emph{Comput. Netw.}, vol.~47, no.~6, pp.
  825--845, Apr. 2005. [Online]. Available:
  \url{http://dx.doi.org/10.1016/j.comnet.2004.09.004}
\BIBentrySTDinterwordspacing

\bibitem{17_22}
and F.~{Nashashibi}, , and E.~{Castelli}, ``Indoor intelligent vehicle
  localization using wifi received signal strength indicator,'' in \emph{2017
  IEEE MTT-S International Conference on Microwaves for Intelligent Mobility
  (ICMIM)}, March 2017, pp. 33--36.

\bibitem{18_3}
C.-S. Hsu, Y.-S. Chen, T.-Y. Juang, and Y.-T. Wu, ``An adaptive wi-fi indoor
  localization scheme using deep learning,'' 08 2018, pp. 132--133.

\bibitem{18_7}
\BIBentryALTinterwordspacing
A.~Mittal, S.~Tiku, and S.~Pasricha, ``Adapting convolutional neural networks
  for indoor localization with smart mobile devices,'' in \emph{Proceedings of
  the 2018 on Great Lakes Symposium on VLSI}, ser. GLSVLSI '18.\hskip 1em plus
  0.5em minus 0.4em\relax New York, NY, USA: ACM, 2018, pp. 117--122. [Online].
  Available: \url{http://doi.acm.org/10.1145/3194554.3194594}
\BIBentrySTDinterwordspacing

\bibitem{5_2}
Z.~Li, W.~Trappe, Y.~Zhang, and B.~Nath, ``Robust statistical methods for
  securing wireless localization in sensor networks,'' in \emph{IPSN 2005.
  Fourth International Symposium on Information Processing in Sensor Networks,
  2005.}, April 2005, pp. 91--98.

\bibitem{9_6}
\BIBentryALTinterwordspacing
M.~Azizyan, I.~Constandache, and R.~Roy~Choudhury, ``Surroundsense: Mobile
  phone localization via ambience fingerprinting,'' in \emph{Proceedings of the
  15th Annual International Conference on Mobile Computing and Networking},
  ser. MobiCom '09.\hskip 1em plus 0.5em minus 0.4em\relax New York, NY, USA:
  ACM, 2009, pp. 261--272. [Online]. Available:
  \url{http://doi.acm.org/10.1145/1614320.1614350}
\BIBentrySTDinterwordspacing

\bibitem{17_3}
\BIBentryALTinterwordspacing
X.~Zhou, D.~Guo, and X.~Teng, ``Magspider: A localization-free approach for
  constructing global indoor map for navigation purpose,'' in \emph{Proceedings
  of the ACM Turing 50th Celebration Conference - China}, ser. ACM TUR-C
  '17.\hskip 1em plus 0.5em minus 0.4em\relax New York, NY, USA: ACM, 2017, pp.
  44:1--44:10. [Online]. Available:
  \url{http://doi.acm.org/10.1145/3063955.3064000}
\BIBentrySTDinterwordspacing

\bibitem{12_8}
\BIBentryALTinterwordspacing
S.~Sen, B.~Radunovic, R.~R. Choudhury, and T.~Minka, ``You are facing the mona
  lisa: Spot localization using phy layer information,'' in \emph{Proceedings
  of the 10th International Conference on Mobile Systems, Applications, and
  Services}, ser. MobiSys '12.\hskip 1em plus 0.5em minus 0.4em\relax New York,
  NY, USA: ACM, 2012, pp. 183--196. [Online]. Available:
  \url{http://doi.acm.org/10.1145/2307636.2307654}
\BIBentrySTDinterwordspacing

\bibitem{14_2}
\BIBentryALTinterwordspacing
J.~Dong, Y.~Xiao, M.~Noreikis, Z.~Ou, and A.~Yl\"{a}-J\"{a}\"{a}ski, ``imoon:
  Using smartphones for image-based indoor navigation,'' in \emph{Proceedings
  of the 13th ACM Conference on Embedded Networked Sensor Systems}, ser. SenSys
  '15.\hskip 1em plus 0.5em minus 0.4em\relax New York, NY, USA: ACM, 2015, pp.
  85--97. [Online]. Available: \url{http://doi.acm.org/10.1145/2809695.2809722}
\BIBentrySTDinterwordspacing

\bibitem{kalman1960}
R.~E. Kalman, ``A new approach to linear filtering and prediction problems,''
  \emph{ASME Journal of Basic Engineering}, 1960.

\bibitem{2_1}
F.~{Gustafsson}, F.~{Gunnarsson}, N.~{Bergman}, U.~{Forssell}, J.~{Jansson},
  R.~{Karlsson}, and P.~. {Nordlund}, ``Particle filters for positioning,
  navigation, and tracking,'' \emph{IEEE Transactions on Signal Processing},
  vol.~50, no.~2, pp. 425--437, Feb 2002.

\bibitem{5_10}
S.~Thrun, W.~Burgard, and D.~Fox, \emph{Probabilistic Robotics (Intelligent
  Robotics and Autonomous Agents)}.\hskip 1em plus 0.5em minus 0.4em\relax The
  MIT Press, 2005.

\bibitem{4_6}
J.~Hightower and G.~Borriello, ``Particle filters for location estimation in
  ubiquitous computing: A case study,'' in \emph{UbiComp 2004: Ubiquitous
  Computing}, N.~Davies, E.~D. Mynatt, and I.~Siio, Eds.\hskip 1em plus 0.5em
  minus 0.4em\relax Berlin, Heidelberg: Springer Berlin Heidelberg, 2004, pp.
  88--106.

\bibitem{8_5}
\BIBentryALTinterwordspacing
O.~Woodman and R.~Harle, ``Pedestrian localisation for indoor environments,''
  in \emph{Proceedings of the 10th International Conference on Ubiquitous
  Computing}, ser. UbiComp '08.\hskip 1em plus 0.5em minus 0.4em\relax New
  York, NY, USA: ACM, 2008, pp. 114--123. [Online]. Available:
  \url{http://doi.acm.org/10.1145/1409635.1409651}
\BIBentrySTDinterwordspacing

\bibitem{16_3}
J.~Dong, Y.~Xiao, Z.~Ou, Y.~Cui, and A.~Yla-Jaaski, ``Indoor tracking using
  crowdsourced maps,'' in \emph{2016 15th ACM/IEEE International Conference on
  Information Processing in Sensor Networks (IPSN)}, April 2016, pp. 1--6.

\bibitem{14_10}
\BIBentryALTinterwordspacing
R.~Gao, M.~Zhao, T.~Ye, F.~Ye, Y.~Wang, K.~Bian, T.~Wang, and X.~Li, ``Jigsaw:
  Indoor floor plan reconstruction via mobile crowdsensing,'' in
  \emph{Proceedings of the 20th Annual International Conference on Mobile
  Computing and Networking}, ser. MobiCom '14.\hskip 1em plus 0.5em minus
  0.4em\relax New York, NY, USA: ACM, 2014, pp. 249--260. [Online]. Available:
  \url{http://doi.acm.org/10.1145/2639108.2639134}
\BIBentrySTDinterwordspacing

\bibitem{15_3}
J.~Dong, Y.~Xiao, Z.~Ou, and A.~Ylä-Jääski, ``Utilizing internet photos for
  indoor mapping and localization - opportunities and challenges,'' in
  \emph{2015 IEEE Conference on Computer Communications Workshops (INFOCOM
  WKSHPS)}, April 2015, pp. 636--641.

\bibitem{sfm}
\BIBentryALTinterwordspacing
C.~Wu, ``Visualsfm : A visual structure from motion system.'' [Online].
  Available: \url{http://ccwu.me/vsfm/}
\BIBentrySTDinterwordspacing

\bibitem{15_9}
S.~Chen, M.~Li, K.~Ren, and C.~Qiao, ``Crowd map: Accurate reconstruction of
  indoor floor plans from crowdsourced sensor-rich videos,'' in \emph{2015 IEEE
  35th International Conference on Distributed Computing Systems}, June 2015,
  pp. 1--10.

\bibitem{15_4}
J.~Niu, B.~Wang, L.~Cheng, and J.~J. P.~C. Rodrigues, ``Wicloc: An indoor
  localization system based on wifi fingerprints and crowdsourcing,'' in
  \emph{2015 IEEE International Conference on Communications (ICC)}, June 2015,
  pp. 3008--3013.

\bibitem{13_4}
V.~Radu and M.~K. Marina, ``Himloc: Indoor smartphone localization via activity
  aware pedestrian dead reckoning with selective crowdsourced wifi
  fingerprinting,'' in \emph{International Conference on Indoor Positioning and
  Indoor Navigation}, Oct 2013, pp. 1--10.

\bibitem{5_7}
S.~Willis and S.~Helal, ``Rfid information grid for blind navigation and
  wayfinding,'' in \emph{Ninth IEEE International Symposium on Wearable
  Computers (ISWC'05)}, Oct 2005, pp. 34--37.

\bibitem{7_7}
------, ``Rfid information grid for blind navigation and wayfinding,'' in
  \emph{Ninth IEEE International Symposium on Wearable Computers (ISWC'05)},
  Oct 2005, pp. 34--37.

\bibitem{7_8}
E.~D'Atri, C.~M. Medaglia, A.~Serbanati, U.~B. Ceipidor, E.~Panizzi, and
  A.~D'Atri, ``A system to aid blind people in the mobility: A usability test
  and its results,'' in \emph{Systems, 2007. ICONS '07. Second International
  Conference on}, April 2007, pp. 35--35.

\bibitem{9_9}
\BIBentryALTinterwordspacing
U.~B. Ceipidor, C.~Medaglia, A.~Serbanati, G.~Azzalin, M.~Barboni, F.~Rizzo,
  and M.~Sironi, ``Sesamonet: an rfid-based economically viable navigation
  system for the visually impaired,'' \emph{International Journal of RF
  Technologies: Research and Applications}, vol.~1, no.~3, pp. 214--224, 2009.
  [Online]. Available: \url{https://doi.org/10.1080/17545730903039806}
\BIBentrySTDinterwordspacing

\bibitem{4_4}
\BIBentryALTinterwordspacing
L.~M. Ni, Y.~Liu, Y.~C. Lau, and A.~P. Patil, ``Landmarc: Indoor location
  sensing using active rfid,'' \emph{Wireless Networks}, vol.~10, no.~6, pp.
  701--710, Nov 2004. [Online]. Available:
  \url{https://doi.org/10.1023/B:WINE.0000044029.06344.dd}
\BIBentrySTDinterwordspacing

\bibitem{5_1}
K.~Wendlandt, M.~Berhig, and P.~Robertson, ``Indoor localization with
  probability density functions based on bluetooth,'' in \emph{2005 IEEE 16th
  International Symposium on Personal, Indoor and Mobile Radio Communications},
  vol.~3, Sept 2005, pp. 2040--2044 Vol. 3.

\bibitem{14_3}
X.~Zhao, Z.~Xiao, A.~Markham, N.~Trigoni, and Y.~Ren, ``Does {BTLE} measure up
  against {WiFi}? a comparison of indoor location performance,'' in
  \emph{European Wireless 2014; 20th European Wireless Conference}, May 2014,
  pp. 1--6.

\bibitem{16_5}
Y.~Zhuang, J.~Yang, Y.~Li, L.~Qi, and N.~El-Sheimy, ``Smartphone-based indoor
  localization with bluetooth low energy beacons,'' \emph{Sensors}, vol.~16,
  no.~5, 2016.

\bibitem{14_4}
S.~Liu, Y.~Jiang, and A.~Striegel, ``Face-to-face proximity estimationusing
  bluetooth on smartphones,'' \emph{IEEE Transactions on Mobile Computing},
  vol.~13, no.~4, pp. 811--823, April 2014.

\bibitem{17_21}
X.~{Hou} and T.~{Arslan}, ``Monte carlo localization algorithm for indoor
  positioning using bluetooth low energy devices,'' in \emph{2017 International
  Conference on Localization and GNSS (ICL-GNSS)}, June 2017, pp. 1--6.

\bibitem{92_2}
D.~P. Huttenlocher, W.~J. Rucklidge, and G.~A. Klanderman, ``Comparing images
  using the hausdorff distance under translation,'' in \emph{Proceedings 1992
  IEEE Computer Society Conference on Computer Vision and Pattern Recognition},
  June 1992.

\bibitem{12_15}
B.~Taylor, D.~J. Lee, D.~Zhang, and G.~Xiong, ``Smart phone-based indoor
  guidance system for the visually impaired,'' in \emph{2012 12th International
  Conference on Control Automation Robotics Vision (ICARCV)}, Dec 2012, pp.
  871--876.

\bibitem{14_14}
Y.~Huang, Z.~Xu, R.~Wang, and D.~Chen, ``Grib: Gesture recognition interaction
  with mobile devices for blind people,'' in \emph{2014 IEEE International
  Conference on Computer and Information Technology}, Sept 2014, pp. 604--609.

\bibitem{14_8}
M.~B. Dehkordi, A.~Frisoli, E.~Sotgiu, and C.~Loconsole, ``Pedestrian indoor
  navigation system using inertial measurement unit,'' 2014.

\bibitem{5_13}
E.~Foxlin, ``Pedestrian tracking with shoe-mounted inertial sensors,''
  \emph{IEEE Computer Graphics and Applications}, vol.~25, no.~6, pp. 38--46,
  Nov 2005.

\bibitem{10_10}
A.~R.~J. Ruiz, F.~S. Granja, J.~C.~P. Honorato, and J.~I.~G. Rosas,
  ``Pedestrian indoor navigation by aiding a foot-mounted imu with rfid signal
  strength measurements,'' in \emph{2010 International Conference on Indoor
  Positioning and Indoor Navigation}, Sept 2010, pp. 1--7.

\bibitem{19_5}
B.~{Li}, J.~P. {Muñoz}, X.~{Rong}, Q.~{Chen}, J.~{Xiao}, Y.~{Tian},
  A.~{Arditi}, and M.~{Yousuf}, ``Vision-based mobile indoor assistive
  navigation aid for blind people,'' \emph{IEEE Transactions on Mobile
  Computing}, vol.~18, no.~3, pp. 702--714, March 2019.

\bibitem{98_1}
\BIBentryALTinterwordspacing
R.~Jacobson, ``Cognitive mapping without sight: four preliminary studies of
  spatial learning,'' \emph{Journal of Environmental Psychology}, vol.~18,
  no.~3, pp. 289 -- 305, 1998. [Online]. Available:
  \url{http://www.sciencedirect.com/science/article/pii/S0272494498900986}
\BIBentrySTDinterwordspacing

\bibitem{98_3}
R.~G. R.M.~Kitchin, M.~Blades, ``{Pedestrian dead reckoning: A basis for
  personal positioning},'' \emph{"Progress in Human Geography"}, "1998".

\bibitem{9_4}
A.~Mulloni, D.~Wagner, I.~Barakonyi, and D.~Schmalstieg, ``Indoor positioning
  and navigation with camera phones,'' \emph{IEEE Pervasive Computing}, vol.~8,
  no.~2, pp. 22--31, April 2009.

\bibitem{5_6}
B.~S. Tjan, P.~J. Beckmann, R.~Roy, N.~Giudice, and G.~E. Legge, ``Digital sign
  system for indoor wayfinding for the visually impaired,'' in \emph{2005 IEEE
  Computer Society Conference on Computer Vision and Pattern Recognition
  (CVPR'05) - Workshops}, June 2005, pp. 30--30.

\bibitem{9_3}
J.~COUGHLAN and R.~MANDUCHI, ``Functional assessment of a camera phone-based
  wayfinding system operated by blind and visually impaired users,''
  \emph{International journal of artificial intelligence tools : architectures,
  languages, algorithms}, vol.~18, June 2009.

\bibitem{1_2}
R.~Jirawimut, P.~Ptasinski, V.~Garaj, F.~Cecelja, and W.~Balachandran, ``A
  method for dead reckoning parameter correction in pedestrian navigation
  system,'' in \emph{IMTC 2001. Proceedings of the 18th IEEE Instrumentation
  and Measurement Technology Conference. Rediscovering Measurement in the Age
  of Informatics (Cat. No.01CH 37188)}, vol.~3, 2001, pp. 1554--1558 vol.3.

\bibitem{7_12}
C.~S. Kallie, P.~R. Schrater, and G.~E. Legge, ``Variability in stepping
  direction explains the veering behavior of blind walkers,'' \emph{J Exp
  Psychol Hum Percept Perform}, vol.~33, February 2007.

\bibitem{Inertial_Navigation}
A.~Noureldin, T.~B. Karamat, and J.~Georgy, \emph{Fundamentals of Inertial
  Navigation, Satellite-based Positioning and their Integration}, 1st~ed.\hskip
  1em plus 0.5em minus 0.4em\relax Springer-Verlag Berlin Heidelberg, an
  optional note.

\bibitem{9_2}
K.~Kunze, P.~Lukowicz, K.~Partridge, and B.~Begole, ``Which way am i facing:
  Inferring horizontal device orientation from an accelerometer signal,'' in
  \emph{2009 International Symposium on Wearable Computers}, Sept 2009, pp.
  149--150.

\bibitem{3_3}
D.~Mizell, ``Using gravity to estimate accelerometer orientation,'' in
  \emph{Seventh IEEE International Symposium on Wearable Computers, 2003.
  Proceedings.}, Oct 2003, pp. 252--253.

\bibitem{7_9}
K.~Kunze and P.~Lukowicz, ``Using acceleration signatures from everyday
  activities for on-body device location,'' in \emph{2007 11th IEEE
  International Symposium on Wearable Computers}, Oct 2007, pp. 115--116.

\bibitem{ASEN3200}
P.~AxeIrad and D.~Lawrence, ``Lecture notes in orbital mechanics attitude
  dynamics and control,'' Spring 2006.

\bibitem{Quaternion}
\BIBentryALTinterwordspacing
E.~of~Mathematics, ``Quaternion.'' [Online]. Available:
  \url{https://www.encyclopediaofmath.org/index.php/Quaternion}
\BIBentrySTDinterwordspacing

\bibitem{8_2}
X.~Yun, E.~R. Bachmann, and R.~B. McGhee, ``A simplified quaternion-based
  algorithm for orientation estimation from earth gravity and magnetic field
  measurements,'' \emph{IEEE Transactions on Instrumentation and Measurement},
  vol.~57, no.~3, pp. 638--650, March 2008.

\bibitem{14_5}
V.~Renaudin, C.~Combettes, and F.~Peyret, ``Quaternion based heading estimation
  with handheld mems in indoor environments,'' in \emph{2014 IEEE/ION Position,
  Location and Navigation Symposium - PLANS 2014}, May 2014, pp. 645--656.

\bibitem{11_4}
P.~Goyal, V.~J. Ribeiro, H.~Saran, and A.~Kumar, ``Strap-down pedestrian
  dead-reckoning system,'' in \emph{2011 International Conference on Indoor
  Positioning and Indoor Navigation}, Sept 2011, pp. 1--7.

\bibitem{5_11}
F.~Ichikawa, J.~Chipchase, and R.~Grignani, ``Where's the phone? a study of
  mobile phone location in public spaces,'' in \emph{2005 2nd Asia Pacific
  Conference on Mobile Technology, Applications and Systems}, Nov 2005, pp.
  1--8.

\bibitem{15_5}
C.~Combettes and V.~Renaudin, ``Comparison of misalignment estimation
  techniques between handheld device and walking directions,'' in \emph{2015
  International Conference on Indoor Positioning and Indoor Navigation (IPIN)},
  Oct 2015, pp. 1--8.

\bibitem{15_5_2}
\BIBentryALTinterwordspacing
J.~Janardhanan, G.~Dutta, and V.~Tripuraneni, ``Attitude estimation for
  pedestrian navigation using low cost mems accelerometer in mobile
  applications, and processing methods, apparatus and systems,'' Apr.~8 2014,
  uS Patent 8,694,251. [Online]. Available:
  \url{https://www.google.ch/patents/US8694251}
\BIBentrySTDinterwordspacing

\bibitem{15_5_3}
\BIBentryALTinterwordspacing
A.~ALI, H.~Chang, J.~Georgy, Z.~Syed, and C.~Goodall, ``Method and apparatus
  for determination of misalignment between device and pedestrian,'' Jul.~24
  2014, wO Patent App. PCT/CA2014/000,040. [Online]. Available:
  \url{https://www.google.com/patents/WO2014110672A1?cl=en}
\BIBentrySTDinterwordspacing

\bibitem{12_12}
\BIBentryALTinterwordspacing
Z.~Syed, J.~Georgy, C.~Goodall, A.~Noureldin, N.~El-Sheimy, and M.~Atia,
  ``Methods of attitude and misalignment estimation for constraint free
  portable navigation,'' Sep.~27 2012, uS Patent App. 13/426,884. [Online].
  Available: \url{https://www.google.com/patents/US20120245839}
\BIBentrySTDinterwordspacing

\bibitem{15_5_4}
\BIBentryALTinterwordspacing
M.~Chowdhary, M.~Sharma, A.~Kumar, S.~Dayal, and M.~Jain, ``Method and
  apparatus for determining walking direction for a pedestrian dead reckoning
  process,'' May~22 2014, uS Patent App. 13/682,684. [Online]. Available:
  \url{https://www.google.com/patents/US20140142885}
\BIBentrySTDinterwordspacing

\bibitem{15_5_6}
M.~Kourogi and T.~Kurata, ``A method of pedestrian dead reckoning for
  smartphones using frequency domain analysis on patterns of acceleration and
  angular velocity,'' in \emph{2014 IEEE/ION Position, Location and Navigation
  Symposium - PLANS 2014}, May 2014, pp. 164--168.

\bibitem{5_14}
L.~Fang, P.~J. Antsaklis, L.~A. Montestruque, M.~B. McMickell, M.~Lemmon,
  Y.~Sun, H.~Fang, I.~Koutroulis, M.~Haenggi, M.~Xie, and X.~Xie, ``Design of a
  wireless assisted pedestrian dead reckoning system - the navmote
  experience,'' \emph{IEEE Transactions on Instrumentation and Measurement},
  vol.~54, no.~6, pp. 2342--2358, Dec 2005.

\bibitem{9_14}
N.~Wang, E.~Ambikairajah, S.~J. Redmond, B.~G. Celler, and N.~H. Lovell,
  ``Classification of walking patterns on inclined surfaces from accelerometry
  data,'' in \emph{2009 16th International Conference on Digital Signal
  Processing}, July 2009, pp. 1--4.

\bibitem{5_12}
F.~Cavallo, A.~M. Sabatini, and V.~Genovese, ``A step toward gps/ins personal
  navigation systems: real-time assessment of gait by foot inertial sensing,''
  in \emph{2005 IEEE/RSJ International Conference on Intelligent Robots and
  Systems}, Aug 2005, pp. 1187--1191.

\bibitem{10_12}
I.~Skog, J.~O. Nilsson, and P.~Händel, ``Evaluation of zero-velocity detectors
  for foot-mounted inertial navigation systems,'' in \emph{2010 International
  Conference on Indoor Positioning and Indoor Navigation}, Sept 2010, pp. 1--6.

\bibitem{14_7}
K.~Abdulrahim, T.~Moore, C.~Hide, and C.~Hill, ``Understanding the performance
  of zero velocity updates in mems-based pedestrian navigation,'' vol.~5, 03
  2014.

\bibitem{9_13}
J.~Borenstein and L.~Ojeda, ``Heuristic reduction of gyro drift in vehicle
  tracking applications.''

\bibitem{10_11}
A.~R. Jiménez, F.~Seco, J.~C. Prieto, and J.~Guevara, ``Indoor pedestrian
  navigation using an ins/ekf framework for yaw drift reduction and a
  foot-mounted imu,'' in \emph{2010 7th Workshop on Positioning, Navigation and
  Communication}, March 2010, pp. 135--143.

\bibitem{9_10}
A.~R. Jimenez, F.~Seco, C.~Prieto, and J.~Guevara, ``A comparison of pedestrian
  dead-reckoning algorithms using a low-cost mems imu,'' in \emph{2009 IEEE
  International Symposium on Intelligent Signal Processing}, Aug 2009, pp.
  37--42.

\bibitem{2_3}
{ Harvey Weinberg}, ``{Using the ADXL202 in Pedometer and Personal Navigation
  Applications},''
  \url{http://www.analog.com/media/en/technical-documentation/application-notes/513772624AN602.pdf},
  2002.

\bibitem{17_5}
X.~Teng, D.~Guo, Y.~Guo, X.~Zhou, Z.~Ding, and Z.~liu, ``Ionavi: An
  indoor-outdoor navigation service via mobile crowdsensing,'' \emph{ACM
  Transactions on Sensor Networks}, vol.~13, 01 2017.

\bibitem{bats}
``{Echolocation},'' \url{http://www.bats.org.uk/pages/echolocation.html}.

\bibitem{91_1}
T.~Ifukube, T.~Sasaki, and C.~Peng, ``A blind mobility aid modeled after
  echolocation of bats,'' \emph{IEEE Transactions on Biomedical Engineering},
  vol.~38, no.~5, pp. 461--465, May 1991.

\bibitem{17_4}
\BIBentryALTinterwordspacing
B.~Zhou, M.~Elbadry, R.~Gao, and F.~Ye, ``Batmapper: Acoustic sensing based
  indoor floor plan construction using smartphones,'' in \emph{Proceedings of
  the 15th Annual International Conference on Mobile Systems, Applications, and
  Services}, ser. MobiSys '17.\hskip 1em plus 0.5em minus 0.4em\relax New York,
  NY, USA: ACM, 2017, pp. 42--55. [Online]. Available:
  \url{http://doi.acm.org/10.1145/3081333.3081363}
\BIBentrySTDinterwordspacing

\bibitem{1_3}
\BIBentryALTinterwordspacing
S.~B. Modi, P.~Chandak, V.~S. Murty, and E.~L. Hall, ``Comparison of three
  obstacle-avoidance methods for a mobile robot,'' 2001. [Online]. Available:
  \url{https://doi.org/10.1117/12.444194}
\BIBentrySTDinterwordspacing

\bibitem{7_2}
\BIBentryALTinterwordspacing
S.~Cardin, D.~Thalmann, and F.~Vexo, ``A wearable system for mobility
  improvement of visually impaired people,'' \emph{The Visual Computer},
  vol.~23, no.~2, pp. 109--118, Feb 2007. [Online]. Available:
  \url{https://doi.org/10.1007/s00371-006-0032-4}
\BIBentrySTDinterwordspacing

\bibitem{7_11}
B.-S. Shin and C.-S. Lim, ``Obstacle detection and avoidance system for
  visually impaired people.''

\bibitem{13_7}
\BIBentryALTinterwordspacing
I.~Ercoli, P.~Marchionni, and L.~Scalise, ``A wearable multipoint ultrasonic
  travel aids for visually impaired,'' \emph{Journal of Physics: Conference
  Series}, vol. 459, no.~1, p. 012063, 2013. [Online]. Available:
  \url{http://stacks.iop.org/1742-6596/459/i=1/a=012063}
\BIBentrySTDinterwordspacing

\bibitem{7_14}
D.-W. Jung, Z.-S. Lim, B.-G. Kim, and N.-K. Kim, ``Multi-channel ultrasonic
  sensor system for obstacle detection of the mobile robot,'' in \emph{2007
  International Conference on Control, Automation and Systems}, Oct 2007, pp.
  2347--2351.

\bibitem{4_3}
D.~Aguerrevere, M.~Choudhury, and A.~Barreto, ``Portable 3d sound / sonar
  navigation system for blind individuals,'' 01 2004.

\bibitem{1_1}
\BIBentryALTinterwordspacing
I.~Ulrich and J.~Borenstein, ``The guidecane-applying mobile robot technologies
  to assist the visually impaired,'' \emph{Trans. Sys. Man Cyber. Part A},
  vol.~31, no.~2, pp. 131--136, Mar. 2001. [Online]. Available:
  \url{http://dx.doi.org/10.1109/3468.911370}
\BIBentrySTDinterwordspacing

\bibitem{14_12}
M.~S. Sadi, S.~Mahmud, M.~M. Kamal, and A.~I. Bayazid, ``Automated walk-in
  assistant for the blinds,'' in \emph{2014 International Conference on
  Electrical Engineering and Information Communication Technology}, April 2014,
  pp. 1--4.

\bibitem{15_13}
M.~Rey, I.~Hertzog, N.~Kagami, and L.~Nedel, ``Blind guardian: A sonar-based
  solution for avoiding collisions with the real world,'' in \emph{2015 XVII
  Symposium on Virtual and Augmented Reality}, May 2015, pp. 237--244.

\bibitem{18_1}
K.~Patil, Q.~Jawadwala, and F.~C. Shu, ``Design and construction of electronic
  aid for visually impaired people,'' \emph{IEEE Transactions on Human-Machine
  Systems}, vol.~PP, no.~99, pp. 1--11, 2018.

\bibitem{94_1}
S.~Shoval, J.~Borenstein, and Y.~Koren, ``Mobile robot obstacle avoidance in a
  computerized travel aid for the blind,'' in \emph{Proceedings of the 1994
  IEEE International Conference on Robotics and Automation}, May 1994, pp.
  2023--2028 vol.3.

\bibitem{14_15}
\BIBentryALTinterwordspacing
J.~C. Fernandez-Diaz, W.~E. Carter, R.~L. Shrestha, and C.~L. Glennie, ``Now
  you see it… now you don’t: Understanding airborne mapping lidar
  collection and data product generation for archaeological research in
  mesoamerica,'' \emph{Remote Sensing}, vol.~6, no.~10, pp. 9951--10\,001,
  2014. [Online]. Available: \url{http://www.mdpi.com/2072-4292/6/10/9951}
\BIBentrySTDinterwordspacing

\bibitem{10_13}
D.~Droeschel, D.~Holz, J.~Stückler, and S.~Behnke, ``Using time-of-flight
  cameras with active gaze control for 3d collision avoidance,'' in \emph{2010
  IEEE International Conference on Robotics and Automation}, May 2010, pp.
  4035--4040.

\bibitem{15_10}
Y.~Peng, D.~Qu, Y.~Zhong, S.~Xie, J.~Luo, and J.~Gu, ``The obstacle detection
  and obstacle avoidance algorithm based on 2-d lidar,'' in \emph{2015 IEEE
  International Conference on Information and Automation}, Aug 2015, pp.
  1648--1653.

\bibitem{8_8}
Y.-C. Chang, H.~Kuwabara, and Y.~Yamamoto, ``Novel application of a laser range
  finder with vision system for wheeled mobile robot,'' in \emph{2008 IEEE/ASME
  International Conference on Advanced Intelligent Mechatronics}, July 2008,
  pp. 280--285.

\bibitem{12_18}
S.~Kumpakeaw, ``Twin low-cost infrared range finders for detecting obstacles
  using in mobile platforms,'' in \emph{2012 IEEE International Conference on
  Robotics and Biomimetics (ROBIO)}, Dec 2012, pp. 1996--1999.

\bibitem{17_9}
N.~{Baklouti} and A.~M. {Alimi}, ``Interval type-2 beta fuzzy neural network
  for wheeled mobile robots obstacles avoidance,'' in \emph{2017 International
  Conference on Advanced Systems and Electric Technologies}, Jan 2017, pp.
  481--486.

\bibitem{9_15}
\BIBentryALTinterwordspacing
I.~Susnea, V.~Minzu, and G.~Vasiliu, ``Simple, real-time obstacle avoidance
  algorithm for mobile robots,'' in \emph{Proceedings of the 8th WSEAS
  International Conference on Computational Intelligence, Man-machine Systems
  and Cybernetics}, ser. CIMMACS'09.\hskip 1em plus 0.5em minus 0.4em\relax
  Stevens Point, Wisconsin, USA: World Scientific and Engineering Academy and
  Society (WSEAS), 2009, pp. 24--29. [Online]. Available:
  \url{http://dl.acm.org/citation.cfm?id=1736097.1736102}
\BIBentrySTDinterwordspacing

\bibitem{15_11}
A.~M. Alajlan, M.~M. Almasri, and K.~M. Elleithy, ``Multi-sensor based
  collision avoidance algorithm for mobile robot,'' in \emph{2015 Long Island
  Systems, Applications and Technology}, May 2015, pp. 1--6.

\bibitem{17_8B}
J.~Bai, S.~Lian, Z.~Liu, K.~Wang, and D.~Liu, ``Smart guiding glasses for
  visually impaired people in indoor environment,'' \emph{IEEE Transactions on
  Consumer Electronics}, vol.~63, no.~3, pp. 258--266, August 2017.

\bibitem{7_15}
A.~J. Davison, I.~D. Reid, N.~D. Molton, and O.~Stasse, ``Monoslam: Real-time
  single camera slam,'' \emph{IEEE Transactions on Pattern Analysis and Machine
  Intelligence}, vol.~29, no.~6, pp. 1052--1067, June 2007.

\bibitem{17_2}
B.-S. Lin, C.-C. Lee, and P.-Y. Chiang, ``Simple smartphone-based guiding
  system for visually impaired people,'' \emph{Sensors}, 2017.

\bibitem{93_1}
D.~P. Huttenlocher, G.~A. Klanderman, and W.~J. Rucklidge, ``Comparing images
  using the hausdorff distance,'' \emph{IEEE Transactions on Pattern Analysis
  and Machine Intelligence}, vol.~15, no.~9, pp. 850--863, Sep 1993.

\bibitem{0_2}
M.~D. Gavrila, ``Pedestrian detection from a moving vehicle,'' \emph{Springer,
  Berlin, Heidelberg}, 2000.

\bibitem{98_5}
T.~Kalinke, C.~Tzomakas, and W.~V. Seelen, ``A texture-based object detection
  and an adaptive model-based classification,'' in \emph{in Procs. IEEE
  Intelligent Vehicles Symposium‘98}, 1998, pp. 341--346.

\bibitem{89_3}
\BIBentryALTinterwordspacing
G.~Marola, ``Using symmetry for detecting and locating objects in a picture,''
  \emph{Comput. Vision Graph. Image Process.}, vol.~46, no.~2, pp. 179--195,
  May 1989. [Online]. Available:
  \url{http://dx.doi.org/10.1016/0734-189X(89)90168-0}
\BIBentrySTDinterwordspacing

\bibitem{0_3}
\BIBentryALTinterwordspacing
I.~Ulrich and I.~R. Nourbakhsh, ``Appearance-based obstacle detection with
  monocular color vision,'' in \emph{Proceedings of the Seventeenth National
  Conference on Artificial Intelligence and Twelfth Conference on Innovative
  Applications of Artificial Intelligence}.\hskip 1em plus 0.5em minus
  0.4em\relax AAAI Press, 2000, pp. 866--871. [Online]. Available:
  \url{http://dl.acm.org/citation.cfm?id=647288.721755}
\BIBentrySTDinterwordspacing

\bibitem{15_12}
M.~C. Kang, S.~H. Chae, J.~Y. Sun, J.~W. Yoo, and S.~J. Ko, ``A novel obstacle
  detection method based on deformable grid for the visually impaired,''
  \emph{IEEE Transactions on Consumer Electronics}, vol.~61, no.~3, pp.
  376--383, Aug 2015.

\bibitem{17_11}
M.~C. Kang, S.~H. Chae, J.~Y. Sun, S.~H. Lee, and S.~J. Ko, ``An enhanced
  obstacle avoidance method for the visually impaired using deformable grid,''
  \emph{IEEE Transactions on Consumer Electronics}, vol.~63, no.~2, pp.
  169--177, May 2017.

\bibitem{6_11}
H.~Wang, Q.~Chen, and W.~Cai, ``Shape-based pedestrian/bicyclist detection via
  onboard stereo vision,'' in \emph{The Proceedings of the Multiconference on
  "Computational Engineering in Systems Applications"}, vol.~2, Oct 2006, pp.
  1776--1780.

\bibitem{98_6}
K.~Konolige, ``Small vision systems: Hardware and implementation,'' in
  \emph{Robotics Research}, Y.~Shirai and S.~Hirose, Eds.\hskip 1em plus 0.5em
  minus 0.4em\relax London: Springer London, 1998, pp. 203--212.

\bibitem{stereoVision}
``{Stereo Vision Geometry},''
  \url{http://inside.mines.edu/~whoff/courses/EENG512}.

\bibitem{14_11}
F.~Navarro, S.~Cancino, and E.~Estupinan, ``Depth estimation for visually
  impaired people using mobile devices,'' in \emph{2014 IEEE 5th Latin American
  Symposium on Circuits and Systems}, Feb 2014, pp. 1--4.

\bibitem{12_16}
\BIBentryALTinterwordspacing
A.~Rodr{\'i}guez, J.~J. Yebes, P.~F. Alcantarilla, L.~M. Bergasa,
  J.~Almaz{\'a}n, and A.~Cela, ``Assisting the visually impaired: Obstacle
  detection and warning system by acoustic feedback,'' \emph{Sensors (Basel)},
  vol.~12, no.~12, pp. 17\,476--17\,496, Dec 2012, sensors-12-17476[PII].
  [Online]. Available:
  \url{http://www.ncbi.nlm.nih.gov/pmc/articles/PMC3571849/}
\BIBentrySTDinterwordspacing

\bibitem{91_3}
J.~Borenstein and Y.~Koren, ``The vector field histogram-fast obstacle
  avoidance for mobile robots,'' \emph{IEEE Transactions on Robotics and
  Automation}, vol.~7, no.~3, pp. 278--288, Jun 1991.

\bibitem{5_5}
S.~Meers and K.~Ward, ``A substitute vision system for providing 3d perception
  and gps navigation via electro-tactile stimulation,'' 03 2018.

\bibitem{9_16}
M.~A. Ibarra-Manzano, D.~L. Almanza-Ojeda, M.~Devy, J.~L. Boizard, and J.~Y.
  Fourniols, ``Stereo vision algorithm implementation in fpga using census
  transform for effective resource optimization,'' in \emph{2009 12th Euromicro
  Conference on Digital System Design, Architectures, Methods and Tools}, Aug
  2009, pp. 799--805.

\bibitem{94_3}
R.~Zabih and J.~Woodfill, ``Non-parametric local transforms for computing
  visual correspondence,'' in \emph{Computer Vision --- ECCV '94}, J.-O.
  Eklundh, Ed.\hskip 1em plus 0.5em minus 0.4em\relax Berlin, Heidelberg:
  Springer Berlin Heidelberg, 1994, pp. 151--158.

\bibitem{13_9}
M.~Samadi, M.~F. Othman, and S.~H.~M. Amin, ``Stereo vision based robots: Fast
  and robust obstacle detection method,'' in \emph{2013 9th Asian Control
  Conference (ASCC)}, June 2013, pp. 1--5.

\bibitem{16_8}
S.~Aymaz and T.~Cavdar, ``Ultrasonic assistive headset for visually impaired
  people,'' in \emph{2016 39th International Conference on Telecommunications
  and Signal Processing}, June 2016, pp. 388--391.

\bibitem{16_9}
L.~B. Neto, F.~Grijalva, V.~R. M.~L. Maike, L.~C. Martini, D.~Florencio,
  M.~C.~C. Baranauskas, A.~Rocha, and S.~Goldenstein, ``A kinect-based wearable
  face recognition system to aid visually impaired users,'' \emph{IEEE
  Transactions on Human-Machine Systems}, vol.~47, no.~1, pp. 52--64, Feb 2017.

\bibitem{15_15}
F.~Lan, G.~Zhai, and W.~Lin, ``Lightweight smart glass system with audio aid
  for visually impaired people,'' in \emph{TENCON 2015 - 2015 IEEE Region 10
  Conference}, Nov 2015, pp. 1--4.

\bibitem{16_10}
H.~Saleous, A.~Shaikh, R.~Gupta, and A.~Sagahyroon, ``Read2me: A cloud-based
  reading aid for the visually impaired,'' in \emph{2016 International
  Conference on Industrial Informatics and Computer Systems (CIICS)}, March
  2016, pp. 1--6.

\bibitem{17_12}
P.~A. Zientara, S.~Lee, G.~H. Smith, R.~Brenner, L.~Itti, M.~B. Rosson, J.~M.
  Carroll, K.~M. Irick, and V.~Narayanan, ``Third eye: A shopping assistant for
  the visually impaired,'' \emph{Computer}, vol.~50, no.~2, pp. 16--24, Feb
  2017.

\bibitem{16_11}
M.~Poggi and S.~Mattoccia, ``A wearable mobility aid for the visually impaired
  based on embedded 3d vision and deep learning,'' in \emph{2016 IEEE Symposium
  on Computers and Communication (ISCC)}, June 2016, pp. 208--213.

\bibitem{16_12}
D.~Munteanu and R.~Ionel, ``Voice-controlled smart assistive device for
  visually impaired individuals,'' in \emph{2016 12th IEEE International
  Symposium on Electronics and Telecommunications (ISETC)}, Oct 2016, pp.
  186--190.

\bibitem{16_13}
H.~Rashid, A.~S. M.~R. Al-Mamun, M.~S.~R. Robin, M.~Ahasan, and S.~M.~T. Reza,
  ``Bilingual wearable assistive technology for visually impaired persons,'' in
  \emph{2016 International Conference on Medical Engineering, Health
  Informatics and Technology (MediTec)}, Dec 2016, pp. 1--6.

\bibitem{16_14}
Z.~Liu, Y.~Luo, J.~Cordero, N.~Zhao, and Y.~Shen, ``Finger-eye: A wearable text
  reading assistive system for the blind and visually impaired,'' in \emph{2016
  IEEE International Conference on Real-time Computing and Robotics (RCAR)},
  June 2016, pp. 123--128.

\bibitem{16_15}
F.~Ahmad, Tanveerulhaq, I.~Ishaq, D.~Ali, and M.~F. Riaz, ``Bionic kinect
  device to assist visually impaired people by haptic and voice feedback,'' in
  \emph{2016 International Conference on Bio-engineering for Smart Technologies
  (BioSMART)}, Dec 2016, pp. 1--4.

\bibitem{17_13}
T.~Froneman, D.~van~den Heever, and K.~Dellimore, ``Development of a wearable
  support system to aid the visually impaired in independent mobilization and
  navigation,'' in \emph{2017 39th Annual International Conference of the IEEE
  Engineering in Medicine and Biology Society (EMBC)}, July 2017, pp. 783--786.

\bibitem{17_14}
H.~Peiris, C.~Kulasekara, H.~Wijesinghe, B.~Kothalawala, N.~Walgampaya, and
  D.~Kasthurirathna, ``Eyevista: An assistive wearable device for visually
  impaired sprint athletes,'' in \emph{2016 IEEE International Conference on
  Information and Automation for Sustainability (ICIAfS)}, Dec 2016, pp. 1--6.

\bibitem{17_15}
H.~C. Wang, R.~K. Katzschmann, S.~Teng, B.~Araki, L.~Giarré, and D.~Rus,
  ``Enabling independent navigation for visually impaired people through a
  wearable vision-based feedback system,'' in \emph{2017 IEEE International
  Conference on Robotics and Automation (ICRA)}, May 2017, pp. 6533--6540.

\bibitem{14_13}
S.~Garg, R.~K. Singh, R.~R. Saxena, and R.~Kapoor, ``Doppler effect: Ui input
  method using gestures for the visually impaired,'' in \emph{2014 Texas
  Instruments India Educators' Conference (TIIEC)}, April 2014, pp. 86--92.

\bibitem{17_16}
J.~Monteiro, J.~P. Aires, R.~Granada, R.~C. Barros, and F.~Meneguzzi, ``Virtual
  guide dog: An application to support visually-impaired people through deep
  convolutional neural networks,'' in \emph{2017 International Joint Conference
  on Neural Networks (IJCNN)}, May 2017, pp. 2267--2274.

\bibitem{15_16}
S.~Gupta, I.~Sharma, A.~Tiwari, and G.~Chitranshi, ``Advanced guide cane for
  the visually impaired people,'' in \emph{2015 1st International Conference on
  Next Generation Computing Technologies (NGCT)}, Sept 2015, pp. 452--455.

\bibitem{15_17}
N.~B. James and A.~Harsola, ``Navigation aiding stick for the visually
  impaired,'' in \emph{2015 International Conference on Green Computing and
  Internet of Things (ICGCIoT)}, Oct 2015, pp. 1254--1257.

\bibitem{17_17}
S.~Sharma, M.~Gupta, A.~Kumar, M.~Tripathi, and M.~S. Gaur, ``Multiple distance
  sensors based smart stick for visually impaired people,'' in \emph{2017 IEEE
  7th Annual Computing and Communication Workshop and Conference (CCWC)}, Jan
  2017, pp. 1--5.

\bibitem{16_16}
D.~Jayashree, K.~A. Farhath, R.~Amruthavarshini, and S.~Pavithra, ``Voice based
  application as medicine spotter for visually impaired,'' in \emph{2016 Second
  International Conference on Science Technology Engineering and Management
  (ICONSTEM)}, March 2016, pp. 56--60.

\bibitem{17_18}
H.~Jiang, T.~Gonnot, W.~J. Yi, and J.~Saniie, ``Computer vision and text
  recognition for assisting visually impaired people using android
  smartphone,'' in \emph{2017 IEEE International Conference on Electro
  Information Technology (EIT)}, May 2017, pp. 350--353.

\bibitem{15_14}
K.~C. Liu, C.~H. Wu, S.~Y. Tseng, and Y.~T. Tsai, ``Voice helper: A mobile
  assistive system for visually impaired persons,'' in \emph{2015 IEEE
  International Conference on Computer and Information Technology; Ubiquitous
  Computing and Communications; Dependable, Autonomic and Secure Computing;
  Pervasive Intelligence and Computing}, Oct 2015, pp. 1400--1405.

\bibitem{15_18}
S.~M.~R. Bagwan and L.~J. Sankpal, ``Visualpal: A mobile app for object
  recognition for the visually impaired,'' in \emph{2015 International
  Conference on Computer, Communication and Control (IC4)}, Sept 2015, pp.
  1--6.

\bibitem{12_19}
M.~Modzelewski and E.~B. Kaiser, ``Hand gesture recognition interface for
  visually impaired and blind people,'' in \emph{2012 IEEE/ACIS 11th
  International Conference on Computer and Information Science}, May 2012, pp.
  201--206.

\bibitem{16_17}
P.~A. Dhulekar, N.~Prajapatr, T.~A. Tribhuvan, and K.~S. Godse, ``Automatic
  voice generation system after street board identification for visually
  impaired,'' in \emph{2016 International Conference on Global Trends in Signal
  Processing, Information Computing and Communication (ICGTSPICC)}, Dec 2016,
  pp. 91--96.

\bibitem{17_19}
K.~Matsuda and K.~Kondo, ``Towards an accurate route guidance system for the
  visually impaired using 3d audio,'' in \emph{2017 IEEE 6th Global Conference
  on Consumer Electronics (GCCE)}, Oct 2017, pp. 1--2.

\bibitem{17_20}
R.~S. Dasila, M.~Trivedi, S.~Soni, M.~Senthil, and M.~Narendran, ``Real time
  environment perception for visually impaired,'' in \emph{2017 IEEE
  Technological Innovations in ICT for Agriculture and Rural Development
  (TIAR)}, April 2017, pp. 168--172.

\end{thebibliography}

	\newpage
	\appendix
\begin{appendices}
	
\begin{center}
\begin{tabular}{ |c|c| } 
 \hline
 Abbreviation & Explanation  \\ 
 \hline
 GPS &  Global  Positioning  System  \\ 
 WSN & Wireless  Sensor Network  \\ 
 GNSS & Global Navigation Satellite System \\
 TDOA & Time Difference of Arrival \\
 ETA & Electronic Travel Aid \\
 RFID & Radio Frequency Identification \\
 AP & Access Point \\
 MAC & Media Access Control \\
 TOA & Time of Arrival \\
 AOA & Angle of Arrival \\
 RSS & Received Signal Strength \\
 LOS & Line of Sight \\
 RF & Radio Frequency \\
 UWB & Ultra Wide Band \\
 DOA & Direction of Arrival \\
 NN & Nearest Neighbor \\
 KNN & K Nearest Neighbor \\
 WKNN & Weighted K Nearest Neighbor \\
 SVM & Support Vector Machine \\
 CFR & Channel Frequency Response \\
 EKF & Extended Kalman Filter \\
 UKF & Unscented Kalman Filter \\
 EPF & Extended  Particle  Filter \\
 BLE & Bluetooth  Low  Energy \\
 PDR & Pedestrian Dead Reckoning \\
 CTP & Conventional Terrestrial Pole \\
 DCM & Direction Cosine Matrix \\
 PCA & Principle Component Analysis \\
 FLAM & Forward and Lateral Accelerations Modeling \\
 FIS & Frequency  analysis  of  Inertial  Signals \\
 IMU &  Inertial  Measuring  Units \\
 MEMS & Micro  Electro-Mechanical  Systems \\
 ZUPT & Zero Velocity Update \\
 ZARU & Zero Angular Rate Update \\
 ZUNA & Zero Unrefined Acceleration Update \\
 HDR & Heuristic  Drift  Reduction \\
 LIDAR & Light Detection and Ranging \\
 VFH & Vector  Field  Histogram \\
 RANSAC & Random Sample Consensus \\
 \hline
\end{tabular}
\end{center}
	\end{appendices}
	
	\end{document}